\documentclass[preprint,11pt,5p,twocolumn]{elsarticle}
\usepackage{amssymb, nccmath}
\usepackage{url,hyperref}
\usepackage{graphicx}
\usepackage{subcaption} 
\usepackage{mathtools}
\usepackage{amsmath}
\usepackage{blkarray}
\usepackage{rotating}
\usepackage{svg}
\usepackage{enumerate}
\usepackage{enumitem}
\usepackage{algorithmic}
\usepackage{booktabs}
\usepackage{algorithm}
\usepackage{color}
\usepackage{xcolor}
\usepackage{multirow}
\usepackage{lineno}
\usepackage{picins}
\usepackage[utf8]{inputenc}
\usepackage{makecell}
\usepackage{multicol}
\definecolor{mypink3}{cmyk}{0, 0.47808, 0.10429, 0.012}
\journal{Elsevier}

\usepackage{amssymb}
\usepackage{multicol}
\usepackage{fancyhdr}

\begin{document}
\begin{frontmatter}
\title{Equilibrium in the Computing Continuum through Active Inference}

\author{Boris Sedlak\corref{cor1}}
\ead{boris.sedlak@dsg.tuwien.ac.at}
\author{Victor Casamayor Pujol}
\ead{v.casamayor@dsg.tuwien.ac.at}
\author{Praveen Kumar Donta}
\ead{pdonta@dsg.tuwien.ac.at}
\author{Schahram Dustdar}
\ead{dustdar@dsg.tuwien.ac.at}
\address{Distributed Systems Group, TU Wien, 1040 Vienna, Austria}
\cortext[cor1]{Corresponding author}

\begin{abstract}
Computing Continuum (CC) systems are challenged to ensure the intricate requirements of each computational tier.
Given the system's scale, the Service Level Objectives (SLOs) which are expressed as these requirements, must be broken down into smaller parts that can be decentralized. 
We present our framework for collaborative edge intelligence enabling individual edge devices to (1) develop a causal understanding of how to enforce their SLOs, and (2) transfer knowledge to speed up the onboarding of heterogeneous devices. Through collaboration, they (3) increase the scope of SLO fulfillment.
We implemented the framework and evaluated a use case in which a CC system is responsible for ensuring Quality of Service (QoS) and Quality of Experience (QoE) during video streaming. Our results showed that edge devices required only ten training rounds to ensure four SLOs; furthermore, the underlying causal structures were also rationally explainable. The addition of new types of devices can be done a posteriori, the framework allowed them to reuse existing models, even though the device type had been unknown. Finally, rebalancing the load within a device cluster allowed individual edge devices to recover their SLO compliance after a network failure from 22\% to 89\%.
\end{abstract}

\begin{keyword}
Active Inference, Computing Continuum, Scalability, Edge Intelligence, Transfer Learning, Equilibrium 
\end{keyword}
\end{frontmatter}

\section{Introduction}


Computing Continuum (CC) systems as envisioned in~\cite{Beckman2020,dustdar_distributed_2023,tarneberg_6g_2022} are large-scale distributed systems composed of multiple computational tiers. Each tier serves a unique purpose, e.g., providing latency-sensitive services (i.e., Edge), or an abundance of virtual, scalable resources (i.e., Cloud). However, the requirements that each tier must fulfill are equally diverse, as they span a wide variety of edge devices and fog nodes. Assume that requirements would be ensured in the cloud, e.g., by analyzing metrics and reconfiguring individual devices, massive amounts of data would have to be transferred. Also, if edge devices fail to provide their service to a satisfying degree, the latency for detecting and resolving this would be high.

Given the scale of the CC, requirements must be decentralized; this means, that the logic to evaluate requirements must be transferred to the component that they concern. Cloud-level requirements, i.e., Service Level Objectives (SLOs), may thus be broken down into smaller parts that are ensured by the respective components.
To contribute to high-level goals, each device optimizes its service according to its scope.
This allows SLOs to span the entire CC, also called Deep SLOs~\cite{casamayor-pujol_fundamental_2023}.
%
While it is one challenge to segregate and disseminate SLOs, ensuring them is another. Requirements are versatile and may change over time, every component must itself discover how its SLOs are related to its actions. For this to happen, the device could use Machine Learning (ML) techniques to discover causal relations between its environment and SLO fulfillment~\cite{sedlak_designing_2023}. This promotes the usage of Active Inference (ACI)~\cite{friston_free_2023}, a concept from neuroscience that describes how the brain continuously predicts and evaluates sensory information to model real-world processes. Given these causal models, components could adjust their environment according to preferences (i.e., SLOs).

Ensuring SLOs autonomously (i.e., evaluating the environment to infer adaptations) makes components intelligent~\cite{kokkonen_autonomy_2023}; any system (or subsystem) composed entirely of such intelligent, self-contained components becomes more resilient and reliable. No central logic must be employed to ensure SLOs; thus, higher-level components can rely on SLO fulfillment of underlying components. Ascending from intelligent edge devices, the next level would be intelligent fog nodes; those we see in the ideal position to orchestrate the service of edge devices.
Thereby, edge devices in proximity are bundled into a device cluster, administered by a fog node; whenever the Edge is scaled up with new devices (or device types), existing SLO-compliance models can be exchanged within the cluster. 
While each tier has its own SLOs, their tools for adaptation can have a different scale, e.g., fog nodes would be able to shift computations within clusters, from devices that fail their SLOs. Such operations can consider environmental impacts (e.g., network issues), but also heterogeneous device characteristics. 

To realize this vision, we present our framework for collaborative edge intelligence. Guided by ACI, individual edge devices gradually develop a causal understanding to ensure their SLO. This knowledge is federated through a device cluster; edge devices of arbitrary types reuse existing models to ensure their SLOs. Thus, the entire Edge becomes spanned with SLO-compliant devices, which allows other CC tiers to construct their service on top. By the same method, cluster leaders (i.e., fog nodes) infer how to adjust their environment; each tier may thus achieve an equilibrium for offering compound service.

Hence, the contributions of this paper are:

\begin{itemize}
    \item An ACI-based ML technique that allows CC components to gradually identify causal relations between environmental metrics and SLO fulfillment. Components can thus evaluate SLOs decentralized and update their beliefs according to new observations.
    
    \item The transfer and combination of ML models between heterogeneous devices to accelerate their convergence towards SLO-fulfilling configurations. This simplifies the onboarding of new device types in the Edge.
    
    \item An offloading mechanism that redistributes load within an edge-fog cluster according to devices' capabilities to fulfill SLOs. It improves cluster-wide QoS and QoE by counterbalancing environmental factors.

\end{itemize}

The remaining sections of this paper are organized as follows:
Section~\ref{sec:Background} introduces background knowledge that is a prerequisite for presented concepts. 
Section~\ref{sec:CEI} presents our framework for collaborative edge intelligence.
Section~\ref{sec:Evaluation} contains the prototypical implementation of the framework and the evaluation methodology;
the respective results are presented in Section~\ref{sec:Results}.
Section~\ref{sec:RelatedWork} provides an overview of existing research in this field.
Finally, we concluded our paper with a future scope in Section~\ref{sec:conclusion}.
For readers' convenience, frequently used notations and acronyms are summarized in Section~\textit{Nomenclature}.

\section{Background}\label{sec:Background}
The framework presented in this paper builds heavily on two existing concepts that we adapt for our usage, namely causality and ACI. Although these topics might be known to some readers, we provide this section to ensure a solid understanding of their core aspects and terminology. Furthermore, since both concepts are not native to computer science (or distributed systems), we highlight intersections with these fields as far as possible. 

\subsection{Causality and Causal Network Graphs}
\label{subsec:background-causal}


Causality allows modeling causal relations between events or variables. While spurious correlations can mislead and hide the true causes, causality answers \textit{why} a specific event happened. However, to identify causal relations, specific experiments and consideration of expert knowledge are required. To define a general theory for causality, Pearl~\cite{pearl_causal_2009} proposed Structural Causal Models (SCMs). Such a mathematical model can be expressed through causal graphs, e.g., as Directed Acyclic Graph (DAG). Thus, variables in the graph can be arranged from cause to consequence. 

Causality is a hot topic in research because of its ability to provide explanations for phenomena through interpretable graphical models. This is why many works link causality and machine learning; see~\cite{ganguly_review_2023} for a comprehensive review.
Thereby, causality can also be embedded into distributed systems, e.g., for root cause detection \cite{chen_causeinfer_2019}. As another instance, \textit{Lin et al.}~\cite{lin_microscope_2018} use causal graphs in Cloud computing to detect dependencies within a microservices-based architecture. For such use cases, DAGs are an ideal modeling tool. Interestingly, they monitor SLOs to trigger causal inference over their causal graphs, being able to detect the source of the SLO violation.

Another crucial concept for our work -- or generally for scalability in the CC -- is the Markov Blanket (MB). Consider a Bayesian network (BN) represented as a DAG (e.g., Fig.~\ref{fig:dag-10}): a random variable is conditionally independent of all other variables, given its MB. In other words, the MB of a variable \textit{shields} it from all external variables. In a DAG, the MB of a variable consists of its parents, children, and co-parents. Discovering the structure of BNs and extracting MBs through data is not a simple task, and many works are devoted to that; see~\cite{tsamardinos_time_2003} or~\cite{niculescu-mizil_inductive_2007} for specific techniques, and~\cite{vowels_ya_2021} for a thorough survey on the topic.
Regardless of the system size, MBs can achieve modularity; thus, the system can be managed and controlled on a convenient scale.

Graph-based causal models promise to extend systems with explainability. Inspired by that, our work stems from~\cite{dustdar_distributed_2023,pujol_towards_2021} to build MBs around SLO-governed components. Thus, it becomes possible to isolate system variables that affect SLO fulfillment. On the one hand, this drastically reduces the number of variables required for analysis thanks to conditional independence; the system can thus be managed at scale. On the other hand, it is possible to leverage the BN to explain causal effects between variables in the MB and the SLOs' behavior (e.g., failure).

\subsection{Active Inference}
\label{subsec:background-aci}


In this work, we use ACI to provide devices with the capacity to build causal knowledge on how to fulfill their SLOs.
However, we consider ACI an unknown concept for most readers outside of neuroscience; therefore, we use this section to summarize core concepts of ACI according to \textit{Friston et al.}~\cite{friston_life_2013,kirchhoff_markov_2018,friston_reinforcement_2009,smith_step-by-step_2022,sajid_active_2021,parr_active_2022}. This includes (1) free energy minimization, (2) hierarchical organization of beliefs, (3) action-perception cycles, and (4) Bayesian inference and belief updating.

\subsubsection{Core Concepts}

To interpret observable processes, agents internally generate models that resemble these processes, e.g., a human could reason that it rains due to water drops falling from the sky. 
However, if this generative model and the real-world process diverge, the agent will eventually be ``surprised", e.g., because water drops were actually caused by a neighbor watering her plants.
The discrepancy (or uncertainty) between the agent's understanding of the process and the reality is called Free Energy (FE). In simple terms: the lower the FE, the higher the prediction accuracy.

More formally, in Eq.~\eqref{eq:surprise} \& \eqref{eq:free-energy}, the surprise $\Im (o|m)$ of observation $o$ given model $m$ is the negative log-likelihood of the observation.
The surprise itself is capped by the FE of the model -- expressed as the Kullback-Leibler divergence ($D_{K\!L}$) between approximate posterior probability ($Q$) of the hidden states ($x$) and their exact posterior probability ($P$).
%
\begin{equation}
    \Im (o|m)= -\ln\!\!{\overbrace{P(o|m)}^\text{Model Evidence}}
    \label{eq:surprise}
\end{equation}
%
%
\begin{equation}
    F[Q,o] = \underbrace{D_{K\!L}[Q(x)||P(x|o,m)] + \Im (o|m)}_\textrm{(Variational) Free Energy} \geq \Im (o|m) 
    \label{eq:free-energy}
\end{equation}


Internally, agents organize generative models in hierarchical structures; each level interprets lower-level causes and, based on that, provides predictions to higher levels. For example, suppose (1) it rains with a certain probability, (2) I bring an umbrella. This is commonly known as Bayesian inference and allows agents to use priors (i.e., existing beliefs) to calculate the probability of related events. Thus, decision processes can be segregated into self-contained causal structures (i.e., MBs) that share only a limited number of interface variables. For example, only the weather state (\textit{rainy} or \textit{sunny}) is considered for picking the umbrella; any lower-level observations that determined the agent's perception of the weather (e.g., humidity or illumination) are disregarded.

To decrease their FE, ACI agents repeatedly engage in action-perception cycles by (1) predicting sensory inputs, (2) awaiting (or seeking) the outcome, and (3) updating beliefs. This phase is widely known as predictive coding. 
Afterward, they can actively adjust the environment toward their beliefs.
As the agent's internal models become increasingly accurate, causal relationships between the environment and its preferences (e.g., SLOs) are revealed. Agents' ability to discover causal relations, however, is very dependent on the number and accuracy of observations~\cite{camps-valls_discovering_2023}. Luckily, the CC provides an infinite amount of operational metrics.

\subsubsection{Intersection with Distributed Systems}

While ACI seems a fitting choice to achieve causality in the CC, there is only limited work on this intersection. To date, most (non-theoretic) research on ACI has not been embedded and evaluated in operative distributed systems (e.g., \cite{smith_step-by-step_2022,heins_pymdp_2022}).
To the best of our knowledge, our latest research \cite{sedlak_active_2023} is thus among the few works that embedded ACI into distributed systems; another work that we want to highlight is \textit{Levchuk et al.}~\cite{levchuk_active_2019}, which created a decentralized mechanism for team adaptation.

For the remaining paper, our work in \cite{sedlak_active_2023} provides valuable reference points; precisely, how ACI agents can ensure SLO-compliant device configurations. Those agents operated parallel to ongoing processing and evaluated rational information (i.e., environmental states) to adapt generative models according to prediction errors. We call such a model -- at its core a BN -- an Equilibrium-Oriented SLO-Compliance (EOSC) model.
In the following, we will extend these EOSC models to achieve equilibrium within the CC.

\section{Collaborative Edge Intelligence}\label{sec:CEI}
To ensure SLOs throughout computational tiers, we propose our framework for collaborative edge intelligence that is constructed upon our three main contributions: (1) The continuous model optimization based on ACI, which ensures SLOs (locally) on a device basis; (2) the federation and combination of EOSC model between edge devices, which decreases the overhead of training models for different device types from scratch; and (3) the evaluation of SLOs on a cluster-level, which can rebalance load within the cluster according to environmental factors.

\begin{figure}[!t]
    \centering
    \includegraphics[width=1.0\columnwidth]{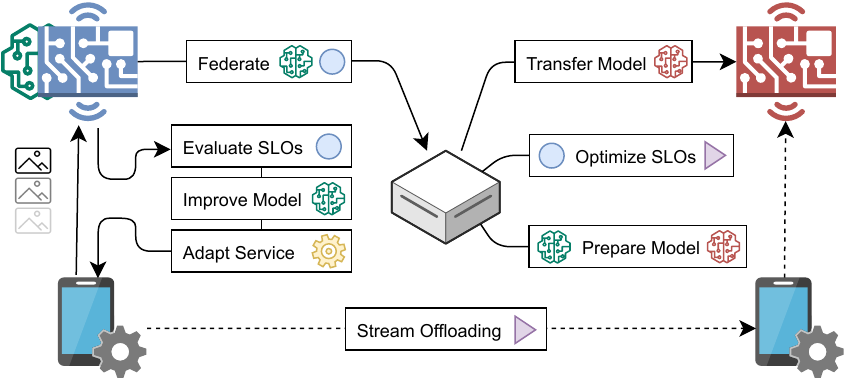}
    \caption{High-level overview of the collaborative edge intelligence framework that continuously improves model evidence, shares this knowledge between edge devices, and optimizes SLO fulfillment within this cluster.}
    \label{fig:high-level}
\end{figure}

These three contributions are described in the respective subsections (\ref{subsec:continuous-model} to \ref{subsec:stream-offloading}); Figure~\ref{fig:high-level} further contains a high-level overview of the framework's capabilities. On the left, it is depicted how SLOs are evaluated to continuously train an ML model and adapt the service accordingly; this model is then federated and combined at a (fog) node, which provides the model to an unknown device type (marked as red). The fog node analyzes the overall SLO fulfillment in the cluster; if it appears beneficial to offload computation from one device to another one (e.g., from the blue to the red one), this is orchestrated by the fog node. Logically, the model transfer and load balancing rely on the SLO fulfillment in the Edge, which is why all three contributions are required to ensure SLOs on multiple tiers (or the entire CC).

\subsection{Continuous Model Optimization}
\label{subsec:continuous-model}

An accurate generative model allows one to explain a system's behavior (e.g., why SLOs were violated), infer how to adapt the system to ensure SLOs, and predict how changes will affect this. Further, prediction errors are always propagated back to the agent so that the model can be improved according to the experienced deviations. In the following, we will first present the representation of the EOSC model and the applied training method. Afterward, this process is integrated into an ACI agent, which uses this process to continuously improve the model accuracy.

\subsubsection{Static Model Training and Inference}
\label{subsubsec:static-model-training}

\begin{figure}[!t]
    \centering
    \includegraphics[width=1.0\columnwidth]{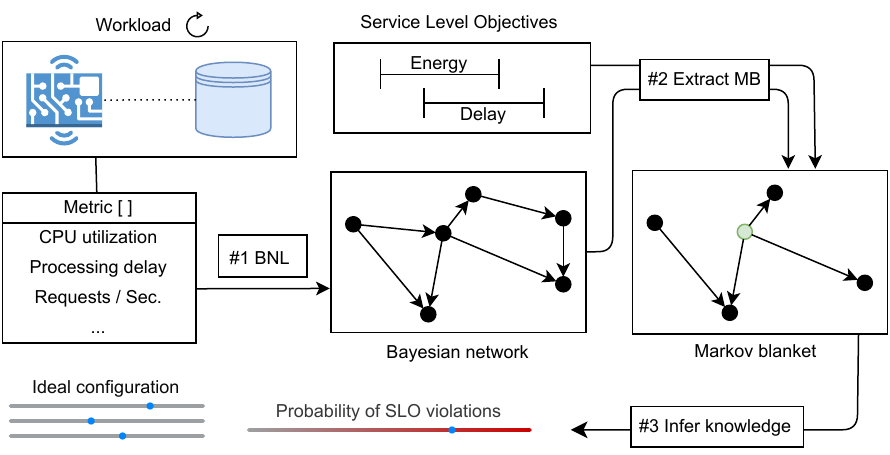}
    \caption{Training a Bayesian Network from processing metrics; this is used to extract the minimum number of variables related to SLO fulfillment and a configuration that satisfies them.}
    \label{fig:static-bnl}
\end{figure}

Within previous work \cite{sedlak_designing_2023}, we presented the idea of obtaining a generative model from processing metrics and inferring system configurations that fulfill SLOs. However, it lacked a formal implementation; this will be the content of this section. Figure~\ref{fig:static-bnl} summarizes our method to train the BN, which is required as a causal structure for our framework:

To report their current state, edge devices produce metrics throughout ongoing processing; this data can be used to create a generative model through Bayesian Network Learning (BNL) (\#1). This reveals (ideally) causal dependencies between variables, including the impact of environmental changes (e.g., increased incoming requests). To decrease the model complexity, we identify a minimum number of variables that are relevant to fulfill system requirements (i.e., SLOs); we call this subset the MB of the BN (\#2). Given the MB, we estimate the probability of SLO violations for different hypothetical scenarios and (\#3) infer the device configuration with the highest statistical compliance level. In the following, we elaborate on these substeps further.

\paragraph{Bayesian Network Learning}
\label{par:bnl}


BNL is an efficient way to generate the most accurate structure from given data; its two main parts are \texttt{STRL} -- structural learning of causal dependencies (i.e., DAG), and \texttt{PARL} -- parameter learning as quantification of variable dependencies.
Structure learning is categorized into constraint-based (e.g., parent-child or grow-shrink) and score-based approaches (e.g., Hill-Climb or genetic algorithm)~\cite{scanagatta2019survey}. In this work, we consider a score-based Hill-Climb Search (HCS) algorithm because of its rapid convergence, low complexity, and efficiency when considering limited attributes. The goal of HCS is to identify the DAG $G_{final}$  from given data $D$, $\exists!\  G_{final} = arg \max\limits_{G \in G^*} score(G,D)$, where $G^*$ indicates the set of possible DAGs, and $score(G,D)$ is calculated using Eq.~(\ref{eq_Score1}) \cite{donta_governance_2023},

\begin{equation}\label{eq_Score1}
    score(S,D) = LL (G,D) \pm \left( \phi(|D|) + ||G||\right)
\end{equation}

where $LL(G,D)$ is computed using Eq.~(\ref{eq_LL}) $\phi(|D|)$ is the Bayesian Information Criterion (BIC) which can be computed as $\phi(|D|) = \frac{1}{2} log(|D|)$, and $||G|| = \sum{(|\mathcal{V}| + |\mathcal{E}|)}$, where $|\mathcal{V}|$ and $|\mathcal{E}|$ denotes number of vertices and edges in DAG $G$, 

\begin{equation}\label{eq_LL}
    LL (G,D) = |D| \times \sum\limits_{i=1}^{|D|} \xi(x_i|Parent(x_i))
\end{equation}

where $|D|$ denotes the number of rows in dataset $D$. $\xi$ denotes entropy; it is calculated as $\sum Pr(x_i|Parent(x_i)) \times \log{(Pr(x_i|Parent(x_i)))}$. 

~

\begin{algorithm}[!t]
	\caption{Hill-Climb search (HCS) algorithm}
	\textbf{INPUT:} $D$, $G_{init}$, $score$ and $\mathcal{C}$ \\
	\text{\textbf{OUTPUT:} $G_{final}$ // Learned Structure} 
    \vspace{-14pt}
	\begin{algorithmic}[1]\label{alg:hcs}
 \STATE $G = G_{init}$, $G_{final} = G_{init}$
	\REPEAT
  \STATE $score = Score(G,D)$
  \FOR{each $(X_i, X_j) \in G \forall \ i,j \leq |G|, i\neq j$}
    \IF{!edge$(X_i, X_j)$}
        \STATE $G_{final} = G_{final} + $ edge$(X_i, X_j)$
        \STATE $score^{'} = Score(G_{final},D)$ using Eq.~{(\ref{eq_Score1})}
        \IF{$score^{'} > score$}
        \STATE $score = score^{'}$
        \ELSE
        \STATE $G_{final} = G_{final} - $ edge$(X_i, X_j)$
        \ENDIF
    \ENDIF
  \ENDFOR
  \UNTIL{$\mathcal{C}$ is satisfied}
  \RETURN $G_{final}$
	\end{algorithmic}
\end{algorithm}

\begin{figure}[t]
\centering
\subfloat[Entire DAG]{\label{subfig:mb-10-filter}\includegraphics[width=.485\columnwidth]{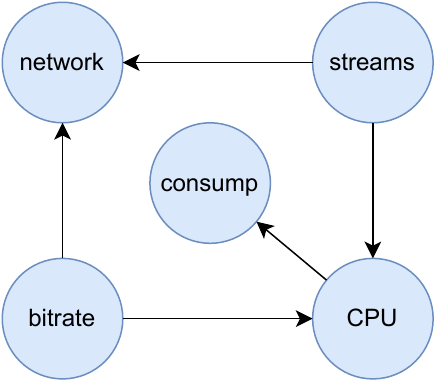}}
\hfill
\subfloat[MB for \textit{network}]{\label{subfig:mb-10-filter-mb}\includegraphics[width=.485\columnwidth]{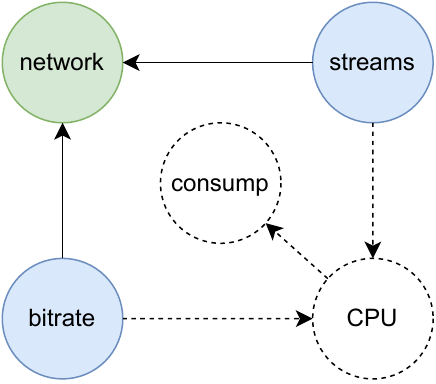}}
\caption{Causal variable relations in the DAG of a trained BN}
\label{fig:dag-10}
\end{figure}

The detailed pseudo-code for the HCS algorithm is presented in Algorithm~\ref{alg:hcs}. HCS starts with an empty graph ($G$) and measures the score using Eq.~(\ref{eq_Score1}). By adding or removing edges between variables, it creates a set of neighboring structures and selects the structure with the highest score. In this way, Algorithm~\ref{alg:hcs} repeats Lines 2-15, until it reaches the maximum score, and chooses the best DAG i.e., $G_{final}$. For a data set with 5 columns, the resulting graph could look like Figure~\ref{subfig:mb-10-filter}. Afterward, as a second step in the BNL process, the variable relations are evaluated: For each node in $G_{final}$, the Conditional Probability Distribution (CPD) is evaluated through Maximum Likelihood Estimation (MLE). The MLE for BNL is calculated using Eq.~(\ref{eq_MLE}),

\begin{equation}\label{eq_MLE}
    MLE(\theta|D) = sup_{\theta} LL(\theta,D)
\end{equation}

where $\theta \in G$, and $LL(\theta,D) = P(D|\theta)$. MLE estimates the model parameters that maximize the likelihood of observed data. This determines the CPD of each node given its parents in the network; the result is stored in the respective Conditional Probability Table (CPT). This concludes parameter learning.

~

\noindent\textit{Lemma 1: } The time complexity of BNL is $\mathcal{O}(n\times 2^\omega)$\\
\textit{Proof:} Calculating time complexity for \texttt{STRL} and \texttt{PARL} is NP-Hard, but there are also some positive tractability results under certain parameterizations \cite{kwisthout2011most,scutari2019learning}.
In general, the time complexity for BNL is $\mathcal{O}(n\times 2^{\omega\ log\ n})$, where $n$ indicates the number of nodes in resulting DAG, and $\omega$ denotes the width of the DAG. However, the algorithm runs similar recursive calls multiple times, and while avoiding identical recursive calls according to \textit{Darwiche}\cite{darwiche2008bayesian}, the BNL needs only $\mathcal{O}(n\times 2^\omega)$ for both structure and parameter learning.

~

The ACI agent will use the presented BNL method for constructing (and later updating) the EOSC model: \texttt{STRL} trains a DAG through HCS; \texttt{PARL} evaluates the conditional variable dependencies through MLE. Together, they can be used to create a BN $model$ from data $D$ as $model = \texttt{PARL}(\texttt{STRL}(D),D)$.

\paragraph{Markov Blanket Selection}
\label{par:mb-selection}

A BN contains by design directed relations and conditional dependencies of random variables; however, to determine the state of an individual node $x$, only a part of the network nodes are influential. This promotes the application of MB \cite{aliferis_short_2010,casamayor_pujol_towards_2021,dustdar_distributed_2023}, which shield a variable from all nodes that are conditionally independent of it. 
Suppose we specify an SLO according to device capabilities (e.g., network throughput $<$ $t$) and evaluate it using a single variable (e.g., \textit{network}), we want to identify metrics related to SLO fulfillment. Namely, these are all variables contained in the MB of \textit{network}; the function $\texttt{MB}(model, network)$ would thus return all blue nodes in Figure~\ref{subfig:mb-10-filter-mb}.

In this context, we distinguish between metrics that statically reflect the system state (e.g., \textit{CPU}), and those that represent a parameterizable variable (e.g., \textit{bitrate}). However, we summarize both using the term "metrics" from a BNL perspective. While static metrics are essential to explain why an SLO is in its current state, only parameterizable ones can be dynamically reconfigured, i.e., they are the possible action states of the ACI agent. Overall, the sum of metrics in the MB provides a clear understanding of why an SLO is in its current state.

~

\noindent\textit{Lemma 2: } The time complexity for MB selection is $\mathcal{O}(\mathcal{P}\times 2^n)$\\
\textit{Proof:} MB of a target variable usually involve three stages \cite{gao2016efficient}. (i) Identify a parent-child (PC) set, which takes more computation i.e., approximately $\mathcal{O}(\mathcal{P}\times 2^{n+1})$, where $n$ is the total number of nodes in DAG, and $\mathcal{P}$ denotes size of conditioned set while searching PC. (ii) Identify \texttt{spouses}: worst case complexity to identify spouses is $\mathcal{O}(\mathcal{C} \times (n-\mathcal{C}))$, where $\mathcal{C}$ is maximum size PC set for $n$. (iii) Extract \texttt{MB}: Finally, to extract MB from previous step in worst case, it needs $\mathcal{O}(\mathcal{S}\mathcal{K} \times \mathcal{C})$, where $\mathcal{S}$, and $\mathcal{K}$ denotes number of spouses, and maximum size spouses set from DAG, respectively. Overall, $\mathcal{O}(\mathcal{P}\times 2^{n+1}) + (\mathcal{C} \times (n-\mathcal{C})) + (\mathcal{S}\mathcal{K} \times \mathcal{C})$, asymptotically $\mathcal{O}(\mathcal{P}\times 2^n)$ is the worst case time complexity for MB selection.

\paragraph{Knowledge Extraction}
\label{par:knowledge-extraction}

There exist two main categories of algorithms for extracting knowledge from BNs, namely Approximate Inference (AxI) and Exact Inference (EI). Given a BN and system requirements (i.e., SLOs), we seek to extract probabilities of SLO violations under different environmental states. This mechanism works equally for different CC tiers; an edge device, for example, could use its BN to answer $P(network > t)$, with $t$ being a custom threshold. For dynamic reconfiguration, we require inference to be (1) accurate, (2) converge reliably, and (3) fast for large networks. We argue that EI and, in particular, Variable Elimination (VE) \cite{zhang_simple_1994} as an instance, fulfill these constraints. In the following, VE is explained:

For a BN with a node set $\{v_1,v_2,v_3,v_4\} \in V$, VE accepts a list of target variables $T=\{v_1,v_2\}$, variable assignments $A=[(v_3:a_3)]$, and an elimination order $O=\{v_4,v_3\}$. The query provides the conditional probabilities of the variables $T$ given assignments $A$. Each variable must either be eliminated or within the target set, thus $\forall v \in V, v \in T \oplus v \in O$.
VE iterates over $O$ and eliminates variables from $V$ while updating the beliefs of the remaining nodes; $V$ thus eventually contains only $T$. In the given case, $v_4$ is eliminated first and $v_3$ second; the difference is the assignment of $v_3$, which introduces evidence in the form of $P(\{v_1,v_2\}|v_3=a_3)$.
While the elimination order has no functional consequence, it is relevant for the efficiency of VE and thus its scalability.

For the following example, recall the DAGs from Figure~\ref{fig:dag-10}: We construct a QoS SLO that is fulfilled if \textit{network} is below $t$ and infer the probability of SLO violations for different variable assignments. To decrease the complexity, we execute VE only on $mb = \texttt{MB}(model,network)$, the node list $V$ thus equals $\{network,streams,bitrate\}$. For later usage, we call VE through $\texttt{INFERENCE}(m_x,T,A)$, where $m_x$ can be any subset of the BN. We execute $\texttt{INFERENCE}$ with $mb$, $T=\{network\}$, $A=[(streams:2),(bitrate:720)]$, and arbitrary $O$. The result contains all conditional probabilities of $network$ given the variable assignment; from which we can extract $P(network>t)$.

~

This will be our central mechanism for identifying the probabilities of SLO violations given a system state.
If an SLO is violated due to an environmental change, e.g., higher $streams$ and thus exceeded $network$, we can compare possible configurations and provide the one with the highest probability of fulfilling the SLO.
In the given example, only \textit{bitrate} can be parameterized (i.e., configured); to fulfill the $network$ SLO, the corresponding measure could thus be to decrease $bitrate$.
This matches our envisioned level of intelligence, i.e., "understanding a situation and reacting according to needs", and neatly fits the principles of elastic computing \cite{dustdar_principles_2011}.

~

\noindent\textit{Lemma 3: } The time complexity for knowledge extraction is $\mathcal{O}(n\times \tau^{\nu+1})$\\
\textit{Proof:} The complexity of knowledge extraction is primarily determined by the VE process, whereas other stages are linear or constant. The VE requires $\mathcal{O}((\digamma+n)\times \tau^\nu)$ \cite{ren2022bayesian}. Here, $\digamma$ is the number of factors in the model $m$, $n$ is the number of nodes in $V$, $\nu$ is the induced width of the elimination order i.e. $|O|$ (for example, $|O|$=2 for our running case study), and $k$ is the maximum cardinality of a variable in $V$. Since $m < n$ and $m$ is chosen as small as possible, we can consider VE's complexity as $\mathcal{O}(n\times \tau^{\nu+1})$. This equals the overall knowledge extraction. 

~

\noindent\textit{Theorem 1: } The time complexity for \textit{Static Model Training and Inference} is $\mathcal{O}(n\times \tau^{\nu+1})$\\
\textit{Proof:} The complexity of \textit{Static Model Training and Inference} depends on BNL, MB selection, and knowledge extraction. The complexity of BNL, MB selection and knowledge extraction are $\mathcal{O}(n\times 2^\omega)$, $\mathcal{O}(\mathcal{P}\times 2^n)$, and $\mathcal{O}(n\times \tau^{\nu+1})$, respectively (according to \textit{Lemma~1} -- \textit{Lemma~3}).  So, the complexity for \textit{Static Model Training and Inference} is $\mathcal{O}(n\times 2^\omega + \mathcal{P}\times 2^n + n\times \tau^{\nu+1})$. We assume, $k>2$, $\mathcal{P}<n$ and $\nu < n$, so the asymptotic complexity of \textit{Static Model Training and Inference} can be concluded as $\mathcal{O}(n\times \tau^{\nu+1})$.


\subsubsection{Active Inference Cycle}
\label{subsub:aci-cycle}

\begin{figure*}[t]
    \centering
    \includegraphics[width=0.9\textwidth]{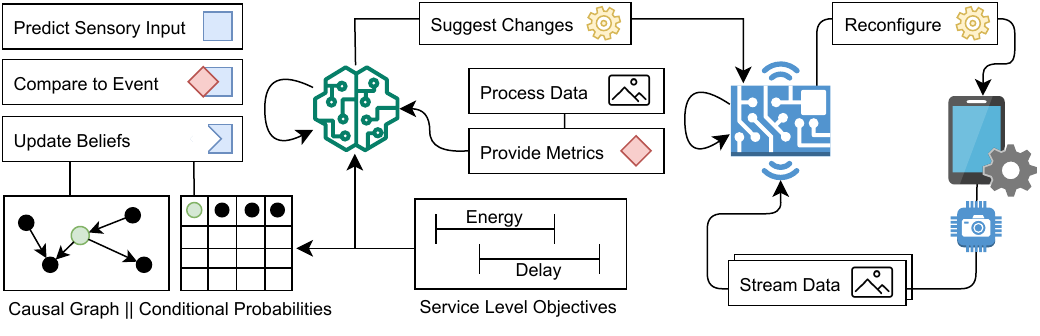}
    \caption{Overview of the Active Inference cycle -- learning how to fulfill SLOs by adapting the generative process}
    \label{fig:aci-overview}
\end{figure*}

The tools presented in the last section allowed creating a BN from processing metrics, extracting an MB, and inferring system configurations that fulfill given SLOs. Supposed there is sufficient data available, BNL can be a one-time process; however, there are two fundamental issues: (1) data shifts, which likely occur after some time, will inevitably distort the accuracy of the ML model, and (2) it is impractical to empirically evaluate how an exponential number of system configuration impacts SLO fulfillment. Large and complex systems, such as the CC, require a different approach, one that creates and updates a model incrementally according to new observations, and at the same time draws conclusions for unknown parameter combinations from existing data.

To evaluate this parameter space of configurations, we extend the ACI agents from \cite{sedlak_designing_2023} to interpolate between empirically evaluated combinations; further, to maintain the model's FE low, the agent continuously updates conditional probabilities of variable relation according to new observations. In theory, our ACI agents can be employed at any CC tier; nevertheless, the running example in this paper is focused on intelligent edge devices, which collaborate under the supervision of a fog node. Thus, we raise the granularity of intelligence from the Edge to the Fog. The causality filter employed by our framework allows to manage each layer alike -- to ensure SLO fulfillment on a higher scale, any upper layers can build upon the SLO compliance of lower levels. In the following, we will present the different tasks and subtasks executed by an Edge-based ACI agent; this includes training and updating the BN, as well as evaluating its scope of actions according to a set of behavioral factors. Based on that, the agent decides if and how to modify the system, which will again be reflected through system metrics.

\paragraph{Agent and Operation}

The ACI agent operates parallel to regular device tasks, e.g., serving clients. Although regular operation, model training, and inference are logically separated, they take place on the same physical device; Figure~\ref{fig:aci-overview} contains a visual representation of that: assume an edge device that continuously performs a workload, e.g., processing data provided by clients. The agent observes the device state and the environment through metrics; thus, it can evaluate whether processing complies with SLOs, e.g., if a request was finished with $delay < t$. From that data, the agent creates a BN (as in Section~\ref{subsubsec:static-model-training}), where conditional probabilities reflect the SLO fulfillment under a discrete environmental state. Then, the agent starts with predictive coding, i.e., forecasting whether future events will fulfill SLOs, comparing the expectation with actual observations, and updating the BN accordingly.

After each iteration, the agent infers how to modify the system configuration to optimize local SLO fulfillment. Following that approach, the ACI agent can create a generative model from scratch
or update a BN according to new observations by following its sensing-acting loop. Thus, it is possible to cancel out data shifts, which might, e.g., be the result of a model transfer from one edge device type to another. ACI can therefore perform the fine-tuning that is required after such an operation.

\paragraph{Model Boundaries}
\label{par:model-boundaries}

The trained model reflects the characteristics of the workload and the environment; in other words, the accuracy will be higher in parameter spaces that are more exploited by the agent. Environmental states that were not present during model learning, or only to a degree that did not allow to identify causal relations, can thus only be treated to a limited extent. This determines the boundaries of the generative model, which manifest in terms of temporal or hierarchical depth. In this context, \textit{Parr et al.}~\cite{parr_active_2022} provided a guideline for model design.

The temporal depth reflects the timely horizon of predictions, e.g., predictions that cover a short period (i.e., a low number of observations) have lower precision than long-term averages. According to \cite{sedlak_designing_2023}, it is a natural choice to align the length of the ACI cycle to the frequency of new samples coming in. For example, if an edge device controls a production engine that produces items in batches of 500 ms, the ACI cycle (i.e., predicting, comparing, and updating beliefs) must be completed within this timeframe.
On the other hand, large and complex BNs, i.e., hierarchically deep ones, require higher computational effort to execute inference queries, while sparse graphs may fail to capture all causal relations~\cite{vowels_ya_2021}; predictions will thus be increasingly inaccurate because dependencies within the environment were not revealed. Nevertheless, executing inference queries on the MB of a target variable (i.e., a subsection of the graph), is a viable way to decrease the query complexity.

Regardless of whether dense or sparse, the DAG and the CPTs, i.e., the main parts of the BN, are under constant optimization; they are the priors of the model -- the initial assumptions that will be updated according to prediction errors to form posterior beliefs.
For example, assume an edge device that processes video streams, for which the ACI agent trains a model (as presented in Figure~\ref{fig:aci-overview}); variables in the BN are related, as shown in Figure~\ref{subfig:mb-10-filter}. Whether the maximum \textit{network} capacity (i.e., a QoS SLO) is reached, is determined by the bitrate and the number of video streams; these, on the other hand, determine the CPU utilization of the device and the energy consumption.
While some variable relations can be updated at will, e.g., the impact of \textit{CPU} $\rightarrow$ \textit{consumption}, others are tied to environmental limitations. For example, whether \textit{bitrate} violates an SLO constructed on \textit{network}, is dependent on the device characteristics. In the latter case, \textit{network} is not immutable but only updated as the environment changes, e.g., because the network interface is upgraded.

\paragraph{Free Energy Minimization}
\label{par:high-level-loop}

To create an accurate model, the ACI agent operates in cycles; each cycle processes a $batch$ of observations that reflects the environmental state, including the latest system configuration. The agent continuously evaluates the $batch$, updates its $model$, and chooses which system configuration ($c_{next} $) to choose for the next iteration.
%
Throughout these cycles, the ACI agent has one central goal: decreasing the FE, or in other words, minimizing surprise of predictions. Therefore, we will first present how we calculate surprise and then embed it into the high-level loop executed by the agent. 

For calculating the surprise for $batch$ and $model$ we present Algorithm~\ref{alg:surprise-calculation}; this can be seen as an implementation of Eq.~\eqref{eq:surprise}. To decrease the complexity, we limit the calculation to variables that directly reflect SLO fulfillment ($V_{SLO}$),
and execute \texttt{INFERENCE} only on the MB of $V_{SLO}$ (Line 2).
This node set is further filtered (Line 5) to contain only the evidence variables ($ev$) that impact the outcome of $var$; afterward, in Line 7, each $row$ in the $batch$ is filtered to contain only these variables. In Lines 8 \& 9, the probability of observing $var$, i.e., the state of the SLO, given the environment ($evidence$) is first inferred and then appended as $log\_likelihood$. For each $var$, the $cpt$ from $model$ is considered, from which $k$ -- the number of states -- can be extracted as a representation of model complexity. \texttt{CPT} is as a helper function to get the CPT for a $var$ in $model$. Together with $n$ -- the number of observations -- the BIC is calculated (Line 14). After calculating the surprise for each $var \times row$, this overall sum is returned.

\begin{algorithm}[t]
\caption{\texttt{SURPRISE} for model and batch}\label{alg:surprise-calculation}
\begin{algorithmic}[1]

\REQUIRE $model$, $batch$, $V_{SLO}$
\ENSURE $\Im$ // surprise over all observations

\STATE $\Im \gets 0$
\STATE $mb \gets \texttt{MB}(model, V_{SLO})$

\FOR{\textbf{each} $var$ \textbf{in} $V_{SLO}$}
    \STATE $log\_likelihood \gets 0$
    \STATE $ev \gets \texttt{MB}(model,var)$

    \FOR{\textbf{each} $row$ \textbf{in} $batch$}
        \STATE $evidence \gets row \cap ev$
        \STATE $p \gets \texttt{INFERENCE}(mb, var, evidence)$
        \STATE $log\_likelihood \gets log\_likelihood + \log(p)$
    \ENDFOR

    \STATE $cpt \gets \texttt{CPT}(model, var)$
    \STATE $k \gets |cpt|$ // number of states in the CPT
    \STATE $n \gets |batch|$
    \STATE $bic \gets (-2) \times log\_likelihood + k \times \log(n)$
    \STATE $\Im \gets \Im + bic$
\ENDFOR

\RETURN $\Im$

\end{algorithmic}
\end{algorithm}

The surprise has a special role within our ACI cycle, as it determines when and how BNL takes place; consider therefore Algorithm~\ref{alg:iterate-function}, which shows the high-level loop executed by the ACI agent. At the beginning of each iteration, the agent ensures that there exists a model, otherwise, it creates an initial structure from $batch$ (Lines 1 \& 2). Notice, that \texttt{STRL} and \texttt{PARL} accept now another parameter -- $model$ -- which allows to update the DAG and CPTs of $model$ according to $batch$. Whether \texttt{STRL} or \texttt{PARL} is executed (Lines 7-11) is determined by the surprise magnitude ($s$). If $s$ exceeds the median surprise of the last 10 rounds ($m_{10}$) by a custom factor $h$, \texttt{STRL} is applied; otherwise, if $s$ exceeds $m_{10}$, \texttt{PARL} is applied. This distinction is necessary because \texttt{STRL} and \texttt{PARL} have quite different runtimes, as we will reveal in Section~\ref{sec:results-and-discussion}. Finally, in Lines 12 \& 13, the agent evaluates possible system configurations and determines which one it will use for the following iteration. We will explain these two functions in the next two paragraphs.

\begin{algorithm}[t]
\caption{An Iteration in the ACI Cycle }\label{alg:iterate-function}
\begin{algorithmic}[1]

\REQUIRE $model$, $batch$, $\Im$, $h$, $V_{SLO}$
\ENSURE $c_{\text{next}}$ // Next configuration

\IF{$model$ = $\emptyset$}
    \STATE $model \gets \texttt{PARL}(\texttt{STRL}(\emptyset, batch), batch)$
\ENDIF

\STATE $s \gets$ \texttt{SURPRISE}($model$, $batch$, $V_{SLO}$)
\STATE $\Im \gets \Im \cup \{s\}$
\STATE $m_{10} \gets median (\Im_{10})$ // over the last 10 values

\IF{$s > (m_{10} \times h)$}
    \STATE $model \gets \texttt{STRL}(model, batch)$
\ELSIF{$s > m_{10}$}
    \STATE $model \gets \texttt{PARL}(model, batch)$
\ENDIF

\STATE $K \gets \texttt{CALCULATE\_FACTORS}(model)$
\STATE $c_{\text{next}} \gets \texttt{BEST\_CONFIGURATION}(K)$
\RETURN $c_{\text{next}}$
\end{algorithmic}
\end{algorithm}

~

\paragraph{Behavioral Factors}
\label{par:behavioral}

The behavior of the ACI agent, i.e., how it selects between possible system configurations, is determined by three major factors: The pragmatic value ($pv$) defines how well the device fulfilled client expectations, e.g., if a streamed video's resolution is satisfactory. The risk assigned ($ra$) determines how likely the system will fail its service, e.g., if the stream packets are delivered on time. Lastly, the information gain ($ig$) represents the agent's expectation of how much it can improve model accuracy. The $ig$ is directly related to surprise minimization, whereas $pv$ and $ra$ reflect the agent's capability to fulfill SLOs. To separate concerns, we divide SLOs according to their characteristics: $pv$ represents QoE requirements, while $ra$ contains QoS requirements. Combined, these three factors determine the behavior of the agent; in the following, we will calculate each of them.

To infer the optimal device configuration (i.e., the one with the highest SLO fulfillment), it would be necessary to evaluate a potentially exponential amount of parameter combinations. As discussed before, this is impractical. To that extent, the agent limits itself to finding the Bayes-optimal configuration ~\cite{ghio_bayes-optimal_2023}, i.e., the optimal under current knowledge. Therefore, the ACI agent first infers the assignment for known parameter combinations ($c_k$) that were empirically evaluated and then interpolates between these values to span the entire parameter space. Calculating $pv$ and $ra$ is similar to Algorithm~\ref{alg:surprise-calculation} (Lines 5-8): It requires a subset $V_Q \subseteq V_{SLO}$ -- either QoS or QoE SLOs -- which is used as $ev \gets \texttt{MB}(model,V_Q)$. For each $row$ in $c_k$, $evidence$ is constructed equally, so that $\texttt{INFERENCE}(mb, V_Q, evidence)$ provides the joint probability of all QoS or QoE violations.



\begin{equation}
    ig (c) = e + \left(\frac{\tilde{\Im}_{c}}{\bar{\Im}}\right) \times 100
    \label{eq:ig}
\end{equation}

In accordance with \cite{sedlak_active_2023}, high surprise indicates high information insight and, hence, possible improvement of the model precision. 
However, from an agent's perspective, is it worth abandoning a supposedly satisfactory configuration (in terms of $pv$ and $ra$) to search for a global optimal one? 
This presents a tradeoff between exploration of unknown areas and the tendency to stick to exploited areas; multi-agent systems commonly model this through hyperparameters (e.g., \cite{levchuk_active_2019}). 
In our case, we calculate the $ig$ of a configuration $c \in c_k$ as presented in Eq.~\eqref{eq:ig} \cite{sedlak_active_2023}: it compares the median surprise ($\tilde{\Im}_{c}$) for $c$ with the overall mean surprise ($\bar{\Im}$). Configurations with high $\tilde{\Im}_{c}$ will thus be preferred by the ACI agent.

\paragraph{Parameter Space}
\label{par:parameter-space}

\begin{figure}[t] 
\centering
\subfloat[Interpolation for $pv$]{\label{subfig:interpolation}\includegraphics[width=.485\columnwidth]{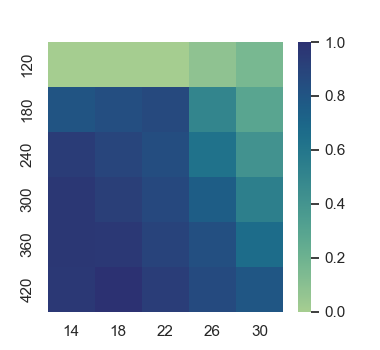}}
\hfill
\subfloat[$ig$ after 1 round]{\label{subfig:ig-boost}\includegraphics[width=.485\columnwidth]{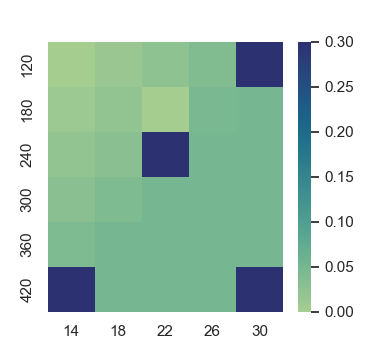}}
\caption{Matrices of behavioral factors used by the ACI agent}
\label{fig:matrices-demo}
\end{figure}

By the presented means, the ACI agent calculates the behavioral factors for all entries in $c_k$ and summarizes them as $K$ (Line 12 of Algorithm~\ref{alg:iterate-function}). For the next step, imagine two configuration parameters $\{fps, pixel\}$ with their combinations arranged in a $[fps \times pixel]$ 2D matrix. After calculating $K$, the unknown spaces in the parameter matrix are filled by performing linear interpolation\footnotemark.
As a resulting example, we provide the matrix depicted in Figure~\ref{subfig:interpolation}. 
Later, in Section~\ref{subsec:implementation}, the agent will interpolate within a 3D parameter space.

\footnotetext{In fact, this will be done using the Python \href{https://scipy.org/}{scypy} package, which triangulates the input data through a convex hull, and performs on each triangle linear barycentric interpolation.}

Contrarily to $pv$ and $ra$, the agent does not apply interpolation to estimate the $ig$ of an unknown parameter configuration. Instead, in the absence of observations for $c$, it assumes that $ig(c) = \max(\Im)$. Further, it remains to introduce a hyperparameter from Eq.~\ref{eq:ig}, namely $e$. To improve the interpolation of $pv$ and $ra$, the agent initially focuses on key positions of the possible configurations. Figure~\ref{subfig:ig-boost} illustrates that tendency; the visually highlighted blocks are increased by $e=0.3$. When calculating the behavioral factors, the ACI agent thus initially focuses on these cornerstones to set up the interpolation; after visiting $c$, it subtracts $e$ from $ig(c)$.

To summarize the possible risks but also benefits that emerge from a configuration $c$, we combine the three factors under a common one ($u$) that is calculated as $u_c = pv_c + ra_c + ig_c$. The ACI agent compares the common factors of all possible configurations and selects the highest-scoring one (Line 13 of Algorithm~\ref{alg:iterate-function}).
By repeating this cycle, the agent gradually develops an understanding of which areas in the parameter space are more likely to fulfill SLOs, e.g., the left-bottom area in Figure~\ref{subfig:interpolation}.

~

\noindent\textit{Theorem 2. } The time complexity for Active Inference Cycle is $\mathcal{O}(n\times 2^\omega)$\\
\textit{Proof. } The time complexity for the Active Inference Cycle is determined by combining the complexities of Agent and Operation, Model Boundaries, FE Minimization, Behavioral Factors, and Parameter Space. Except for FE Minimization, all other stages are computed using constant i.e., $\mathcal{O}(1)$ or linear time i.e., $\mathcal{O}(n)$. FE minimization takes $\mathcal{O}(n^3)$ (according to Zuker's RNA prediction approach \cite{lei2012cpu}). Algorithm~\ref{alg:surprise-calculation} makes a key role in FE minimization and it is mainly depends on number of $V_{SLO}$ and batch size i.e., $|batch|$. So, the asymptotic complexity for Algorithm~\ref{alg:surprise-calculation} is approximately $\mathcal{O}(V_{SLO}\times batch)$ (Note: Constant time excluded.). The complexity of Algorithm~\ref{alg:iterate-function} depends on three functions such as Surprise calculation (\texttt{SURPRISE}), Structure learning (\texttt{STRL}), and Parameter (\texttt{PARL}) learning.
According to \textit{Lemma 1} and \textit{Darwiche}, we can conclude asymptotic complexity for ACI cycle iteration when $model = \emptyset$ and $s> (m_{10}\times h)$ for all cases is determined as  $\mathcal{O}(2\times n\times 2^\omega)$. In summary, the time complexity for active inference cycle through iteration will take $\approx\mathcal{O}(n^3) + \mathcal{O}(V_{SLO}\times batch) + \mathcal{O}(n\times 2^\omega)$, and asymptotically i.e., $\approx\mathcal{O}(n\times 2^\omega)$.

~
 
This concludes the agent's continuous model optimization, which maintains an up-to-date model of a processing task (i.e., the generative process). The high accuracy in the EOSC model allows the ACI agent to infer (Bayes-)optimal device configurations, which ensures QoS and QoE of ongoing operation. In the following, we will now focus on the collaboration between the Edge-based agents.

\subsection{Knowledge Transfer within the Cluster}
\label{subsec:knowledge-transfer}


By now, we presented ACI agents that can create generative models from scratch or update a model according to new observations. However, if we assume a cluster of nearby devices that process similar workloads, training EOSC models for every device seems redundant. Also, if we aim to extend the cluster with more devices (i.e., scaling up horizontally), model training delays the time until devices operate according to requirements. Instead, we envision the federation of knowledge between these edge devices by exchanging EOSC models within the device cluster.
Such a transfer learning approach appears to be a straightforward process if the models were trained in the exact same environment \cite{wu_online_2017}. However, the reality is that the Edge is composed of multiple heterogeneous device types; the resulting models thus reflect the characteristics of the device it was trained on, i.e., its capability to cope with SLOs depends on the processing hardware. 
For example, a multi-core device is certainly capable of processing multiple video streams, while a single-core one is not. Furthermore, the behavior of the ACI agent (i.e., which device configurations it favors) and the workload patterns (e.g., high demand by clients) determine which areas of the parameter space are more or less exploited.

\begin{figure}[t]
    \centering
    \includegraphics[width=.95\columnwidth]{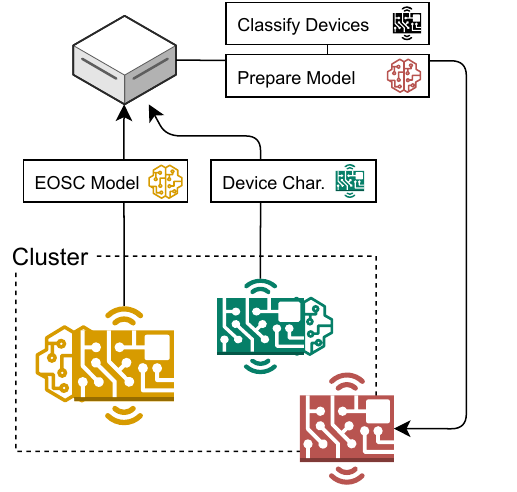}
    \caption{Transfer learning between devices in an Edge cluster according to hardware characteristics}
    \label{fig:knowledge-transfer}
\end{figure}

Whenever a new device (type) joins a cluster, the question is whether there exists a device within the cluster whose environment and characteristics match the newly-joint device's.
Consider Figure~\ref{fig:knowledge-transfer}, the yellow and green devices were already present in the cluster and shared their EOSC models and device characteristics (e.g., hardware specs or environmental factors) with the cluster leader (i.e., standing hierarchically above the device cluster). As the red device joins the cluster, the characteristics are compared to select a fitting model. In cases where the characteristics of multiple devices are similar, their models are merged and provided to the newly-joint device.
Thus, the red device builds its EOSC model on top of existing knowledge in the federation.

In the following, we will dive deeper into this transfer-learning process by answering (1) how models are federated between devices, (2) how hardware characteristics are compared to select a model, and (3) how models are combined to fit the target device. 

\subsubsection{Cluster-wide Model Exchange}

Exchanging EOSC models has two directions: (1) receiving -- when joining a device cluster it might be preferable to adopt an existing model rather than training one, and (2) providing -- any device might itself share its model with devices that join the cluster.
The selection of a fitting model, however, can happen on any trusted device; we assume for this task either a cluster leader (i.e., an outstanding device that was elected due to its capabilities) or a powerful fog node.
To provide an estimation, these models are supposedly smaller than 2 MB, as measured in \cite{sedlak_designing_2023,sedlak_active_2023}. 

When making the architectural decision (i.e., either cluster leader or fog node), there are various factors to consider, among them: network scale, cost, geographic location, and availability.
In cases where the cluster would be small (e.g., 10 devices), an edge device (e.g., chosen from Table~\ref{tab:device-list}) could cope with collecting and preparing EOSC models; however, for larger clusters (e.g., 1000 devices), regular edge devices might fail to do so.
In any case, a strong factor for using fog nodes is their high availability, as fog nodes could reliably cache a high number of EOSC models from various edge devices. Either choice, they assume the same responsibilities, which is why we call them simply \textit{leader node}.
This leader node periodically collects the EOSC models of all devices registered in the cluster, as well as their hardware characteristics. Based on this information, models will be provided for new device types.

\subsubsection{Model Comparison and Selection}
\label{subsubsec:model-comparison}

Transferring a EOSC model to a newly-joint device raises two questions: First, is the transfer of an existing model more efficient than learning the model from scratch? And second, how to choose the most convenient model for the new device? Of course, the second question assumes that the device type is unknown and the cluster does not contain the respective trained models so far. The first question will be answered and discussed as a result of this article; the second question, however, requires building a hypothesis around how to choose a model. 

The dynamism within the training environment has a decisive impact on the resulting model: applications with a stable number of user requests do not suffer many dynamics, while applications that are linked to specific events (i.e., disaster management) can experience extremely different requirements.
However, we assume that environmental factors are out of our hands -- we are unaware of the dynamics of the environment in which the device is set. 
Due to that, we focus on the device characteristics when transferring models between edge devices. To that extent, we get inspiration from the work of \textit{Casamayor et al.}~\cite{pujol_intelligent_2023}, which allows classification of heterogeneous characteristics of the devices found in a cluster, namely their CPU and GPU capacity. This means that we relatively classify the CPU capacity ($p$) of the devices in the cluster in a range $[p_{min},p_{max}]$, and their GPU capacity ($g$) from $[g_{min},g_{max}]$. Given that there are numerous edge devices without GPU, it is possible to set $g_{min}=0$. To make this more tangible, in Section~\ref{subsec:implementation}, we present a list of edge devices whose hardware is classified accordingly.
Finally, we define each device's capacities as $dc = p + g$. To estimate the similarity of device characteristics and to identify a device with a matching model, the leader node selects the device(s) with the closest integer $dc$.

Indeed, there exist other methods to classify edge devices' capacities, as well as to build the proximity or distance score between them. However, we are choosing the presented one because it is simple and can be computed quickly. This is important because devices are categorized relative to each other; due to that, the leader node must perform the classification repeatedly whenever a new device joins the cluster. 

\subsubsection{Combination and Preparation of Models}
\label{subsubsection:combination}

Heterogeneous edge devices differ in terms of hardware characteristics; therefore, we identified models that were trained on comparable devices. 
However, with the presented mechanism, there would frequently occur situations in which there is not exactly one device that trumps all others. For example, consider a device with type $t_x$ that wants to join a device cluster; there are already numerous device types present, among them $t_a$ and $t_b$. The leader node classifies the device capabilities as $dc_a = 3, dc_b = 5, \text{ and } dc_x = 4$. Which EOSC model should now be provided to $t_x$, the one trained on $t_a$ or on $t_b$? And in case $dc_a = 2 \text{ and } dc_b = 7$; is choosing $dc_a$ really the smartest choice here?

What we envision for both cases is merging the models from $t_a$ and $t_b$, thus creating a new model $t_{ab}$ that presents the intersection between these two. In the second case, where $dc_x$ does not exactly fall between $dc_a$ and $dc_b$, this is done proportionately. Therefore, what is required is a mechanism to combine EOSC models -- still BNs at their cores. To date, merging BNs is an ongoing research field that still presents various limitations \cite{vagnoli_updating_2022,vanis_novel_2023}; in most cases, it is coupled to conditions that the models must fulfill.
Due to this, we limit our work to merging CPTs.

As long as two models $m_a$ and $m_b$ contain the same structure (i.e., their DAGs are identical) and their CPTs have the same cardinality (i.e., variable states), this can be done as follows: For a random variable $r$ and its $\texttt{CPT}(m,r)$, each table cell's expected value ($P$) is calculated as shown in Eq.~\eqref{eq:cpt-cell}; $P_a$ and $P_b$ represent the probabilities in the cells of $m_a$ and $m_b$, the coefficients $w_a$ and $w_b$ reflect the distribution of $dc_x$ between $dc_a$ and $dc_b$. For example, in case $dc_x$ is aligned centrally between them, they take the value $w_a = w_b = 0.5$; otherwise, it is shifted proportionally, but $w_a + w_b = 1$ must always remain true.
\begin{equation}\label{eq:cpt-cell}
    P_x = (w_a \times P_a) + (w_b \times P_b)
\end{equation}

\noindent\textit{Lemma 4. } The time complexity for model preparation is ${\mathcal{O}}((q + v)^{k+1})$.\\ 
\textit{Proof.} Although multiplying table cells is a simple process, it has to be done for every CPT in the BN. Notice, CPTs are multidimensional arrays with $k + 1$ dimensions, where $k$ represents the number of incoming edges for $r$. So depending on the complexity of the BN, which consists of $v$ variables with each up to $q$ states, this can take ${\mathcal{O}}((q + v)^{k+1})$ multiplications for a fully connected graph.

~

If $m_a$ and $m_b$ do not fulfill the requirements for simple multiplication, they would have to undergo a transformation process. Nevertheless, in Section~\ref{subsubsec:practical-limitations}, we apply a workaround to merge BN whose CPTs have different cardinalities. Solving this issue on a general level requires dedicating future work solely to this challenge.

After merging the EOSC models, the leader node provides the resulting model to the newly-joint device; once received, we consider the transfer learning completed. Thereby, it allowed us to decrease the time required for model training or even skip it entirely. The big remaining question is now: how accurately does the resulting model match the characteristics of the new device? This will be evaluated, as it determines the device's capability to ensure SLOs. Nevertheless, consider that any transferred EOSC model can again be supervised by an ACI agent. Hence, even though the model would not match perfectly, the agent can perform the required fine-tuning to ensure model accuracy.

\subsection{Stream Offloading in the Edge-Fog Cluster}
\label{subsec:stream-offloading} 

Regardless of whether trained by an ACI agent or transferred from another device, a EOSC model is a decisive step toward SLO fulfillment. Thus, edge devices are continuously reconfigured to achieve maximum SLO compliance. However, despite our efforts, edge devices are still vulnerable to environmental factors that cannot be controlled, e.g., irregular peaks in client traffic. While a EOSC model can have a hard time finding an SLO-compromising device configuration, idle edge devices in close proximity might be available for offloading computation. Again, to match our desired level of intelligence, this can be achieved through collaboration between the agents. Given that the struggling edge device is part of a device cluster, it is possible to (1) compare the device's capabilities to fulfill their SLOs within their environment, and (2) balance the load accordingly. Notice, that shifting the load within the cluster is a (local) reconfiguration that follows the same rules as in Section~\ref{subsec:continuous-model}; this time, however, on a higher level.

\begin{figure*}[t]
    \centering
    \includegraphics[width=0.85\textwidth]{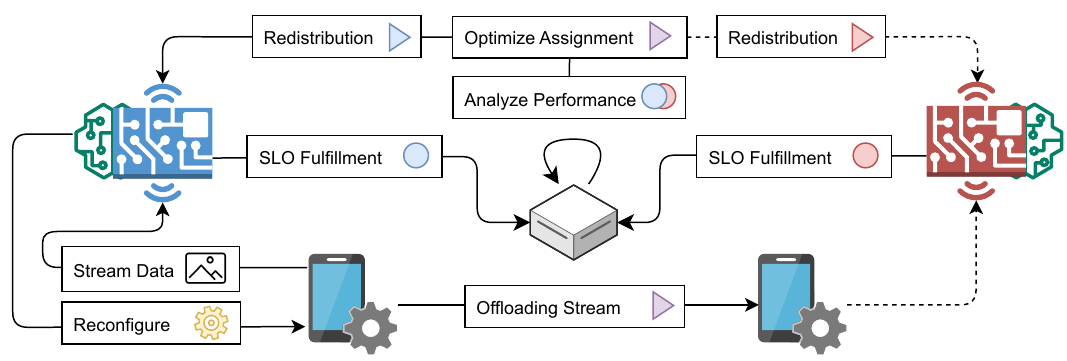}
    \caption{Evaluating SLOs within a cluster of nearby edge devices and reassigning tasks}
    \label{fig:load-balancing}
\end{figure*}

In the following, we describe how to evaluate, analyze, and optimize the cluster-wide SLO compliance; the overall process is visible in Figure~\ref{fig:load-balancing}: The edge devices in the cluster (red \& blue) serve their respective clients, e.g., by processing data, which is subject to dynamic reconfiguration according to the EOSC model. Throughout processing, the edge devices supply their SLO fulfillment to the leader node. Among that, they provide other factors (i.e., as metrics) that potentially impact the fulfillment. Environmental factors (e.g., insufficient hardware, power shortage, or client demand) can thus be contrasted with the devices' capacity to fulfill SLOs. Based on that analysis, the leader reconfigures the cluster (e.g., by redistributing the load) so that QoS and QoE SLOs are optimized within the cluster.

\subsubsection{Cluster-wide Evaluation of SLOs}

To analyze SLO fulfillment on a cluster level, the leader node does not have to evaluate the Edge-based SLOs again -- this is already covered within the Edge tier. Instead, the leader node merely collects the SLO compliance rate per device as a combined factor $f = pv \times ra$. These metrics are collected at the leader node; depending on the desired amount of historical data, the high availability of the Fog would again be beneficial for collecting the data. The question is now how to transfer metrics: Considering the potential size of a device cluster, we opt for a push-based approach, where devices periodically supply their data to the cluster.

Apart from the SLO fulfillment, edge devices provide metrics that reflect their current environmental state. This includes any factors that the leader node should consider; what cannot be quantified can also not serve as a basis for cluster-wide optimization. If a battery-equipped device suffers occasional power shortages, it can report this conditional to the leader node, which adapts the network, e.g., by offloading computations to other devices to decrease its power drain. However, in the event of an entire network outage, devices can be incapable of reporting their state, and another node (e.g., leader) would have to detect this. Other frequent conditions can be general network congestion, including poor latency, jitter, or packet loss, but also devices' geographic location, user density, and peak usage times. Given their impact on the devices' capacity to fulfill SLOs, the leader node will rebalance the environment.

\subsubsection{Analysis \& Optimization per Device}

Optimizing the devices' environments requires methods to draw conclusions between discrete environmental states and their consequential SLO fulfillment. To that extent, we aim -- again -- to identify causal relations between metrics; however, this time on a cluster level. Given a metric set (i.e., reflecting the environmental state) and the respective SLO rates per device, the leader node can construct a BN and infer how environmental changes impact the SLO fulfillment. To accelerate the construction of such a model, the leader node can combine metrics from devices of the same type, or even those that have comparable hardware characteristics (as done in Section~\ref{subsubsec:model-comparison}). 
Although we ascended from an Edge to a cluster level, we still use the same tool for analyzing and adapting the environment -- the EOSC model. However, to make a distinction, we call this new instance a EOSC-F (Fog) model. To make the EOSC-F model training autonomous, it can again be guided by ACI.

Given a trained EOSC-F model (or rather, its DAG), it is evident which environmental factors ($\sigma_{env}$) have a causal impact on SLO fulfillment. This can also help to improve the QoS in the long run, e.g., by pinpointing issues within the infrastructure. However, we aim to ensure SLO fulfillment the moment the QoS or QoE drops; the EOSC-F model can therefore consider the devices' environment and redistribute client load to ensure maximum SLO fulfillment within the cluster. To that extent, we present Algorithm~\ref{alg:client-reassignment}, which distributes a number of streams ($n_{client}$) between the devices ($\Lambda$) in the cluster. Notice, that \texttt{Inference} is again executed only on the variables that relate to SLO fulfillment, i.e., $\texttt{MB}(model, f)$, by filtering the $model$ (Line 2 \& 10). In Lines 6-18, the agent then iteratively assigns clients to the device, whose SLO fulfillment is the least impacted by receiving another stream ($ass[\lambda]+1$). This assumes, that both $ass$ and $\sigma_{env}$ are part of $ev$, i.e., have an impact on SLO fulfillment. To that extent, $\sigma_{env}[\lambda]$ can contain factors like device characteristics. After assigning all streams within the cluster, the assignment can be orchestrated to the clients.



\begin{algorithm}[t]
\caption{Client reassignment algorithm }
\label{alg:client-reassignment}
\begin{algorithmic}[1]

\REQUIRE $model$, $n_{client}, \sigma_{env}$
\ENSURE $ass$ // assignment according to env. state

\STATE $i \gets 0$
\STATE $ev \gets \texttt{MB}(model, f)$

\FOR{each $\lambda \in \Lambda$}
\STATE $ass[\lambda] = 0$
\ENDFOR

\WHILE{$i < n_{clients}$}
\STATE $\delta_{best} = -\infty $ 
\FOR{each $\lambda \in \Lambda$}
\STATE $evidence \gets ev  \cap (\sigma_{env}[\lambda] \cup ass[\lambda] + 1)$
\STATE $\delta \gets \texttt{INFERENCE} (ev, f, evidence)$

\IF{$\delta > \delta_{best}$}
\STATE $\delta_{best} = \lambda$
\ENDIF
\ENDFOR

\STATE $ass[\delta_{best}] \gets ass[\delta_{best}] + 1$
\STATE $i \gets i + 1$
\ENDWHILE
\RETURN $ass$
\end{algorithmic}
\end{algorithm}
~

\noindent\textit{Lemma 5. } The time complexity for client reassignment (Algorithm~\ref{alg:client-reassignment}) is $\mathcal{O}(n^2)$.\\
\textit{Proof. } According to the step count method, the algorithm takes $\Lambda$ interactions for initial assignments. Then, it take $n_{clients}$ times of $\Lambda$. So, the time complexity can be defined as $\mathcal{O}(\Lambda + (n_{clients}\times \Lambda))$, which is $\approx \mathcal{O}((n_{clients}\times \Lambda))$. In worst case $ n_{clients} = \Lambda = n$, then the asymptotic time complexity for the client reassignment algorithm is $\mathcal{O}(n^2)$.

\subsubsection{Orchestration and Redistribution}

As a last step, the new cluster configuration must be enforced; in this case, by informing the pertinent devices of the new assignment. The leader node pushes this information to all edge devices that must alter their configuration. In accordance with Figure~\ref{fig:load-balancing}, this includes all devices that offload or receive clients (red \& blue); thus, the red device redirects clients to the blue device.

To improve the SLO fulfillment rate within the cluster, the assignment considered each device's environment to provide an adequate configuration on a cluster level.
Regardless of whether the QoS was impacted by poor network conditions or by poor hardware, if these conditions are packed as stateful information, the leader node can optimize the cluster accordingly. Thus, it covers heterogeneities between edge devices, which themselves might fail to scale their service given the stress introduced by the environment.

~

\noindent\textit{Theorem 3. } The time complexity for proposed EOSC approach is $\mathcal{O}(n \times \tau^{\nu + 1})$\\
\textit{Proof. } The proposed EOSC consists of three stages including Static Model Training and Inference, Active Inference Cycle, Knowledge Transfer within the Cluster, and Stream Offloading in the Edge-Fog Cluster. It is measured as $(n\times \tau^{\nu+1}) + (n\times 2^\omega) + ((q + v)^{k+1}) + (n^2)$. So, the asymptotic complexity of the proposed work can be concluded as $\mathcal{O}(n\times \tau^{\nu+1})$.

~

This concluded the redistribution of client load, which optimized the overall SLO fulfillment in the cluster according to the EOSC-F model.
To transfer intelligence to the network edge, or even to the level of a cluster or fog node, this section provided various concepts that all had the same goal: ensure SLOs in the respective system. To that extent, it remains to provide a prototypical implementation of the presented ideas, evaluate it according to key aspects, and argue to what extent it is ready for wider adoption. This will be the content of the next two sections.


\section{Evaluation}\label{sec:Evaluation}
 In the following, we describe (1) a CC scenario that requires edge devices to continuously transform video streams; this use case poses various requirements that must be ensured throughout processing. Afterward, we outline (2) our prototype that ensures SLOs through collaborative edge intelligence. Essentially, this is the implementation of the framework presented in this paper. Lastly, we explain (3) the methodology according to which the prototype will be evaluated. Section~\ref{sec:results-and-discussion} will contain the respective results. 
\subsection{Use Case Description}


The CC as a distributed system provides unprecedented opportunities for service providers and clients alike, e.g., in terms of processing or requirements assurance. 
As an example, consider a region with frequent natural disasters where the humanitarian situation should be documented. Therefore, reporters provide video streams in which vulnerable groups, e.g., minors of age, are detected. In the same step, individuals can be counted or visually highlighted; their identities, however, must be preserved. The region suffers from occasional network breakdowns (i.e., this affects access to global resources like the cloud but not internal connectivity); the reporting team thus provides ad hoc networking infrastructure in the form of edge devices, which can be installed in close proximity to the operation area. Reporters equipped with IoT cameras are now capturing their surroundings; the video streams are transformed on edge devices, where they can be cached as long as global internet services are unavailable. Once resumed, the videos are streamed to a cloud platform that provides the content to worldwide consumers. 


\paragraph{Envisioned Solution}

Due to the nature of how such disasters happen, it is impossible to fine-tune the complete streaming architecture beforehand. Therefore, the system is unaware of how to ensure its service (i.e., characterized by a set of SLOs) within this highly dynamic environment. To that extent, we advertise our framework for collaborative edge intelligence as the missing piece:
Edge devices are supervised by ACI agents, which ensure QoS and QoE through their EOSC model. Whenever the computing architecture is extended with new devices (i.e., scaled horizontally), existing models can be transferred to this new device, regardless of whether its device type is known. Therefore, both devices must be part of a cluster, in which the leader node identifies the device whose hardware characteristics are most similar.
Apart from that, the leader node continuously analyzes edge devices' capacity to comply with SLOs; in case some devices are excessively loaded
or suffer from short-term network issues, the assignment between IoT cameras and edge devices is adapted to optimize the cluster-wide SLO fulfillment.


\subsection{Implementation}
\label{subsec:implementation}

While the last part of the use case outlined the envisioned solution, not all of these aspects are implemented and evaluated; in this regard, we focus on the ideas presented in this paper. This especially concerns the three contributions of the presented framework, i.e., the ACI-based model training, the knowledge transfer between heterogeneous devices, and the rebalancing of load according to environmental factors. Aspects such as the bootstrapping of IoT and edge devices and the leader node election (e.g., fog or edge) have already been covered, e.g., by \textit{Murturi et al.}~\cite{murturi_decentralized_2022,dustdar_towards_2020}. The same applies to the cloud-based distribution of video streams. An exception, however, is the privacy-preserving stream transformation; for this, we make use of previously evaluated work~\cite{sedlak_privacy_2023}. To give our evaluation more rigor, we chose this tool over simulating the workload and its impact on SLOs.


\subsubsection{Prototype}
\label{subsubsection:prototype}

We provide the Python-based prototype of our framework in a GitHub repository\footnote{https://github.com/borissedlak/workload/tree/main/FGCS}; it contains all source code we used to implement the three contributions, as well as the EOSC models for each device type. The core logic is separated into two classes: \texttt{Agent} and \texttt{FogNode}. These are the high-level loops executed in the main thread; all other processes (e.g., \texttt{ACI} or \texttt{VideoProcessor}) run in detached threads.
The central library that is applied for training and updating BNs, as well as running inference queries, is \textit{pgmpy}~\cite{ankan_pgmpy_2023}.
\textit{pgmpy} offers ample support of BNL techniques; however our choice is also motivated by personal preference -- the framework's performance must be analyzed under different libraries (e.g., as done by \cite{zhang_comprehensive_2023}).
To improve the portability of our framework and simplify distribution, we provide a docker image\footnote{https://hub.docker.com/repository/docker/basta55/workload/} that can be executed platform independently\footnotemark. The image exposes multiple env variables for configuring the solution, e.g., forcing the \texttt{Agent} to create a EOSC model from scratch or disabling \texttt{ACI} entirely.

The source code also contains the framework for privacy-preserving stream transformation and the ML models for face~\cite{linzaer_ultra-light-fast-generic-face-detector-1mb_2022} and age detection~\cite{rothe_dex_2015}. To improve the reproducibility of results, we cancel out irregularities in the video streams by processing prerecorded videos; these are contained in the same repository. To simulate redirecting IoT devices within the cluster, it thus suffices to open/close processing threads on the edge devices; this simplifies networking. The \texttt{Agent} can thus reconfigure the stream assignment immediately, at the end of every \texttt{ACI} iteration. Because the use case is focused on video streaming and the number of frames per second (\textit{fps}) that are transferred, each iteration lasts up to 1000ms. 

\footnotetext{In fact the docker image is restricted to daemons with a \textit{linux/arm64} architecture. This is the embedded architecture for all device in Table~\ref{tab:device-list}, except for \textit{Laptop} (\textit{x86\_64}).}

\subsubsection{Practical Limitations}
\label{subsubsec:practical-limitations}

Merging BN, as presented in Section~\ref{subsubsection:combination}, is only possible under the specified conditions, which are not always given during the ACI process. The number of states in a CPT, for example, is highly dynamic and extended as new batches of data are received. To merge the EOSC models under such circumstances, we provided a workaround: Instead of merging two BNs ($m_a$ and $m_b$), we extend one of them (e.g., $m_a$). The device that trained $m_b$ maintains a backup of the training data ($d_b$); this we use to update the CPTs of $m_a$ through \texttt{PARL}\footnotemark, i.e., $m_{ab} = \texttt{PARL}(m_a,d_b)$. Notice, that this merges the conditional probabilities of the models, but not the structure; this remains an open question. While the resulting models are valid, we cannot assume that the original training data is always maintained.

\footnotetext{This functionality is natively offered by \textit{pgmpy}; by default, the models are merged proportionally to the number of samples that $m_a$ and $d_b$ contain. This can be fine-tuned by adjusting the \textit{n\_prev\_samples} parameter; we use this to prioritize new observations $batch$ over existing conditional probabilities.}

Another limitation is that the DAG of the $model$ cannot be updated frivolously through \texttt{STRL}; this triggers numerous updates within the CPTs of the BN, which are not supported by default in \textit{pgmpy}. Although \textit{bnlearn}~\cite{scutari_learning_2010} promises these features, we require a package that can be embedded into our Python environment. Therefore, we make use of the following workaround: Instead of updating the DAG of model $m_a$ according to new observations $batch$, we train a new BN with $data = batch \text{ } \cup \text{ } d_a$, where $d_a$ reflects again the backup data. So internally, the ACI agent executes $\texttt{STRL}(model,batch)$ as $\texttt{PARL}(\texttt{STRL}(data),data)$, which likewise updates the CPTs with every execution. We consider it out of scope to solve this limitation within our work.

\subsubsection{Variables and SLOs}
\label{subsubsec:variables-slos}

For the given use case, the agents consider device and application (i.e., video processing) metrics to construct EOSC models. Internally, BNL transforms metrics into model variables, which are used to evaluate conditional probabilities. Table~\ref{tab:metric-list} contains an overview of all captured metrics; each row contains a description, measuring unit, and if it can be used as parameter. Notice, that only parameterizable variables can be adjusted by the ACI agent to optimize SLO fulfillment. For example, \textit{pixel} and \textit{fps} are video stream properties of the IoT device, which are reconfigured by the edge device according to the agent's behavior. The leader node, on the other hand, can only adjust the number of \textit{streams} per device; this, however, was out of scope for an individual device.

\begin{table*}[t!]
  \centering
  \caption{List of metrics captured by the devices, which are turned into variables by ACI}
  \label{tab:metric-list}
  \begin{tabular}{lcclc}
    \toprule
    Name   & Origin & Unit & Description &  Param \\
    \midrule
    \textit{pixel}        & IoT   & num       & number of pixel contained in a frame  & Edge \\
    \textit{fps}           & IoT  & num       & number of frames received per second  & Edge \\
    \textit{bitrate}       & IoT  & num       & number of pixels transferred per second  & No \\
    \textit{cpu}     & Edge        & \%        & utilization of the device CPU  & No \\
    \textit{memory}       & Edge   & \%        & utilization of the system memory  & No \\
    \textit{streams}    & Edge      & num   & number of IoT devices providing data  & Fog \\
    \textit{consumption}   & Edge  & W         & energy pulled by the device  & No \\
    \textit{network}    & Edge & num         & data transferred over network interface  & No \\
    \textit{delay}    & App.       & ms        & processing time per video frame & No \\
    \textit{success}    & App. & T/F       & if a pattern (i.e., face) was detected  & No \\
    \textit{distance}     & App.   & num        & relative object distance  between frames  & No \\
    \midrule
    \textit{$slo\_rate$}    & Edge & \%         & combined SLO Fulfillment rate ($pv \times ra$)   & No \\
    \textit{device\_type}    & Edge & enum         & physical device type   & No \\
    \textit{congestion}    & Edge & num         & network congestion that increases latency  & No \\
    \bottomrule
  \end{tabular}
\end{table*}

The EOSC (or EOSC-F) models can be applied in different computational tiers to ensure each tier's unique requirements; thus, their model variables might not overlap. The edge-based EOSC model contains the upper part of the variables, i.e., from \textit{pixel} to \textit{success}, whereas the cluster-based EOSC-F model treats the lower part. Notice that the metric's origin, i.e., if it was measured from system stats or the application, does not determine where it is used as a variable. From these variables, we construct SLOs that reflect the system state in terms of QoS and QoE. The ACI agent considers this classification when calculating $pv$ and $ra$ (recall Section~\ref{par:behavioral}). In Table~\ref{tab:slos-extracted}, we present four SLOs that must be ensured during edge-based processing and one that is ensured by the cluster's leader node. To simplify the EOSC models, we include the SLO into BNL and remove the source variable, i.e., \textbf{distance} instead of \textit{distance}

\begin{table}[h]
  \centering
  \caption{Extracted SLOs and their classification.}
  \label{tab:slos-extracted}
  \begin{tabular}{lclc}
    \toprule
    SLO & Condition & Tier & Type \\
    \midrule
    \textbf{network}          & $network < 1.6$ MB/s & Edge & QoS\\
    \textbf{in\_time}         & $delay < 1/fps$ & Edge & QoS\\
    \textbf{success}          & $success=True$ & Edge & QoE\\
    \textbf{distance}         & $distance < 50$ & Edge & QoE\\
    \midrule
    \textbf{slo\_rate}        & $\max(slo\_rate)$ & Fog & Both \\
    \bottomrule
  \end{tabular}
\end{table}

We consider the presented SLOs relevant because (1) \textbf{network} ensures that the combined number of video streams does not exceed the networking capabilities, (2) \textbf{in\_time} makes sure that frames are computed within the available time frame, (3) \textbf{success} guarantees maximal privacy preservation, and (4) \textbf{distance} ascertains a smooth trajectory for tracked objects. The maximum \textbf{slo\_rate} speaks for itself. 

\subsubsection{Device Classification}

Video processing is very dependent on the availability of GPU acceleration~\cite{sedlak_privacy_2023}; therefore, we apply multiple edge devices -- with and without GPUs. All devices applied for this work are listed in Table~\ref{tab:device-list}; in the following, we call them by their ID. The other columns contain hardware characteristics and -- complementarily -- the original price of the device. A special instance is \textit{Xavier$_{CPU}$}: while its physical hardware is equal to \textit{Xavier$_{GPU}$}, we disabled the GPU acceleration (i.e., NVIDIA CUDA) to create another device type. Overall, our devices differ greatly in terms of computing capabilities (e.g., missing GPU support or a highly superior CPU with 16 cores); nevertheless, as a whole, these devices compose the heterogeneous edge layer of the CC architecture.

\setlength{\tabcolsep}{5pt}
\begin{table*}[t]
  \centering
  \caption{List of devices used for implementing and evaluating the presented methodology}
  \label{tab:device-list}
  \begin{tabular}{llrlrl}
    \toprule
    Full Device Name & ID & Price\footnotemark & CPU & RAM &  GPU  \\
    \midrule
    ThinkPad X1 Gen 10  & \textit{Laptop} & 1800 € & Intel i7-1260P (16 core)       & 32 GB & Incompatible \\
    Jetson Orin Nano    & \textit{Orin} & 500 €  & ARM Cortex A78 (6 core)  & 8 GB      & Volta (383 core)\\
    Nvidia Jetson Nano  & \textit{Nano} & 150 € &  ARM Cortex A57 (4 core)   & 4 GB      & Incompatible \\
    Jetson Xavier NX    & \textit{Xavier$_{CPU}$} & 300 € & ARM Carmel v8.2 (6 core)       & 8 GB  & Disabled\\
    \midrule
    Jetson Xavier NX    & \textit{Xavier$_{GPU}$} & 300 €  & ARM Carmel v8.2 (6 core)    & 8 GB  & Amp (1024 core)\\
    \bottomrule
  \end{tabular}
\end{table*}

\footnotetext{Price as of October 11th 2023 from https://sparkfun.com/}

As a prerequisite for transfer learning, we classify devices in a cluster according to their hardware characteristics. Although this process is dynamic, i.e., done repeatedly as devices join or leave or leave the cluster, we focus our evaluation on a scenario where the cluster contains all devices from Table~\ref{tab:device-list}, excluding Xavier$_{GPU}$; the latter will be the device joining the cluster. As discussed in Section~\ref{subsubsec:model-comparison}, we classify these devices relative to each other according to their CPU and GPU capabilities; the results are contained in Table~\ref{tab:hw-scalars}. To achieve the desired distance between the scalars, the CPU is aligned between [$1 \leq p \leq 4$] and the GPU between [$0 \leq g \leq 2$].

\begin{table}[h]
  \centering
  \caption{Classification of device hardware and assigned scalar}
  \label{tab:hw-scalars}
  \begin{tabular}{lllc}
    \toprule
    Device ID & CPU [1,4] & GPU [0,2] & $\Sigma$ \\
    \midrule
    \textit{Laptop}         & Very High (4) & None (0) & 4\\
    \textit{Orin}           & High (3)      & High (2) & 5\\
    \textit{Nano}           & Low (1)       & None (0) & 1\\
    \textit{Xavier}$_{CPU}$ & Medium (2)    & None (0) & 2\\
    \midrule
    \textit{Xavier}$_{GPU}$ & Medium (2)    & Low (1)  & 3\\
    \bottomrule
  \end{tabular}
\end{table}


\subsection{Evaluation Methodology}
\label{subsec:evaluation-methodlogy}

The implementation of the use case is thus set up for evaluation. To ensure a solid foundation for our framework, we will target each of the three pillars (i.e., the contributions) individually. The order in which they are evaluated resembles the one used throughout the paper; this makes sense also from a logical point of view because transfer learning and stream offloading rely on the underlying ACI mechanism. In the three paragraphs below, we outline the evaluated aspects and motivate each question. Combined, this represents our evaluation methodology.

\paragraph{\textbf{Active Inference}} Our main interest includes the executability of the ACI agent on edge devices and the extent to which the EOSC model improves the SLO fulfillment within the Edge. Because structure and parameter learning are recurrent factors in the evaluation, we will put emphasis on when they happen. Namely, our questions include:


\begin{itemize}[label=, left=-7pt]
    \item \textit{A-1: Do MBs reduce the complexity of inference?} 
    
    Increasingly large BNs require mechanisms to limit the complexity of a system; otherwise, resource-restricted edge devices may fail to execute the ACI cycle within an induced time frame. The MB, as a potential remedy, could achieve this.
    
    \item \textit{A-2: What is ACI's operational overhead?} 
    
    Training and updating EOSC models directly on edge devices allows them to adapt quickly to system dynamics. However, any overhead introduced by ACI must not disrupt regular device operation, e.g., data processing.

    \item \textit{A-3: How long require ACI agents to ensure SLOs?} 

    To optimize SLO fulfillment, the agent must be able to infer adequate system configuration. However, there is no guarantee after how many ACI iterations the model will converge to the desired accuracy. Hence, we must provide an estimate for this.
    
    \item \textit{A-4-1: Are the produced Bayesian networks interpretable?} 
    
    Large-scale distributed systems, e.g., the CC, require trusted and reliable components as a solid foundation. 
    Given that ACI can provide structures that are empirically verifiable, this promises to increase trust. 
    
    \item \textit{A-4-2: Is the behavior of ACI agents explainable?}

    Being able to understand an agent's decisions allows to justify (or empirically debug) its behavior, e.g., why the agent chose a certain device configuration at a specific time. If agents follow patterns, this also simplifies the configuration of hyperparameters.
    
    \item \textit{A-5: What is the operational impact of including BNL in the ACI cycle?}

    BNL was identified as the dominant factor for the complexity of the ACI cycle; therefore, we must ascertain whether edge devices can perform BNL without limitations.
    Depending on the results, the two processes could be broken up into a federated learning approach, e.g., to execute sub-steps in the Fog. 

    \item \textit{A-6: Can changes in variable distribution be handled?} 

    Real-world generative processes are not guaranteed to stay stable, a small environmental change (e.g., a new client) might suffice to change the SLO outcome. Nevertheless, these changes should be detected and resolved through ACI-based model training. 
    
    \item \textit{A-7: Can SLOs be modified during runtime?} 

    In the CC, edge devices can be administered by entities that stand hierarchically above them; these can change their role in the architecture, or more simply, their SLOs. If the device could not adapt its existing EOSC model, it would have to train from scratch.

\end{itemize}

\paragraph{\textbf{Knowledge Transfer}} After focusing on the training of EOSC models, we are mainly interested in how well the created models can be exchanged with other edge devices, and if this promises to improve the training time. Ideally, we would thus reuse existing knowledge instead of ``rediscovering" it.

\begin{itemize}[label=, left=-5pt]
    \item \textit{K-1: How high is the SLO fulfillment of transferred models compared to ACI-trained ones?} 
    
    Transfer learning can provide ML models (i.e., specific for one device) to other devices. However, it is not guaranteed that a transferred model performs equally to a model specifically trained for a device. For example, the transferred model might be more likely to violate SLOs; hence, this must be examined by comparing the produced results.
    
    \item \textit{K-2: Can knowledge transfer achieve any speedup?} 
    
    Transferring a trained model removes computational overhead (\textit{A-2}) from the recipient; thus, it could decrease the overall energy dedicated to model training, most beneficial for resource-restricted edge devices. Furthermore, this could decrease the time required to ensure SLOs (\textit{A-3}).
    
    \item \textit{K-3: Can merged models decrease the FE compared to choosing a single one?}

    Models with low FE can infer SLO-fulfilling system configurations with higher accuracy.
    Exchanging knowledge within the cluster can include the combination of multiple eligible models. However, can such combined models interpret observations with less surprise compared to a single transferred model?
    

\end{itemize}

\paragraph{\textbf{Stream Offloading}}


To optimize their SLO fulfillment, intelligent edge device continuously adapt their environment. However, if there are environmental factors that are out of their scope (e.g., network failures or hardware limitations), the device cluster can be the remedy to compensate for these issues. In this context, we want to determine whether the SLO fulfillment of individual devices can be recovered through collaboration.

\begin{itemize}[label=, left=-5pt]
    \item \textit{S-1: How is the load distributed among resource-constrained devices?} 
    
    The Edge, as one CC tier, allows clients to request services from nearby edge devices; however, this fosters situations where load is highly unbalanced within the system. This might cause resource-restricted devices to fail their service; once this is detected, the load must be rebalanced within the system.
    
    \item \textit{S-2: Can the CC hierarchical structure optimize local SLO fulfillment?} 
    

    Depending on the scale of SLO failures, individual devices may be incapable of recovering their service through local reconfiguration. Nevertheless, higher entities in the CC (e.g., the device cluster) can evaluate and resolve this by employing their own SLOs.
    
\end{itemize}


\section{Results and Discussion}\label{sec:Results}
\label{sec:results-and-discussion}

In the following, we will evaluate the prototype according to the presented methodology. We structure our results according to the three contributions and the evaluation order used in Section~\ref{subsec:evaluation-methodlogy}; based on the results, we pose derivative questions for future work. At the end of this section, we take a step back (i.e., not focusing on particular questions), look at the results as one coherent framework, and discuss the applicability of our approach.


\subsection{Active Inference}

\vspace{5pt}

\noindent\textit{A-1: Do MBs reduce the complexity of inference?} 

~

\begin{figure}[!t]
    \centering
    \includegraphics[width=1.0\columnwidth]{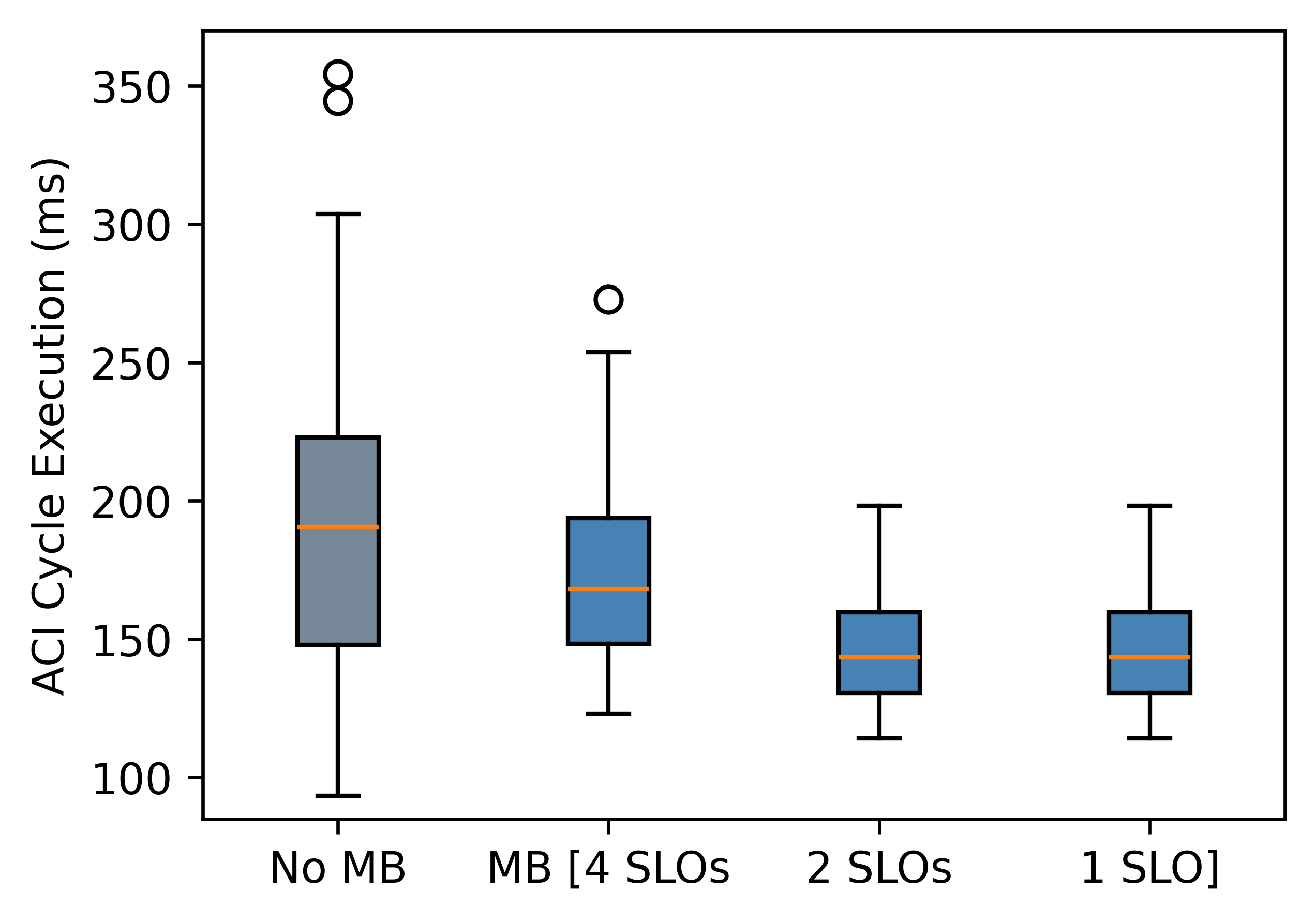}
    \caption{Duration of the ACI cycle depending on the application of an MB and the number of SLOs (\textit{A-1})}
    \label{fig:mb-vs-no-mb}
\end{figure}

To show whether an MB can decrease the ACI cycle duration, we focus on one of its subparts -- the inference. This makes sense since \texttt{INFERENCE} poses the highest algorithmic complexity whenever \texttt{STRL} is not executed. We modify the implementation of Algorithm~\ref{alg:surprise-calculation} (Line 2 \& 8) to execute \texttt{INFERENCE} either (1) on the entire BN including all 4 SLOs, (2) the MB including 4 SLOs, (3) the MB with 2 SLOs, or (4) the MB with 1 SLO. Then, we execute the ACI cycle on \textit{Laptop} and capture the running time of each configuration over a duration of 10 min; this produces 600 observations for each experiment.
Figure~\ref{fig:mb-vs-no-mb} visualizes the time that \textit{Laptop} requires for performing \texttt{INFERENCE}, given the different MB sizes.

We observe two things: (1) applying an MB reduces the median execution type significantly, i.e., from 191 ms (grey) to 159 ms (blue) for 4 SLOs, and (2) decreasing the number of SLOs gradually reduces the execution time further. We thus conclude that MBs can reduce the complexity of VE (\textit{A-1}).

~

\noindent\textit{A-2: What is ACI's operational overhead?}

~

\begin{figure}[!t]
    \centering
    \includegraphics[width=1.0\columnwidth]{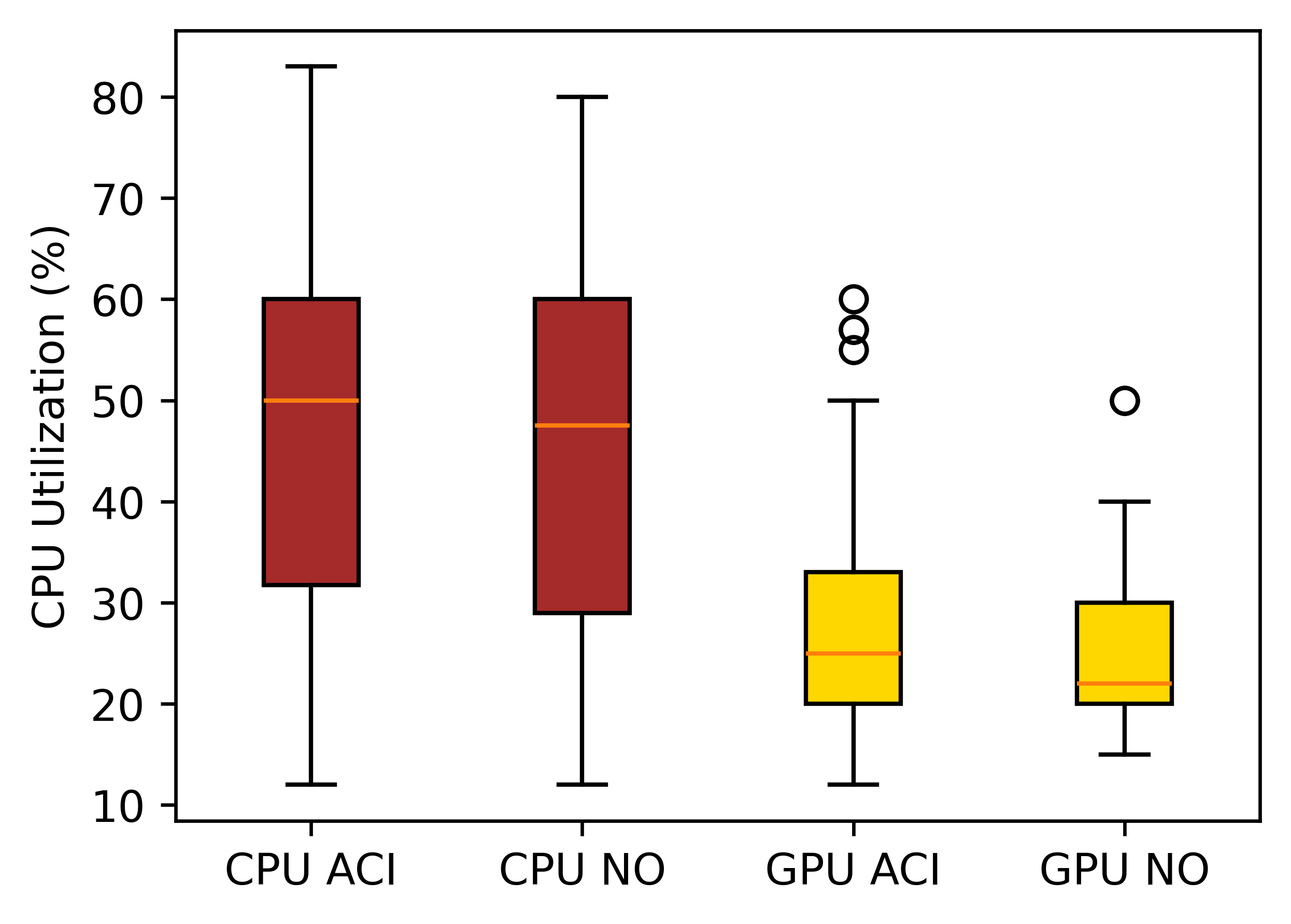}
    \caption{Overhead introduced by the ACI background process when operating on \textit{Xavier}$_{CPU}$ or \textit{Xavier}$_{GPU}$ (\textit{A-2})}
    \label{fig:overhead}
\end{figure}

To evaluate the operational overhead of ACI, we use pre-trained EOSC models for \textit{Xavier$_{CPU}$} and \textit{Xavier$_{GPU}$}. Each device processes 6 video streams. We measure the CPU load (\%) of the two devices with one of these two configurations: (1) ACI enabled, and (2) ACI disabled. We capture the load over 10 min; this produces 600 observations for each experiment. In Figure~\ref{fig:overhead}, we show the CPU load of \textit{Xavier}$_{CPU}$ and \textit{Xavier}$_{GPU}$. The left bar of each device shows the load when operating with ACI and the right one without (i.e., disabled) ACI.

We observe: (1) the CPU load is clearly decreased when processing the videos on a GPU, \textit{Xavier}$_{GPU}$ with ACI enabled presented a 24\% lower load than \textit{Xavier}$_{CPU}$, and (2) the ACI background process introduced a computational overhead of 3\% for both devices (left vs. right bar). Overall, this provides an estimate of the general overhead (\textit{A-2}); however, whether this is acceptable depends on the use case.

~

\noindent\textit{A-3: How long require ACI agents to ensure SLOs?} 

~

\begin{figure}[t]
    \centering
    \includegraphics[width=1.0\columnwidth]{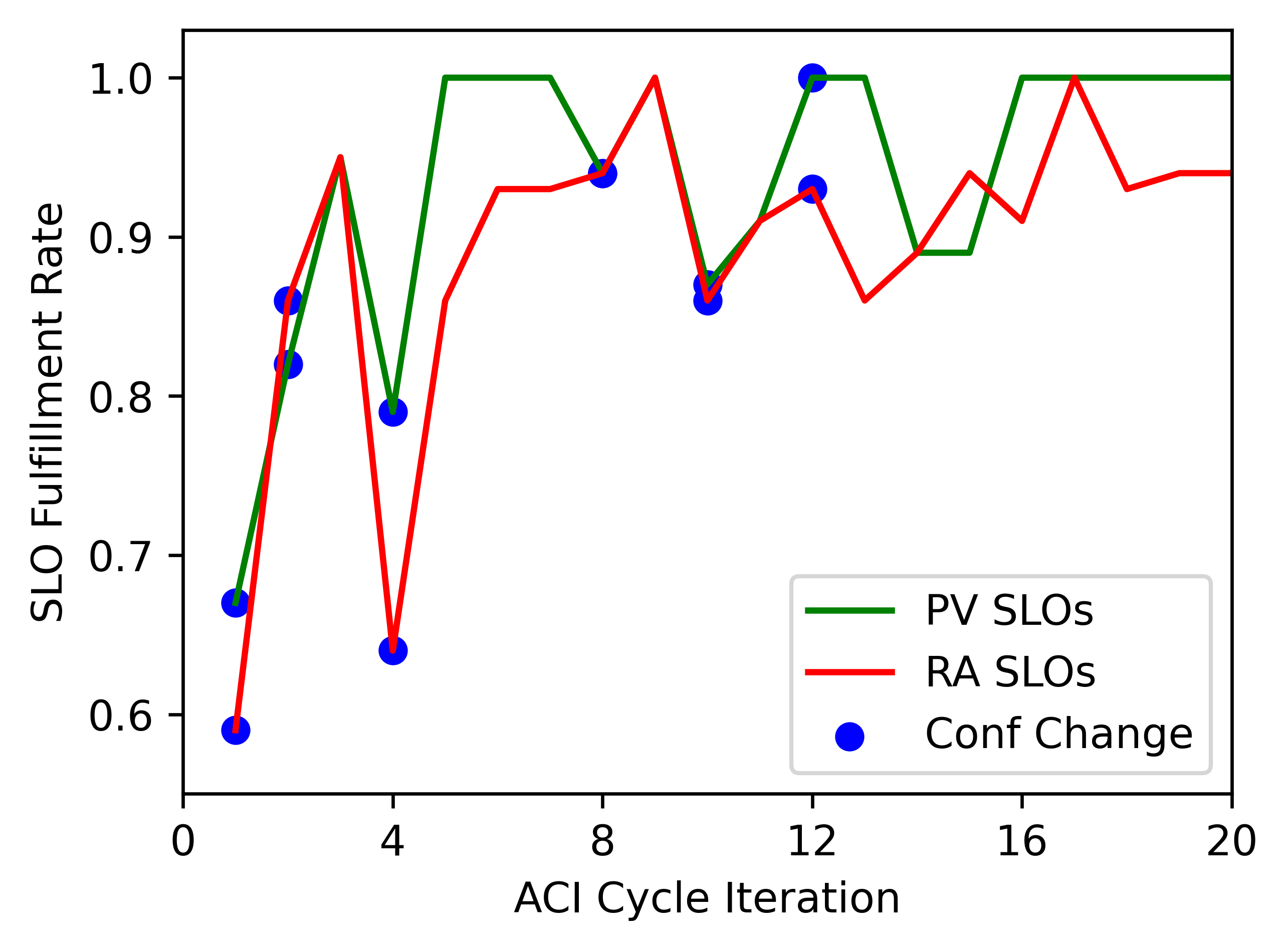}
    \caption{SLO fulfillment rate (split up into $pv$ and $ra$) when operating on a blank \textit{Laptop} client over 20 ACI cycles (\textit{A-3})}
    \label{fig:simple-slo-rate}
\end{figure}

To evaluate the time to train a EOSC model, we count (1) the number of ACI iterations that the agent requires to arrive at a (nearly) optimal device configuration, and (2) how often the agent changes the configuration to get there. The model is trained from scratch; therefore, the ACI agent (i.e., executed on \textit{Laptop}) trains the model over 20 cycles and reports after each cycle (3) the SLO fulfillment according to the selected device configuration. We present the results in Figure~\ref{fig:simple-slo-rate}: The green and red lines represent the SLO fulfillment ($pv$ \& $ra$); whenever the agent reconfigures the edge device, we print a blue dot for both lines in the graph. We observe: (1) the agent requires roughly 7 cycles to converge to a configuration that satisfied SLOs with more than 90\%, which it then maintains in the same range; (2) this state is reached after 3 reconfigurations; and (3) $pv$ and $ra$ showed similar trends in this example. Thus, we answered how long an ACI agent requires to provide an acceptable configuration (\textit{A-3}), both in terms of ACI cycles and the number of reconfigurations.

~

\noindent\textit{A-4-1: Are the produced causal graphs interpretable?} 

~

\begin{figure*}[t]
  \centering
  \subfloat[DAG after 1 round]{\includegraphics[width=0.25\linewidth]{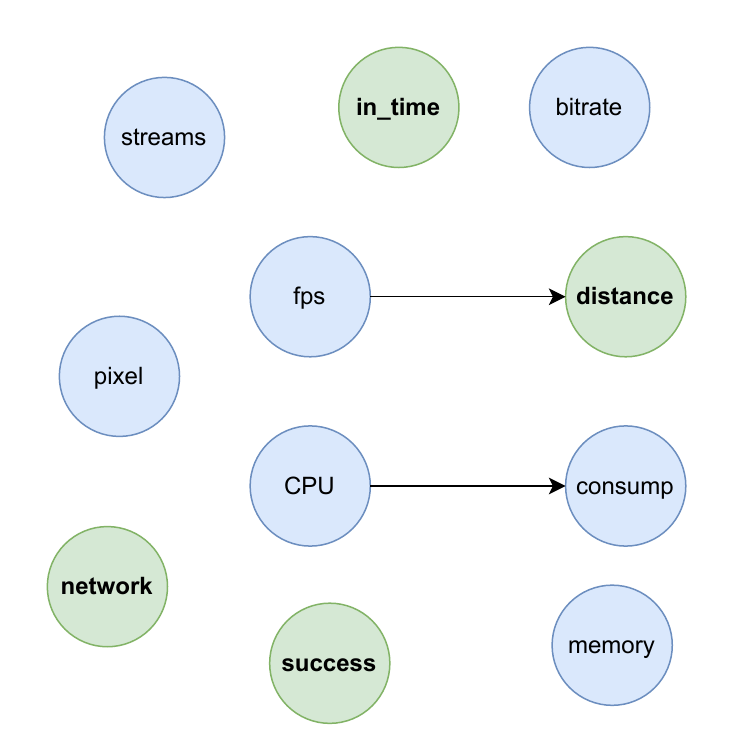}}\hfill
  \subfloat[DAG after 3 round]{\includegraphics[width=0.25\linewidth]{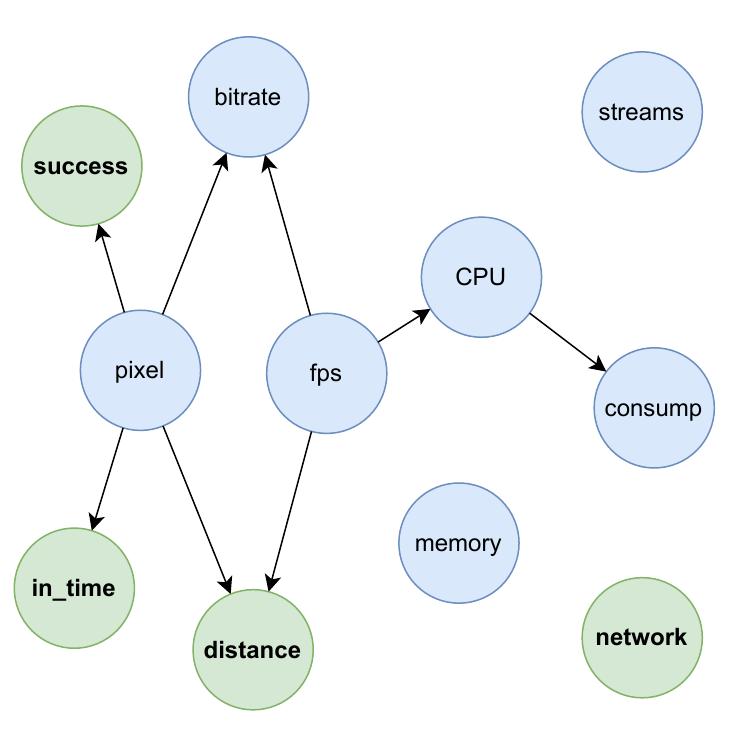}}\hfill
  \subfloat[DAG after 5 rounds]{\includegraphics[width=0.25\linewidth]{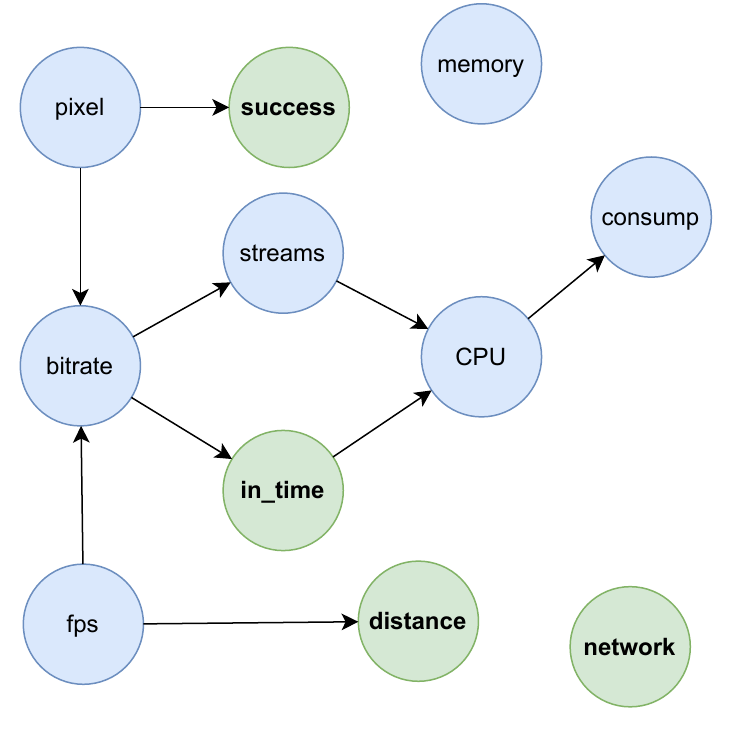}}\hfill
  \subfloat[DAG after 10 rounds]{\includegraphics[width=0.25\linewidth]{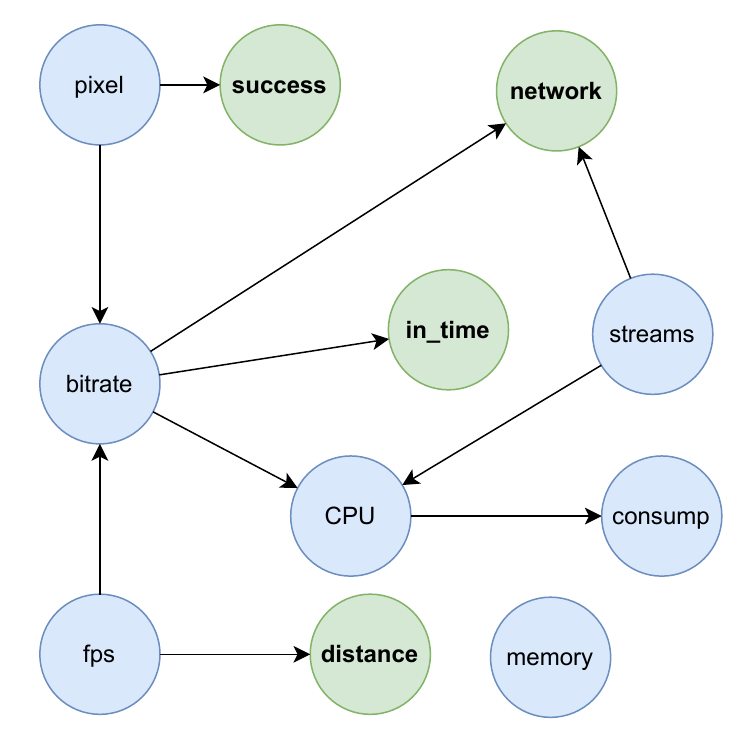}}\hfill
  
  \caption{Progress of the DAG after \{1,3,5,10\} rounds of parameter training when creating a model with ACI on \textit{Laptop} (\textit{A-4-1})}
  \label{fig:dag-process}
\end{figure*}

To discuss the interpretability of created causal structures, we compare the DAGs produced by \texttt{STRL} and highlight at which stage the graph can be empirically explained. We will not consider specific metrics here but interpret the DAGs according to our expert knowledge. On \textit{Laptop}, we train a EOSC model from scratch and extract the DAGs after \{1,3,5,10\} rounds of BNL. Thus, we want to show how the ACI agent discovers (ideally) causal relations between model variables. The results are visible in Figure~\ref{fig:dag-process}: SLO variables (see Table~\ref{tab:slos-extracted}) are colored in green; regular variables in blue. We observe: (1) all SLO variables are influenced by variables that the ACI agent can control, and (2) memory was the only variable that could not be related to any other. After studying the graphs carefully, we could not detect any edge that appears counterintuitive to us; however, this does not prove that they are indeed causal. In total, we claim that the created graph is coherent and the links are understandable (\textit{A-4-1}), but it requires sophisticated experiments to prove causality for each edge. 

~

\noindent\textit{A-4-2: Is the behavior of ACI agents interpretable?}

~

\begin{figure*}[!t]
  \centering
  \subfloat[$pv$ matrix after 1 round]{\includegraphics[width=0.33\linewidth]{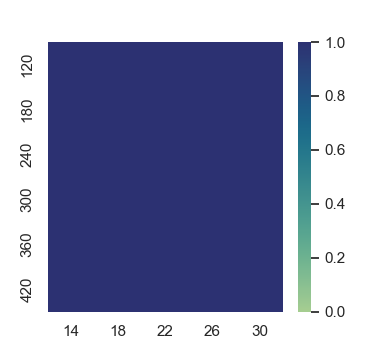}}\hfill
  \subfloat[$ra$ matrix after 1 round]{\includegraphics[width=0.33\linewidth]{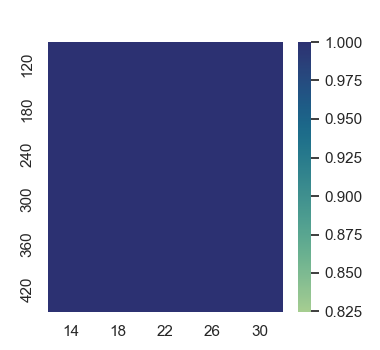}}
  \subfloat[$ig$ matrix after 1 round]{\includegraphics[width=0.33\linewidth]{figures/heatmap/1_ig.png}}\hfill

  \subfloat[$pv$ matrix after 5 rounds]{\includegraphics[width=0.33\linewidth]{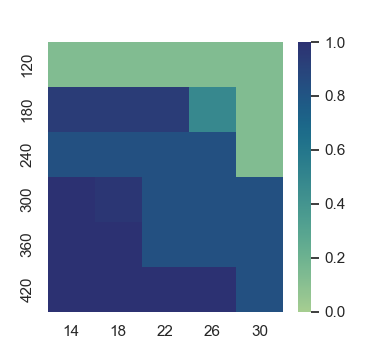}}\hfill
  \subfloat[$ra$ matrix after 5 rounds]{\includegraphics[width=0.33\linewidth]{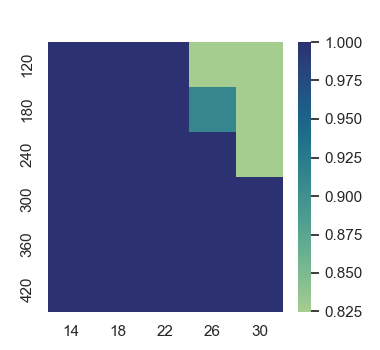}}
  \subfloat[$ig$ matrix after 5 rounds]{\includegraphics[width=0.33\linewidth]{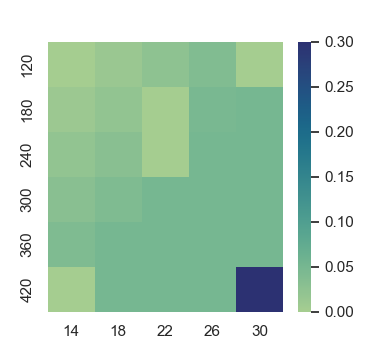}}\hfill

  \subfloat[$pv$ matrix after 50 rounds]{\includegraphics[width=0.33\linewidth]{figures/heatmap/50_pv.png}}\hfill
  \subfloat[$ra$ matrix after 50 rounds]{\includegraphics[width=0.33\linewidth]{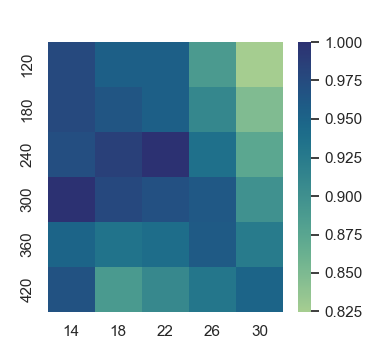}}
  \subfloat[$ig$ matrix after 50 rounds]{\includegraphics[width=0.33\linewidth]{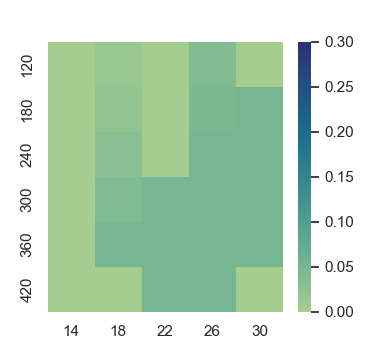}}\hfill

  \caption{Behavioral factors (i.e., $pv$, $ra$, and $ig$) interpolated by the ACI agent to evaluate possible device configurations (\textit{A-4-2})}
  \label{fig:matrices}
\end{figure*}

Complementarily, we were interested in how the behavior of the ACI agent could be interpreted. In Figure~\ref{fig:matrices} we present three matrices for each behavioral factor (i.e., $pv$, $ra$, and $ig$). We executed the ACI agent on \textit{Laptop} and extracted the matrices after \{1,5,50\} iterations. The first row presents the agent's initial assumptions on how the parameters are related to SLO fulfillment ($pv$ \& $ra$) and which rows provide the most insight ($ig$).

We observe: (1) the $ig$ is initially high at corner points in the parameter space (as discussed in Section~\ref{par:parameter-space}), which are visited in the first ACI iterations -- this is evident because at round 5 only one cell with $e=0.3$ remains; (2) the interpolation improves as transitions in the heatmap become smoother (from top to bottom); (3) the highest SLO fulfillment is at $pixel = 300, fps=14$; and (4) the agent develops clear preferences in terms of $pv$ (i.e., bottom-left corner), while the optimal $ra$ is located in the center of the parameter space. Areas to avoid would be, e.g., $pixel = 120$, because image detection requires more detail, or $fps > 22$ because the processing time frame shrinks. Overall, we argue that the visualizations allow understanding the agent's behavior (\textit{A-4-2}).

~

\noindent\textit{A-5: What is the operational impact of including BNL in the ACI cycle?}

~

\begin{figure}[!t]
    \centering
    \begin{subfloat} 
    \centering
    \includegraphics[width=0.94\columnwidth]{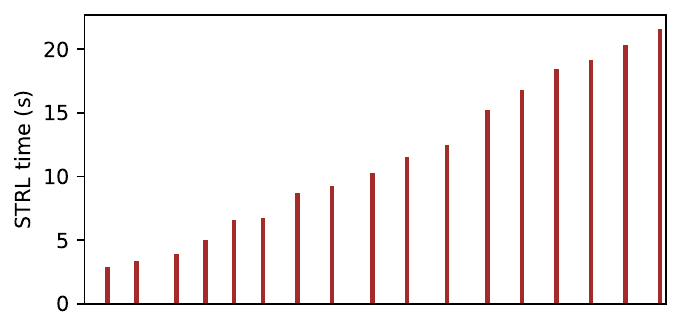}
  \end{subfloat}
  \vspace{-6pt}
  \begin{subfloat} 
    \centering
    \includegraphics[width=1.0\columnwidth]{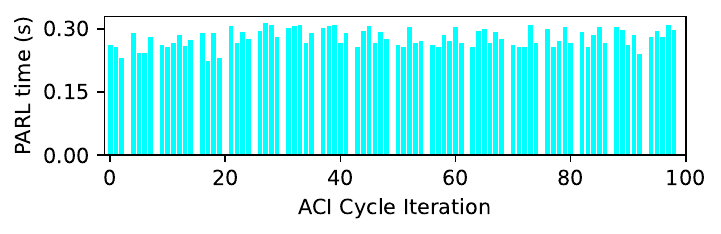}
  \end{subfloat}

    \caption{Duration of structure and parameter learning on the \textit{Xavier$_{GPU}$} when training the BN from scratch (\textit{A-5})}
    \label{fig:bnl-time}
\end{figure}

To answer whether BNL can be applied on regular edge devices, we train a EOSC model on \textit{Xavier$_{GPU}$} and measure the execution time of \texttt{STRL} and \texttt{PARL}, i.e., the BNL sub-steps from Algorithm~\ref{alg:iterate-function}. In Figure~\ref{fig:bnl-time} we visualize the execution time of \texttt{STRL} and \texttt{PARL} over 100 ACI iterations, respectively 1.5 min of operation.
We observe: (1) \texttt{PARL} runs with a stable runtime of around 250ms, (2) the time required for \texttt{STRL} increases as more and more training data becomes available, and (3) running \texttt{STRL} after 100 ACI iterations took more than 20s.
We conclude that \texttt{PARL} might be run on the employed edge device because it can be completed within less than 1000ms (i.e., the time frame for concluding the ACI cycle from Section~\ref{subsubsection:prototype}). However, the runtime of \texttt{STRL} presents an obstacle because the ACI agent might thus have to skip iterations until the ongoing execution of \texttt{STRL} is finished. Hence, it would be advisable to perform \texttt{STRL} on another device (\textit{A-5}) or find a way to decrease the runtime, e.g., by updating the DAG regardless of existing CPTs.

~

\noindent\textit{A-6: Can changes in variable distribution be handled?} 

~

\begin{figure}[t]
\centering
\subfloat[Stream changes]{\includegraphics[width=.542\columnwidth]{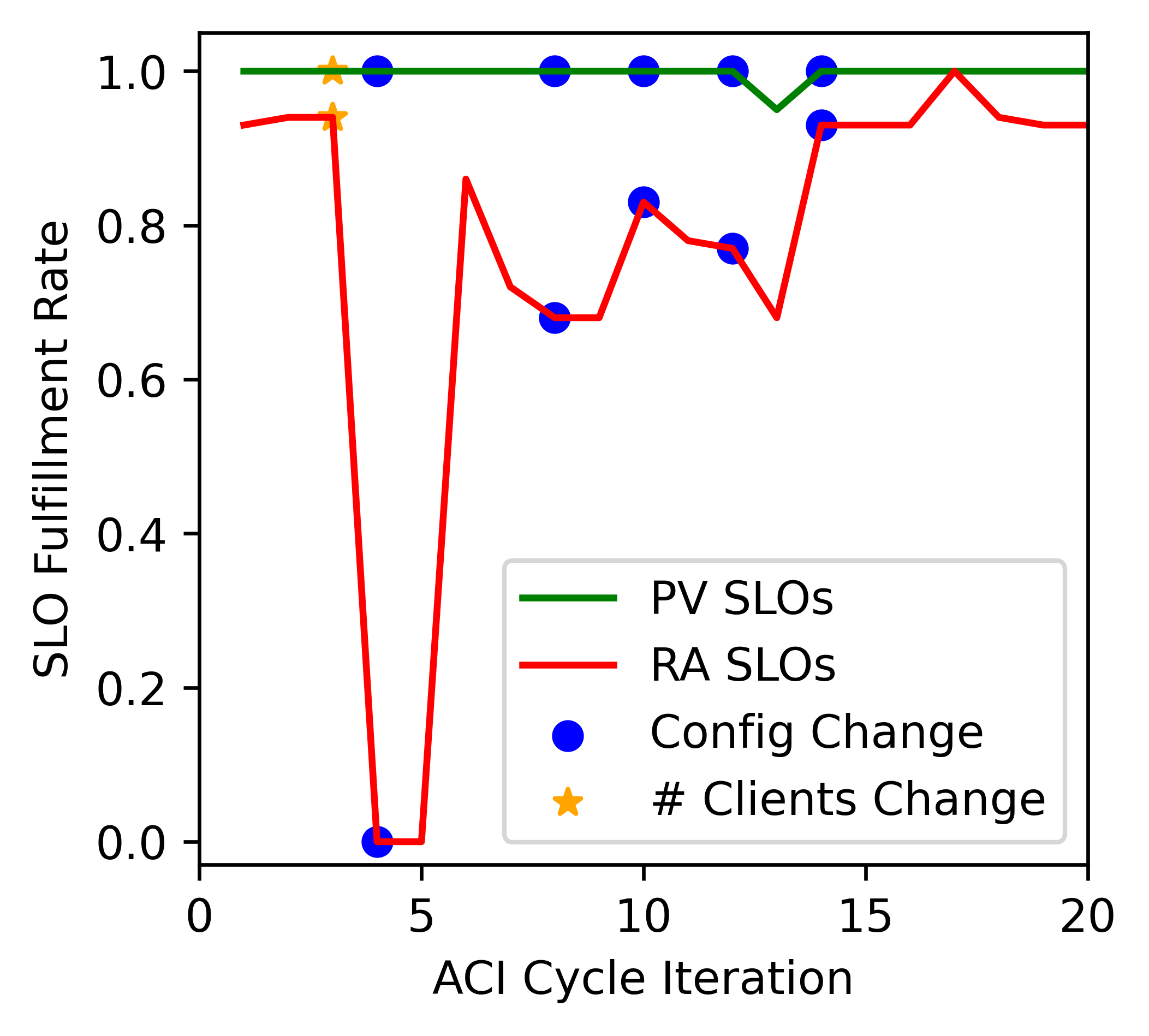}}
\subfloat[Video changes]{\includegraphics[width=.458\columnwidth]{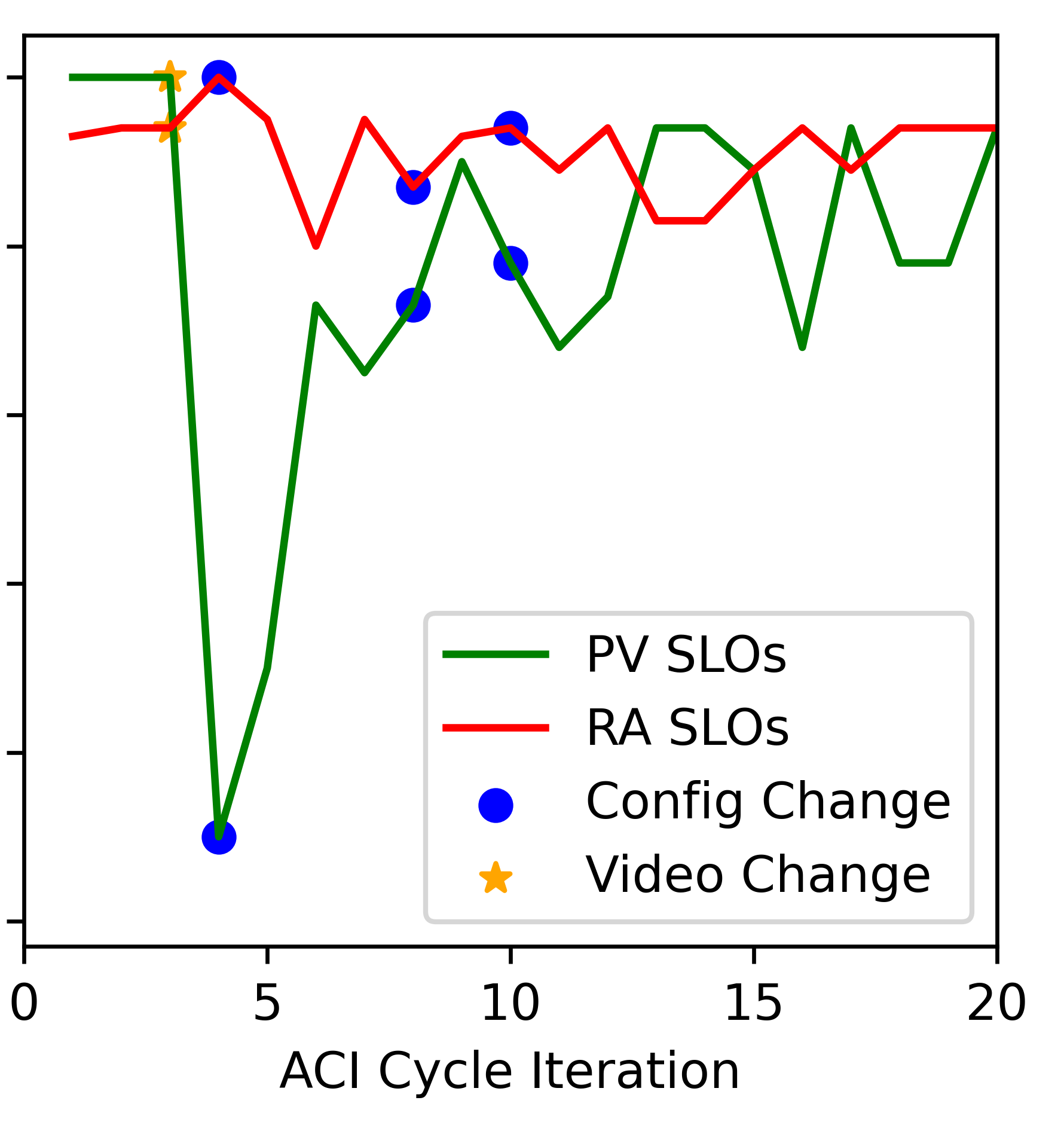}}
\caption{Changes in the variable distribution caused (a) by higher number of video streams or (b) lower video quality (\textit{A-6})}
\label{fig:distribution-change}
\end{figure}

Variable distributions can change due to various external factors; to evaluate how well the system can handle this, we either (1) simulate a peek usage time by increasing the number of processed video streams from 1 to 6, or (2) distort the video content with a Gaussian blur of 5px, which could resemble a foggy video setting. We measure the impact on the SLO fulfillment ($pv$ \& $ra$) over 20 ACI cycles and visualize to what extent the EOSC model is capable of restoring satisfactory (i.e., close to original) SLO rates. Figure~\ref{fig:distribution-change} shows in both subfigures the SLO fulfillment rate of \textit{Laptop}, when the disruptive factor was introduced (i.e., after 3 iterations), and at which points the ACI agent reconfigured the system (blue dots).

We observe: (1) after the stream change, \textit{Laptop} took 11 ACI cycles (incl. 4 reconfigurations) to recover the SLO fulfillment, and (2) the information loss introduced by the video manipulation could not be recovered, although SLO fulfillment was improved as far as possible. Hence, we conclude that the system was able to adapt to changes in the variable distribution (\textit{A-6}); however, only as long as the device can compensate for this factor. In fact, the SLO fulfillment could not be recovered after the video change because the agent did not have an equivalent countermeasure, e.g., increasing the resolution sufficiently.

~

\noindent\textit{A-7: Can SLOs be modified during runtime?} 

~

\begin{figure}
  \centering
  \begin{subfloat} 
    \centering
    \includegraphics[width=\columnwidth]{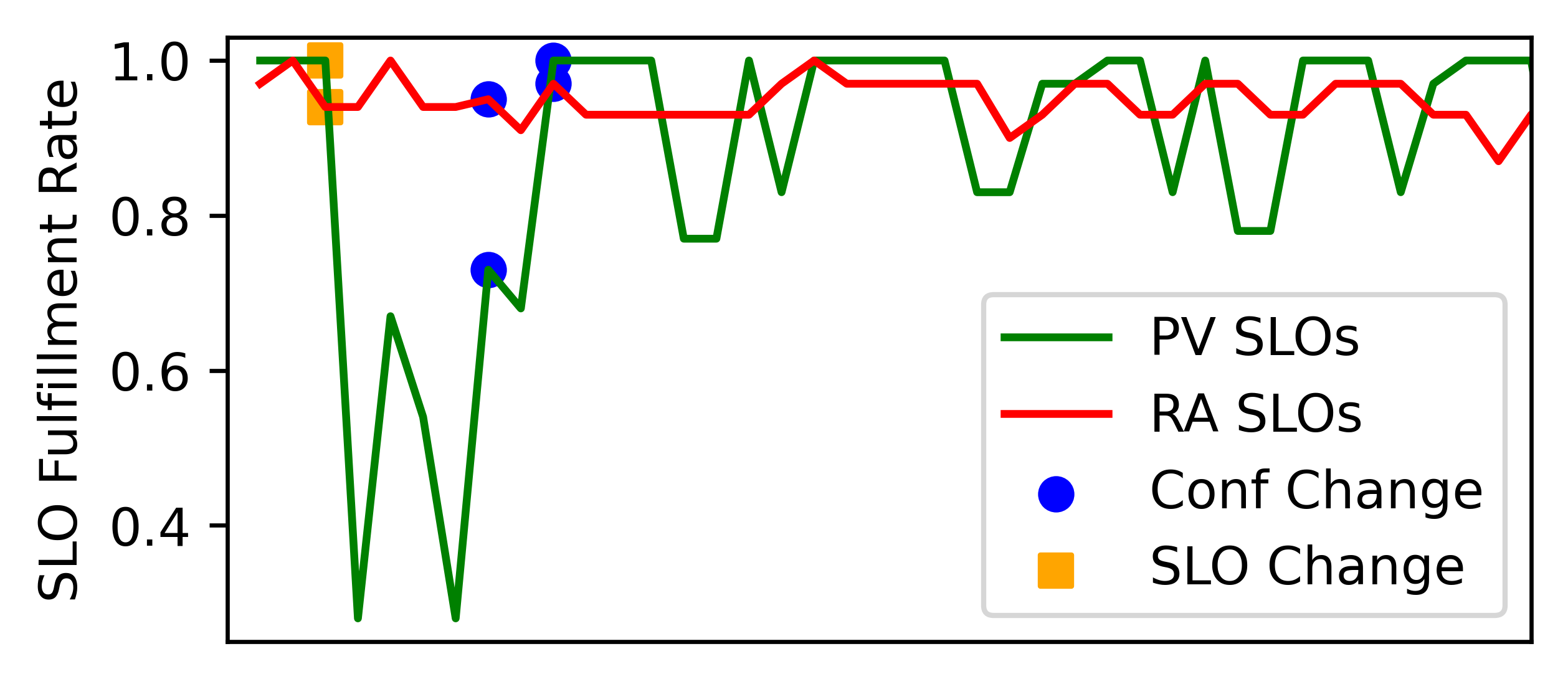}
  \end{subfloat}
  \vspace{-6pt}
  \begin{subfloat} 
    \centering
    \includegraphics[width=1.01\columnwidth]{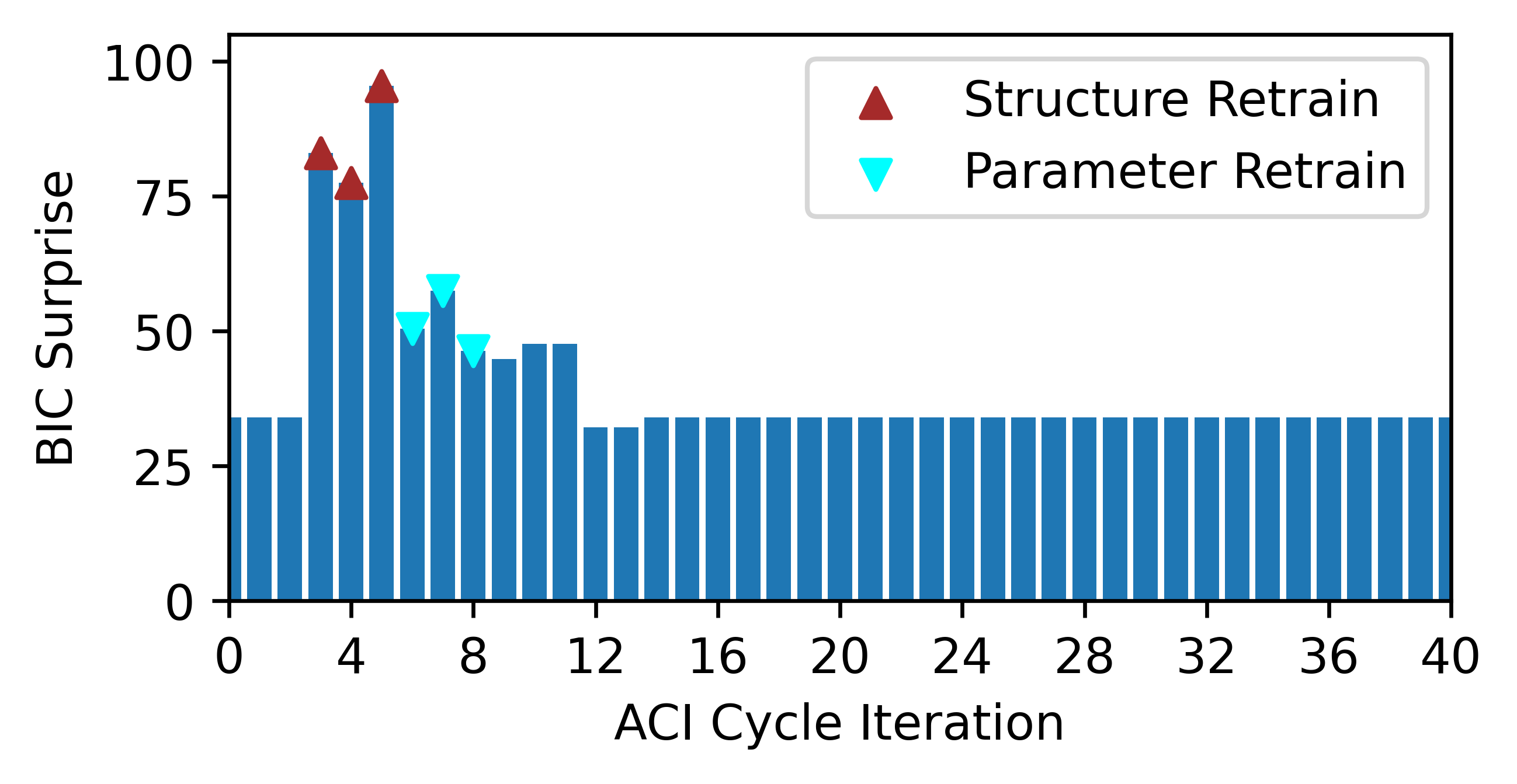}
  \end{subfloat}
  
  \caption{Impact of changing the \textbf{distance} SLO during runtime, combined with the respective surprise values measured (\textit{A-7})}
  \label{fig:combined}
\end{figure}

Requirements might change during operation; to simulate this, we modify the \textbf{distance} SLO from $50$ to $20$ (i.e., clearly stricter) and measure the SLO fulfillment rate before and after the modification. Additionally, we capture the surprise (Algorithm~\ref{alg:surprise-calculation}) to show if SLO outcomes reflected the expectations of the agent. 
Figure~\ref{fig:combined} shows in the upper part the SLO fulfillment rate over 40 ACI cycles; the SLO changes after 3 iterations. The lower part shows the agent's surprise at each round and when \texttt{STRL} or \texttt{PARL} happen. 

We observe: (1) after the SLO change, the agent experienced 9 rounds of high surprise, i.e., $>> 35$, (2) after 2 reconfigurations, the state prior to the SLO change was recovered, although final SLO rates (mean 0.91) are slightly below previous (mean 0.94), and (3) the magnitude of the surprise was decisive for the decision between \texttt{STRL} and \texttt{PARL} (as envisioned in Algorithm~\ref{alg:iterate-function}). However, as known from Figure~\ref{fig:bnl-time}, \texttt{STRL} can exceed the ACI time frame multiple times; hence, the ACI agent is forced to wait for this process to finish. This could be solved, e.g., by offloading \texttt{STRL}. Nevertheless, we conclude that the system was able to handle SLO changes during runtime (\textit{A-7}).

\subsection{Knowledge Transfer}

\vspace{5pt}

\noindent\textit{K-1: How high is the SLO fulfillment of transferred models compared to ACI-trained ones?} 

~

\begin{figure}[!t]
    \centering
    \includegraphics[width=1.0\columnwidth]{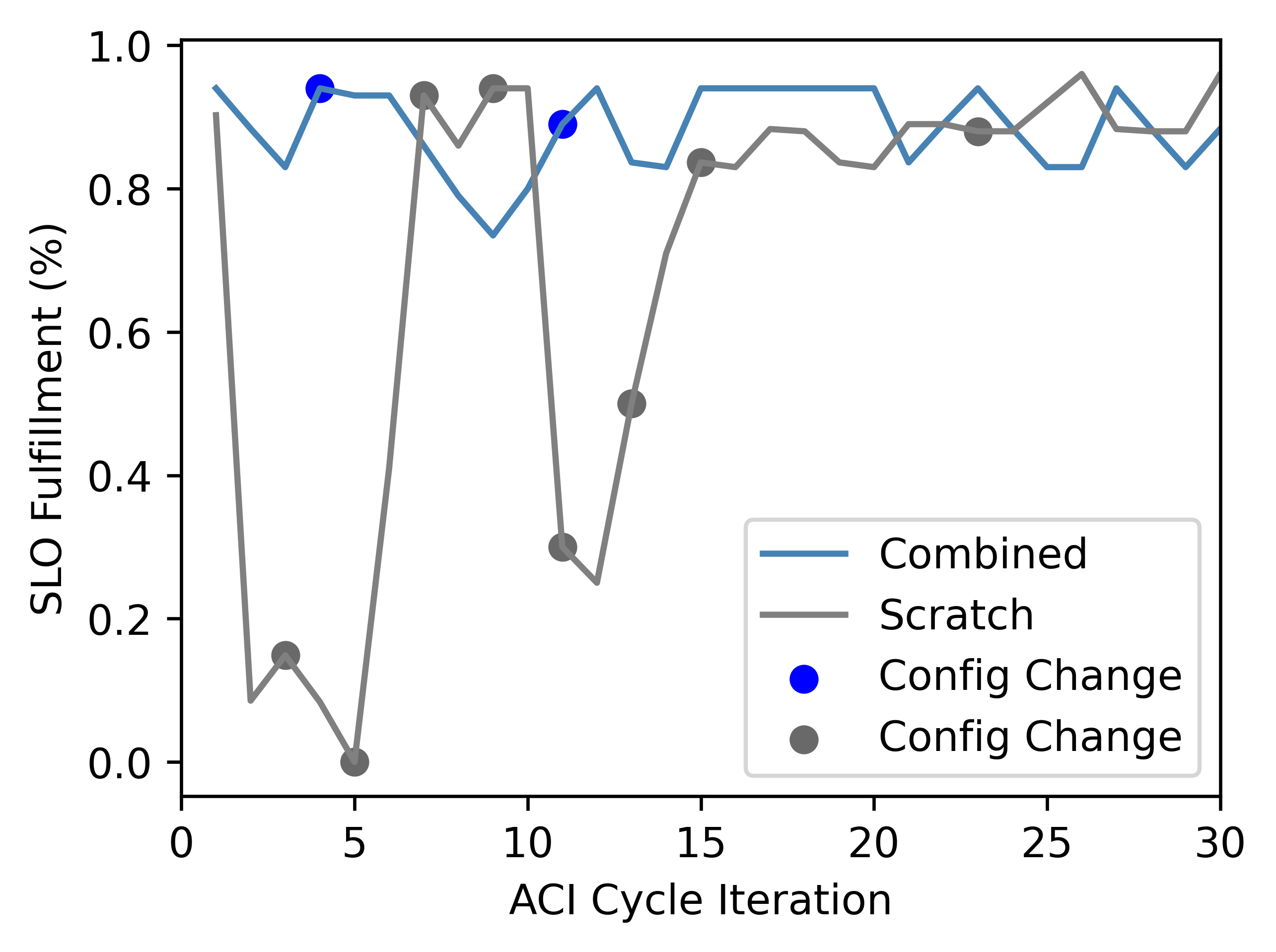}
    \caption{Difference in SLO fulfillment between an agent using a transferred model or training from scratch (\textit{K-1} \& \textit{K-2})}
    \label{fig:slo-kt}
\end{figure}

Transfer learning promises to accelerate model training, but we must ascertain that transferred models perform equally to trained ones. For this, we assume that \textit{Xavier$_{GPU}$} wants to join the device cluster. According to Table~\ref{tab:hw-scalars}, \textit{Laptop} and \textit{Xavier$_{CPU}$} are eligible for providing their model, i.e., their hardware scalars (2 \& 4) are the closest to \textit{Xavier$_{GPU}$} (3). Hence, we merge their EOSC models and transfer the result to \textit{Xavier$_{GPU}$}. Next, we compare the SLO fulfillment of the merged model with a separate run, where a model is trained from scratch. We place both runs into Figure~\ref{fig:slo-kt}; the blue line represents the combined model, and the grey one was trained from scratch. Additionally, we indicate each time the agents changed the configuration.

We observe: (1) the merged model does not face any substantial improvements of its initially high SLO fulfillment; (2) the agent required 14 rounds to arrive at a comparable SLO rate -- this also matches our experience from Figure~\ref{fig:simple-slo-rate}, where \textit{Laptop} required 7 to 16 ACI rounds for training; and (3) the final rates are within the range [0.85,0.95]. From that, we conclude that the results produced by the trained model were comparable to the merged model (\textit{K-1}), and that KT could achieve a speedup of 14 rounds (\textit{K-2}), assuming that the transferred model was ready for usage. Nevertheless, this is only valid for the given setup (i.e., these two devices); it is not possible yet to derive general implications of our approach.

~

\noindent\textit{K-3: Can merged models decrease the FE compared to choosing a single one?}

~

\begin{figure}
  \centering
  \begin{subfigure}{\columnwidth}
    \centering
    \includegraphics[width=1.0\columnwidth]{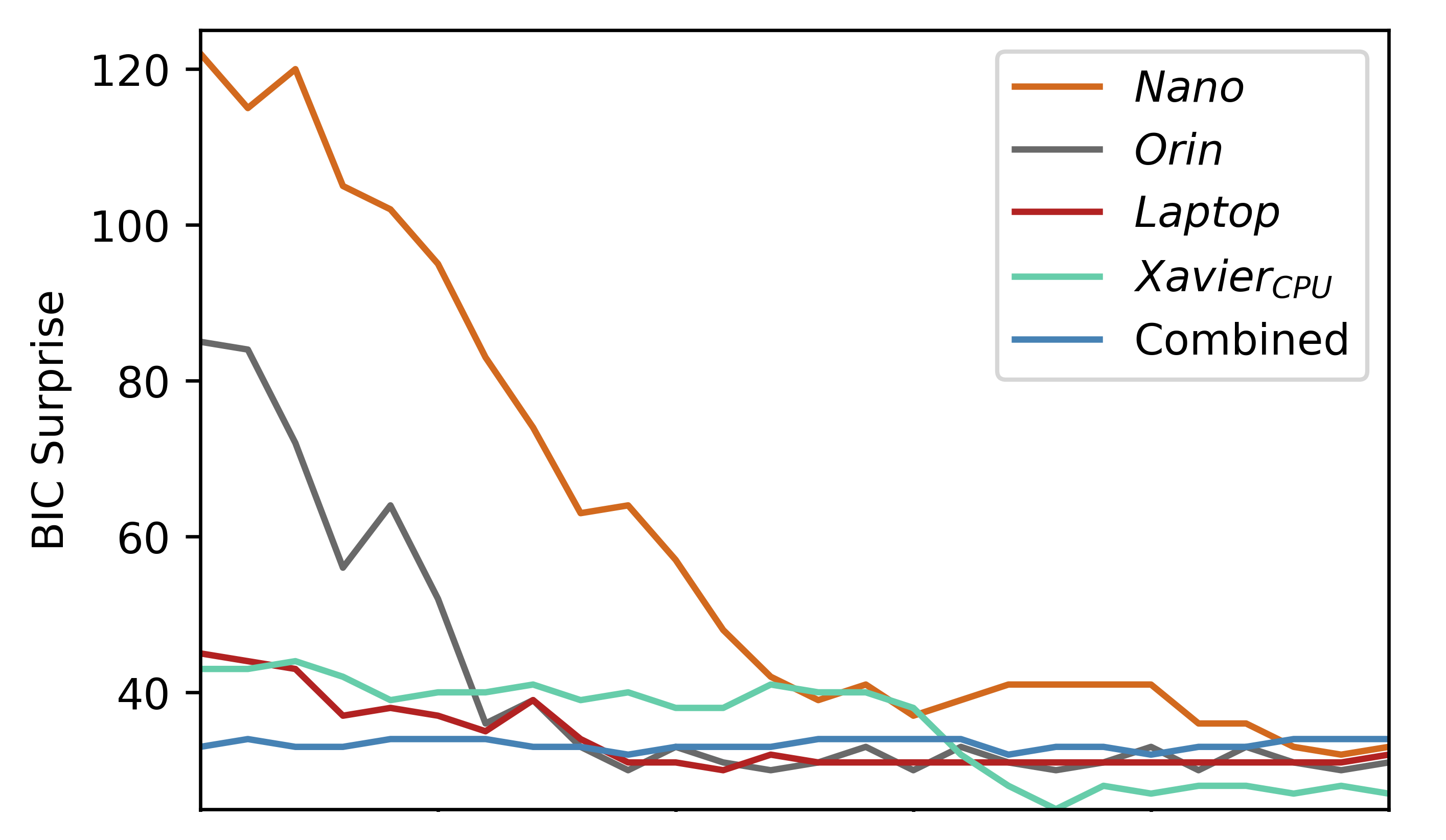}
  \end{subfigure}
  \begin{subfigure}{\columnwidth}
    \centering
    \includegraphics[width=1.0\columnwidth]{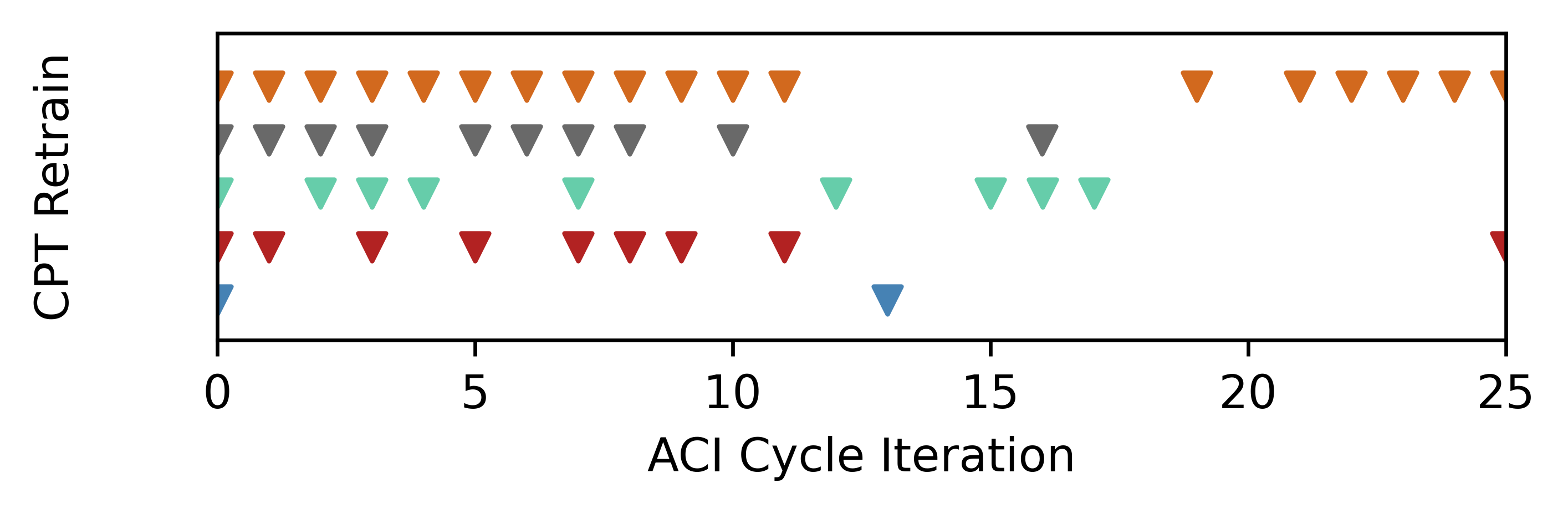}
  \end{subfigure}
  \caption{Overall surprise measured per batch when operating on the \textit{Xavier}$_{GPU}$ with existing models or one combined especially for this device. Paired with the frequency the CPTs are updated through parameter learning (\textit{K-3})}
  \label{fig:surprise-kt}
\end{figure}

As discussed in Section~\ref{subsec:background-aci}, it is hard to estimate the FE of a model, but we consider the fact that surprise is bounded by FE. Although low surprise does not imply low FE, we use it as an indicator:
We transfer a model to \textit{Xavier$_{GPU}$} (merged from \textit{Laptop} and \textit{Xavier$_{CPU}$} as above) and calculate the surprise throughout multiple ACI cycles. This we compare against alternative runs, in which \textit{Xavier$_{GPU}$} uses one of the EOSC models of the other devices (from Table Table~\ref{tab:device-list}). Furthermore, we count the usage of \texttt{PARL}. The results are presented in Figure~\ref{fig:surprise-kt}; each of the colored lines represents one of the respective models, which were copied to \textit{Xavier$_{GPU}$}. The blue line, however, describes the combined model. The lower figure shows for each run when \texttt{PARL} was executed.

We observe: (1) the models trained on \textit{Orin} and \textit{Nano} produced initially very high surprise ($>> 50$), indicating that these models fit \textit{Xavier$_{GPU}$} the least; (2) nevertheless, the agent was able to improve these models and converge to an area where all 5 models provide similar surprise after 25 iterations; (3) the combined model provided initially the best values and only performed \texttt{PARL} twice; and (4) interestingly, although close to each other, the combined model produces after 25 rounds the highest surprise (33), while \textit{Xavier$_{CPU}$} reached 17. This shows, that the frequent retraining performed by the other devices (colored triangles in the lower graph) allowed the other models to surpass \textit{Xavier$_{GPU}$}. This raises the question if it would be advisable to always run \texttt{PARL}, regardless of the surprise magnitude -- Are there even situations when the CPTs should be updated less frequently? Combined, we can answer that the merged model had initially less surprising values (\textit{K-3}); however, frequent retraining may achieve even better results.

\subsection{Stream Offloading}

\vspace{5pt}

\noindent\textit{S-1: How is the load distributed among resource-constrained devices}

~

\begin{figure}[t]
    \centering
    \includegraphics[width=1.0\columnwidth]{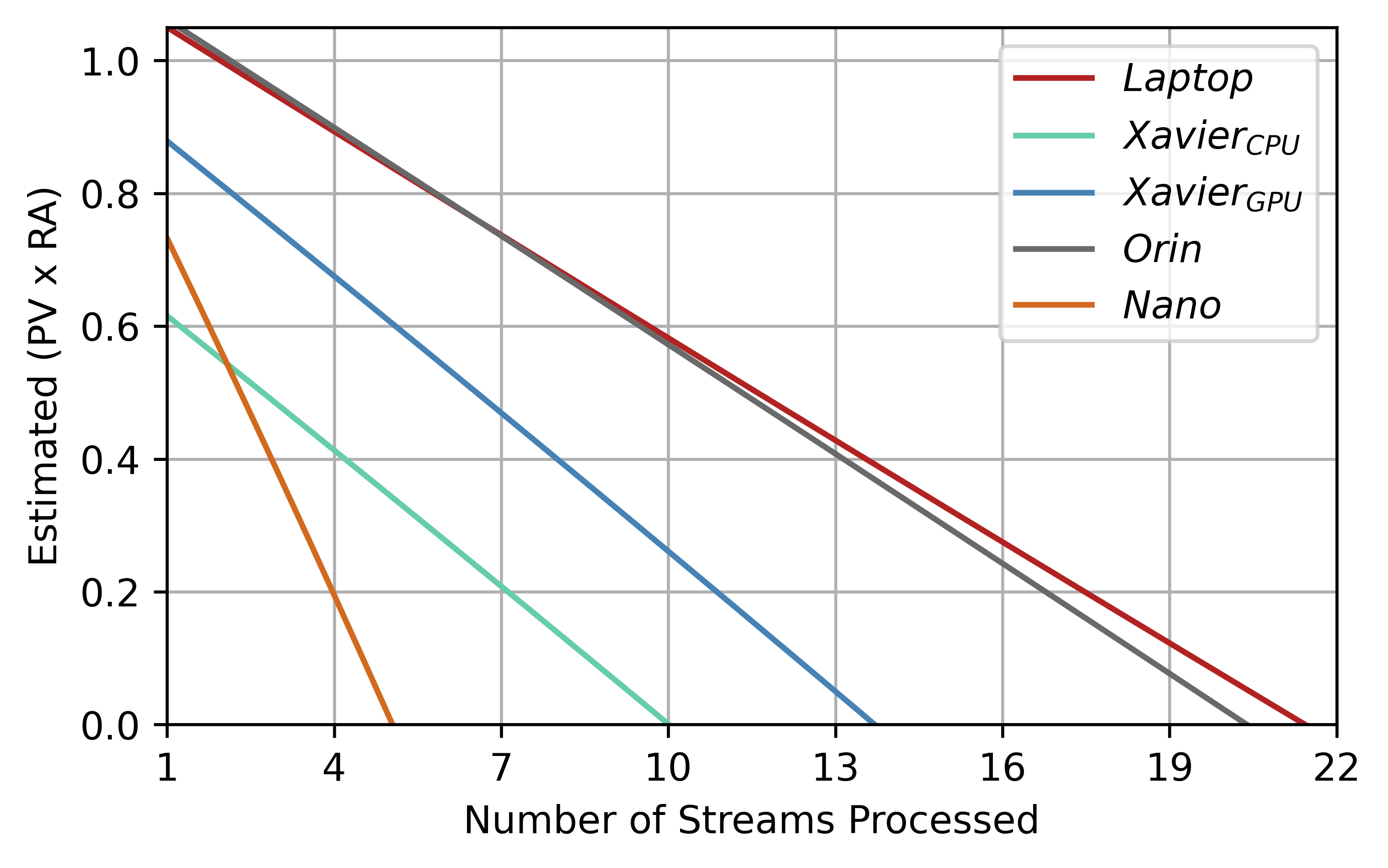}
    \caption{Regression curves between environmental demand for edge devices (\textit{streams}) and their respective SLO fulfillment rates ($pv \times ra$)}
    \label{fig:regression_slo}
\end{figure}

\begin{figure}[t] 
    \centering
    \subfloat[SLO fulfillment per device]{\includegraphics[width=\columnwidth]{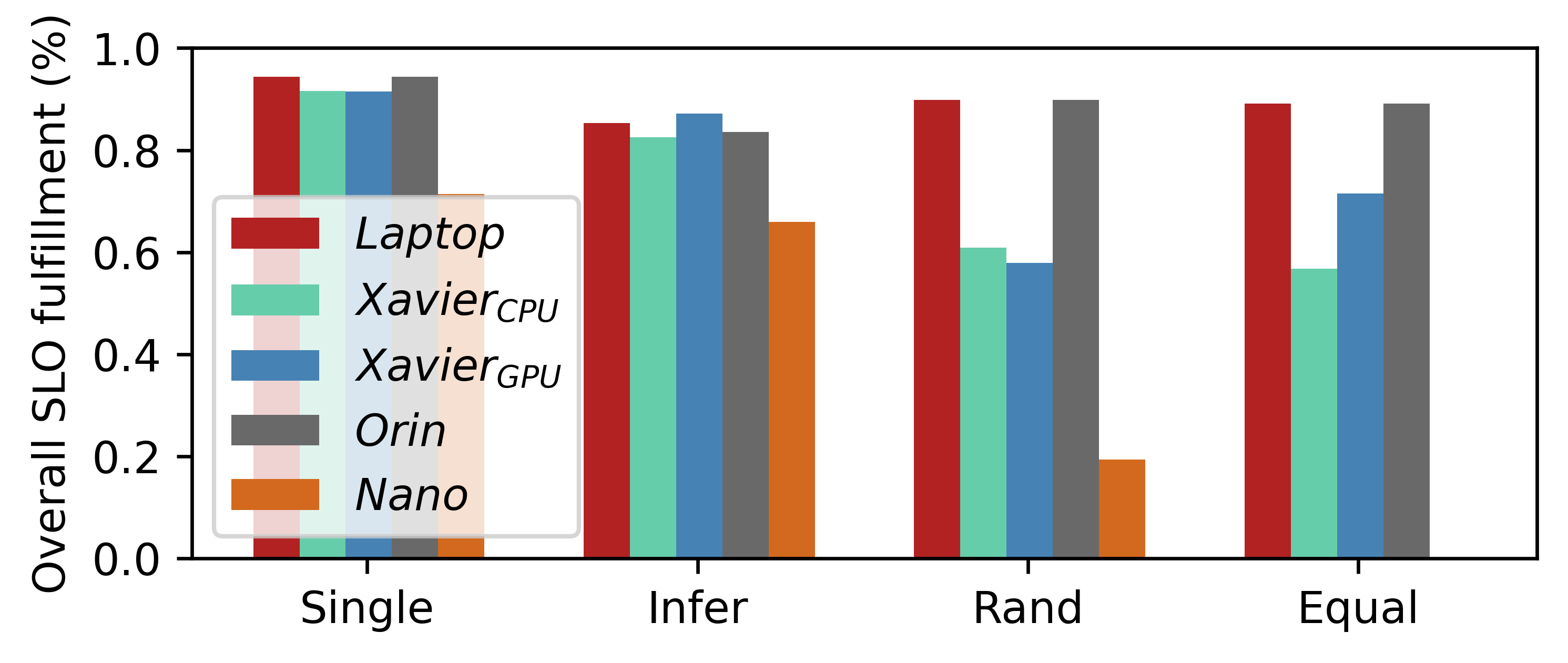}}
    \hfill
    \subfloat[Average and weighted average per batch]{\label{subfig:slo-rebalance-avg}\includegraphics[width=\columnwidth]{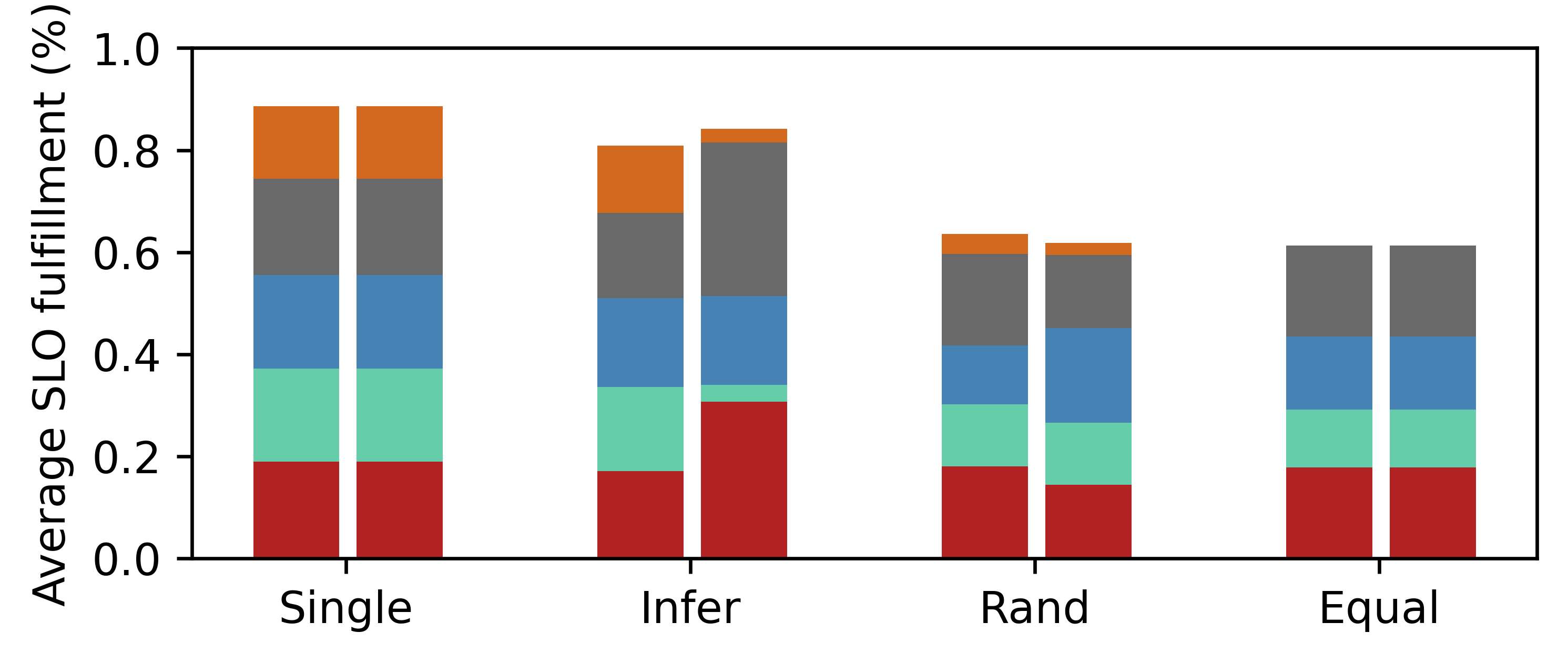}}
    
    \caption{SLO fulfillment within the edge-fog cluster when distributing load according to \textit{Infer}, \textit{Random}, or \textit{Equal}. \textit{Single} is an upper bar for this device constellation (\textit{S-1})}
    \label{fig:slo-rebalance}
\end{figure}

To offload computations within the cluster, we aim to show how low-resource devices are relieved from excessive load. For this, we assume 25 IoT devices that are either assigned \textit{Equal} to the edge devices or \textit{Random}. As an indicator for maximum SLO fulfillment, we added \textit{Single}, where each device processes one stream; Table~\ref{tab:streams-assigned} shows an overview of each scenario's assignment. After operating with \textit{Equal} or \textit{Random}, the leader node starts to optimize the environment, i.e., using the EOSC-F model to distribute the 25 streams depending on the device capabilities (\textit{Infer}). This new assignment is then provided to the edge devices. We thus simulated an offloading or load rebalancing, e.g., \textit{Nano} dropped from 5 (or 3) to 1 stream. In Figure~\ref{fig:slo-rebalance}, we show each device's SLO fulfillment rate per scenario. The left bars of Figure~\ref{subfig:slo-rebalance-avg} show the cluster-wide average of the SLO fulfillment and the right bar the weighted average according to the number of streams ($slo\_rate \times stream$). To get a feeling of the heterogeneous device capabilities, Figure~\ref{fig:regression_slo} provides a regression function that shows how the SLO fulfillment per device is impacted by the number of \textit{streams}.

\setlength{\tabcolsep}{4pt}
\begin{table}[h]
  \centering
  \caption{Streams assigned to each device for evaluated scenarios}
  \label{tab:streams-assigned}
  \begin{tabular}{lcccc}
    \toprule
    Device ID & \textit{Single} & \textit{Equal} & \textit{Rand} & \textit{Infer} \\
    \midrule
    \textit{Laptop} & 1 & 5 & 4 & 9 \\
    \textit{Xavier}$_{GPU}$ & 1 & 5 & 8 & 5 \\
    \textit{Xavier}$_{CPU}$ & 1 & 5 & 5 & 1 \\
    \textit{Orin}   & 1 & 5 & 4 & 9 \\
    \textit{Nano}   & 1 & 5 & 3 & 1 \\
    \midrule
    Sum $\Sigma$ & 5 & 25 & 25 & 25 \\
    \bottomrule
  \end{tabular}
\end{table}

We observe: 
(1) the average SLO fulfillment clearly improved by using \textit{Infer} (0.81) instead of \textit{Random} (0.64) or \textit{Equal} (0.60); (2) this is also reflected by the weighted average (right bars of Figure~\ref{subfig:slo-rebalance-avg}), which puts \textit{Laptop} and \textit{Orin} in focus that processed 9 streams each; (3) the weighted average of \textit{Infer} comes close to the one of \textit{Single} (0.89), even though the cluster processed 25 instead of only 5 streams. From that, we conclude that the intelligent cluster was able to incorporate restricted edge devices (e.g., \textit{Nano}) into the architecture (\textit{S-1}), and that the overall SLO compliance improved by following our approach.

~

\noindent\textit{S-2: Can the CC hierarchical structure optimize local SLO fulfillment? }

~

\begin{figure}[!t]
    \centering
    \includegraphics[width=0.65\columnwidth]{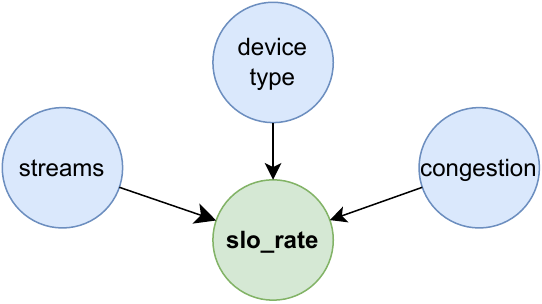}
    \caption{DAG internal to the EOSC-F model}
    \label{fig:dag-fog}
\end{figure}

\begin{figure}[!t]
    \centering
    \includegraphics[width=1.0\columnwidth]{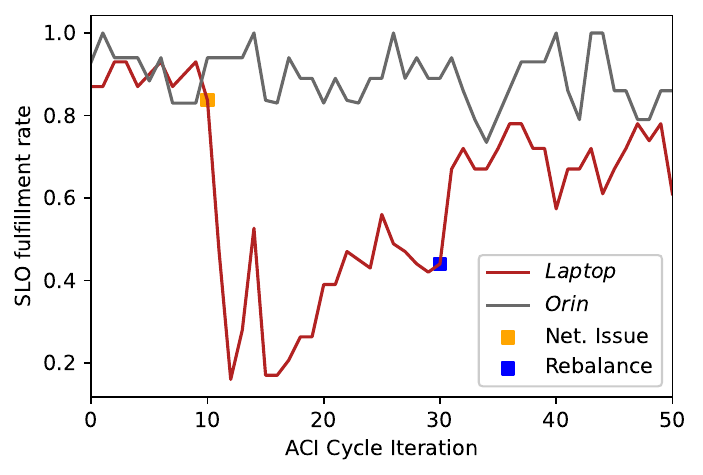}
    \caption{Recovering network congestion by rebalancing the load within the device cluster according to the EOSC-F model; both devices initially processed 5 streams, 3 are offloaded to \textit{Orin} (\textit{S-2})}
    \label{fig:rebalance}
\end{figure}

To improve the SLO fulfillment whenever individual devices lack the required scope, we will resolve such SLO failures within the cluster. Therefore, we consider a condensed device cluster consisting of \textit{Laptop} and \textit{Orin}. \textit{S-1} showed that they have comparable processing capabilities; therefore, it is fair to split 10 streams equally between them. Figure~\ref{fig:dag-fog} provides the DAG internal to the EOSC-F model: Blue nodes are environmental factors, from which only \textit{stream} can be configured (recall Section~\ref{subsubsec:variables-slos}); \textit{slo\_rate} represents the common factor $f = pv \times ra$.
We simulate network congestion\footnotemark for \textit{Orin} -- which the leader node can evaluate through \textit{congestion} -- and redistribute the load according to the EOSC-F model, i.e., $\textit{Orin}=8,\textit{Laptop} = 2$. 
Then, we compare the overall SLO fulfillment before and after offloading; the results are shown in Figure~\ref{fig:rebalance}. The two lines show the SLO fulfillment ($f$) of \textit{Laptop} (red) and \textit{Orin} (blue) over 50 ACI iterations; after 10 rounds, the network gets congested. In round 30, the cluster leader rebalanced the load according to its EOSC-F model; although it is possible to rebalance earlier, we decided to observe the system behavior until manually rebalancing in round 30.

We observe: (1) the network issue crushed the SLO fulfillment of \textit{Laptop} from around 0.9 to a minimum of 0.2 at round 15; (2) the edge device was able to improve the rate in the following 20 iterations by reconfiguration, until reaching a local optimum at 0.43. Further, (3) the cluster-wide SLO compliance was clearly improved through rebalancing, i.e., at round 15 the sum of $f_{\textit{Laptop}} + f_{\textit{Orin}}$ was 1.03, at round 30 it was 1.33, while at round 45 it rose to 1.54. We conclude that the intelligent cluster was able to resolve the introduced network issue (\textit{S-2}) by redistributing the load according to the EOSC-F model. However, to draw general conclusions, we aim to consider a larger range of potential issues.

\footnotetext{Internally, we increase the processing delay according to $congestion$; this increases the overall latency and causes the \textit{in\_time} SLO to fail more likely. The EOSC-F model can then consider $congestion$ as an environmental factor for Algorithm~\ref{alg:client-reassignment}.}

\subsection{Summary and Implications}








%

This section summarizes the presented results and highlights their implications for the applicability of our framework. To that extent, we can report that (1) edge devices were gradually able to ensure local SLO compliance without prior knowledge; it took them 16 rounds to identify factors that impact SLO fulfillment and adapt the environment accordingly, (2) the underlying causal structures and the transitions between device configuration were empirically explainable; this increases traceability and trust of ML models, and (3) shifted variable distributions were canceled out through continuous model retraining; edge devices took 9 rounds to interpret an unprecedented increase in demand, while SLO failures introduced by poor video quality could not be fully recovered. Further, (4) the causality filter based on MBs decreased the complexity of inference and sped up SLO evaluation by 17\%, and (5) our framework introduced a negligible CPU overhead of 3\%, which makes it a suitable choice for resource-restricted devices.

It turned out that (6) BNL, or in particular structure learning, surpassed the given time frame for continuous model adaptation; nevertheless, parameter learning alone took less than 250 ms. Thus, (7) models transferred between nearby devices could be continuously improved, even in cases where they fit poorly; this improves the reusability of models in the heterogeneous Edge, (8) the SLO fulfillment of devices with transferred models equaled the one of self-trained models; this accelerated the distribution of SLO-compliance models within one computational tier by up to 16 rounds, (9) rebalancing the load after a network error increased the overall SLO fulfillment from 1.03 to 1.54; this showed that collaboration within this tier increased the scope of SLO failures that could be covered.
A closing observation is that (10) shifts in the variable distribution showed the same effects on SLO fulfillment as low accuracy after transferring a model to an unknown device type. To our framework, they did not provide any fundamental differences, which is why they could both be resolved through continuous model training.

\section{Related Work}\label{sec:RelatedWork}
This section provides recently published related works that discuss (1) the training and application of causal ML models on the Edge, (2) transfer learning approaches in the CC, and (3) methods of load balancing and computation offloading that are popular across the CC. Following that, we highlight for each of these fields the research gap that our work aims to fill.


\subsection{Causal ML Training on the Edge}
\textit{Sudharsan et al.} \cite{sudharsan_edge2train_2020} developed an \texttt{Edge2Train} model to analyze real-time data on the fly. With Edge2Train, Support Vector Machine (SVM) models are trained offline in edge nodes using real-time IoT.
Adopting causality to Edge2Train can help converge the most efficient training models quickly. 
Diagnosing the root cause of performance degradation in the CC is a challenging issue, and \textit{Chen et al} \cite{chen_causeinfer_2019} use causal inference (\texttt{CauseInfer}) mechanisms to pinpoint the root causes within the system. CauseInfer determines fault propagation paths that can be determined explicitly, without production systems being instrumented. 
A similar approach (called \texttt{Nazar}) is designed by \textit{Hao et al.} in \cite{hao_monitoring_2023}, where they apply mobile devices to diagnose root causes in distributed systems. Further, this approach enhances its training models through cause-specific adaptive mechanisms. Through experiments, Nazar confirmed that training models can be improved due to cause-specific adaptation while monitoring a large number of devices. 

\textit{Lin et al.} introduced \texttt{Microscope} in \cite{lin_microscope_2018}, a micro-service environment to diagnose the possible root causes of abnormal services in distributed systems through causal graphs. \textit{Lin et al.} demonstrate that Microscope can construct a service causal graph in real time and infer the root cause of abnormal services. \textit{Tariq et al.} present the What-If Scenario Evaluator (\texttt{WISE}) tool in \cite{tariq_answering_2008}, which predicts the effect of potential configuration and deployment changes on content delivery networks (CDN). WISE initially learns causal relations among existing response time distributions. Based on the available datasets, it estimates possible future response time distributions. Finally, it allows network designers to express possible deployment scenarios without knowing how variables will affect response time. 

~

There evidently exists work that identifies and applies causal understanding to ensure system requirements; however, with the exception of Nazar~\cite{hao_monitoring_2023}, they treat model training as a one-time process. Hence, drifts (or shifts) in the variable distribution stay undetected. Further, it is impractical to assume that sufficient training data is available to arrive at this causal understanding; this is also the shortcoming of Nazar. Contrarily, our approach, which focuses on ACI, is able to gradually create causal models over multiple iterations (i.e., as new training data becomes available), and continuously ensures model accuracy by updating beliefs according to prediction errors.



\subsection{Transfer Learning in the CC}
\textit{Goyal et al.} present \texttt{MyML}~\cite{goyal_hardware-friendly_2022}, a hardware-friendly model transfer for edge nodes. MyML uses transfer learning to create small, lightweight, custom ML models based on user preferences. This approach is hardware-friendly, bottom-up pruning, which can be utilized on any mobile edge platform because of its ability to handle large, compute-intensive ML models. In addition, systolic array-based edge accelerators are introduced to prevent cloud interactions. \textit{Wu et al.} present a novel approach to online transfer learning for both heterogeneous and homogeneous labels of multi-source domains \cite{wu_online_2017}. This approach is very efficient in online classification, and the weights are dynamically adjusted depending on the source domain. The work fits well into the CC due to the complex heterogeneity of devices within the system. \textit{Hsu et al.} provide a clustering mechanism that considers the similarity of domains and tasks for transfer learning \cite{hsu_learning_2018}. They provided a similarity function for cross-task transfer learning that is based on similarities between domains.

\textit{Xing et al.} introduced a model called \texttt{RecycleML} in \cite{xing_enabling_2018} that enables multi-modality among edge devices, where knowledge is shared by transforming common latent features into their lower layers. Further, it provides task-specific knowledge transfer between models through the retraining of higher layers beyond the latent space shared by both models, thus reducing the need for labeled data. 
\textit{Sharma et al.} proposed a knowledge transfer technique between edge devices to lower computational intensity without losing accuracy and convergence speed \cite{sharma_are_2018}. In this, the student network takes the knowledge from the teacher network to achieve this goal. Using an IoT testbed, \textit{Kolcun et al.} \cite{kolcun_case_2020} evaluated various machine learning classifiers' convergence speed and accuracy. These testbeds considered both data- and resource-specific constraints. The results of each local testbed's training models are transmitted to the gateway to minimize global training model overhead.   

~

Transferring ML models is an important measure for relieving resource-restricted devices from training; teacher devices can therefore consider the context of the student to provide a tailored model. This is an important feature since edge devices have heterogeneous characteristics; however, none of the presented works considered low-level hardware characteristics to identify potential teachers among nearby devices. Further, while it is possible to combine models, the presented techniques are not applicable to the causal structures that we require for decentralized SLO assurance.
To that extent, our framework uses hardware classification to find adequate models within a device cluster and creates a tailored model by merging the conditional probabilities of BNs.


\subsection{SLO-Induced Load Balancing and Offloading}

Elasticity is one of the most effective ways to ensure requirements of dynamic workloads by automatically provisioning or de-provisioning resources based on demand \cite{dustdar_principles_2011}. \texttt{SLOC} is a novel elastic framework developed by \textit{Nastic et al.} in \cite{nastic_sloc_2020}, that allows users to provide and consume cloud resources in an SLO-native manner while guaranteeing performance. Its primary goal is to provide better support for SLOs by exploiting and advancing current elasticity management solutions. Further, \textit{Furst et al.} bring elastic service principles from the cloud to edge computing \cite{furst_elastic_2018}. They evaluated elastic and non-elastic services at the edge while processing images to latency SLOs, and noticed improved service provisioning through elasticity. 


\textit{Tran and Kim} introduce an edge serverless auto-scaling method based on traffic prediction that can be used against a Kubernetes cluster \cite{TRAN2024304}. In this work, system resource usage is optimized to ensure latency SLOs. No additional resources are required to perform this operation; this optimizes the amount of available resources. 
\textit{Hazra et al.} \cite{hazra2022cooperative} proposed efficient heuristic-based transmission scheduling and graph-based computational offloading (\texttt{TSCO}) through mixed linear programming to achieve energy efficiency and minimize latency.  
A single- and multi-task load balancing with a prioritization approach to computing Deep Neural Networks (DNNs) at the edge has been presented by \textit{Karjee et al.} in \cite{karjee_energy_2021}. In these approaches, prioritized tasks are distributed among IoT and edge nodes to balance energy, lower latency, and continue task execution without restarting the system. \textit{Lim and Lee} proposed a load-balancing approach for distributing mobile devices tasks within a cloud-edge continuum using graph coloring \cite{lim_load_2020}. Through this process, computing resources are scaled with increased edge resource utilization. 

A trilayer mobile hybrid hierarchical peer-to-peer (MHP2P) model was proposed by \textit{Duan et al.} in \cite{duan_novel_2022} as a cloudlet for efficient load balancing strategy through mobile edge computing (MEC). MHP2P promises high reliability, scalability, and efficiency in service lookups. Moreover, there is a load-balancing scheme to ensure that MHP2P loads are evenly distributed between MEC servers and queries. \textit{Rao et al.} presented another load-balancing strategy for P2P systems through virtual servers \cite{rao_load_2004}; it presents three basic load-balancing schemes whose main difference is the amount of information required for rearrangement. In \cite{menino_novel_2021}, \textit{Menino} proposed efficient failure detection mechanisms for unstructured overlay networks. This approach aims to identify efficient neighborhood overlays, which dynamically identify and maintain each node in P2P networks.


~

SLOs are an efficient way for modeling and enforcing requirements; thus, high-level SLOs can be segregated and enforced at the respective CC component. Nevertheless, the remaining question is whether the component has the required scope to recover SLO failures (e.g., by offloading computation), but it is impractical to evaluate SLOs in the cloud (e.g., MHP2P). Hence, ad-hoc hierarchical structures could provide a remedy, which \textit{Menino}~\cite{menino_novel_2021} are the only ones to use among the related work. However, they all assume prior knowledge of which variables impact SLO fulfillment.
Contrarily, our approach (1) gradually increases the SLO scope by forming device clusters that can span the entire CC, and (2) evaluates causal relations among environmental variables to shift the load from impacted devices.


\section{Conclusion and Future Work}\label{sec:Conclusion}
\label{sec:conclusion}



This paper presented a novel framework for collaborative and distributed edge intelligence that ensures decentralized SLO fulfillment. It allows CC systems to break down high-level requirements and enforce them at the component that they concern; thereby, we create self-adaptive devices that themselves ensure dynamic requirements. 
For each of its components, the framework is able to develop causal reasoning between environmental factors and SLO fulfillment. 
Resource-restricted devices that cannot create this knowledge themselves were able to exchange and combine causal models according to their hardware characteristics. This accelerates the onboarding of unknown device types and simplifies horizontal scaling within the Edge. Contrarily, any attempt to achieve this centrally would struggle with heterogeneous device characteristics, the induced network latency, and the communication overhead.
To create advanced SLOs with an increased scope, devices collaborated as clusters under the supervision of a Fog node; this forms higher-level components that can again supervise their own set of SLOs. As a consequence, the cluster was able to use its extended environment to resolve SLO violations, e.g., by offloading computation among pertinent devices. Erecting these hierarchical structures allows us to fulfill the intricate requirements of multiple computational tiers.





We provided a prototype of the framework for a distributed video transformation use case and evaluated it according to twelve aspects; the results showed the potential of our approach for ensuring SLOs throughout CC tiers. For future work, we aim to dynamically assemble hierarchical structures and evaluate limitations regarding the number of SLOs and devices that can be managed. Further, this work builds heavily on the analysis of causal relations between SLO fulfillment and environmental factors; however, to claim true causality, dedicated experiments must be integrated into the framework. 
%
%
%
Once this is established, our framework will provide the necessary causal links to tame requirements within the CC.

\section*{Acknowledgement}
Funded by \textit{the European Union (TEADAL, 101070186)}. Views and opinions expressed are however those of the author(s) only and do not necessarily reflect those of the European Union. Neither the European Union nor the granting authority can be held responsible for them.


\section*{Nomenclature}
This section provide a summary of notations (\ref{tab:Notations}) and acronyms (\ref{tab:Acronyms}) used in this paper. 
\begin{table}[!t]
    \centering\footnotesize
    \caption{Frequently used notations.}
    \label{tab:Notations}
    \begin{tabular}{r|l}\hline
       \textbf{Notation}  & \textbf{Meaning} \\\hline
       $D_{KL}$ & Kullback-Leibler divergence\\
        $x$ & Hidden states of a model\\
        $P$ & Exact posterior probability\\
        $Q$ & Approximate posterior probability\\
        $\Im$ & Surprise for observation\\
        $m$ & Model\\
        $m_{ab}$ & Merged BN model  $m_a$ and $m_b$\\
        $v$ & Number of variables in a BN \\
        $o$ & Observation\\ 
        $D$ & Datasets\\
        $G$ & Graph\\
        $G^*$ & Set of possible DAGs for a $G$\\
        $G_{final}$&  DAG\\
        $LL(G,D)$ & Log-likelihood for $G$ of $D$\\
        $\phi(|D|)$ & BIC\\
        $\mathcal{V}$ & Vertex set of $G$\\
        $\mathcal{E}$ & Edge set of $G$ \\ 
        $\xi$ & Entropy\\
        $MLE$ & Maximum Likelihood Estimation\\
        $\omega$ & Width of DAG\\
        $n$ & Number of nodes ($\mathcal{V}$) in DAG \\ 
        $\mathcal{P}$ & Size of conditioned set while
searching PC\\
        $\mathcal{C}$ & max(PC set) for $n$\\
        $\mathcal{S}$ & Number of Spouses from $G_{final}$\\
        $\mathcal{K}$ & Maximum size Spouses from $G_{final}$\\
        $T$ & Target variables\\
        $O$ & Elimination order \\
        $mb$ & Markov blanket\\
        $t$ & Threshold\\
        $pv$ & Pragmatic value\\
        $ra$ & Risk assigned\\
        $ig$ & Information gain\\
        $u$ & Combined behavioral factors\\
        $f$ & Combined SLO fulfillment\\
        $K$ & Behavioral factors of known configurations\\
        $ev$ & Evidence variable list\\
        $V_{SLO}$ & SLO target variables list\\
        $V_Q$ & Subset of $V_{SLO}$, either QoS or QoE\\
        $e$ & Constant for visiting state more frequently\\
        $h$ & Factor for choosing between \texttt{STRL} and \texttt{PARL}\\
        $\tilde{\Im}_{c}$ & Median surprise for configuration $c$\\ 
        $\bar{\Im}$ & Mean overall surprise\\
        $c$ & Configuration\\
        $q$ & Number of states for a variable r in a BN\\
        $p$ & CPU capacity\\
        $g$ & GPU capacity\\
        $dc$ & Device capacity\\
        $w$ & Coefficient for relative device capacity\\
        $k_r$ & Incoming edges for a variable $r$ in a BN \\
        $r$ & Random variable \\
        $\sigma_{env}$ & Environment factors\\
        $\Lambda$ & Physical devices\\
        $n_{clients}$ & Number of streams\\
        $d_a$ & backup data to create $m_a$\\
        
    \hline
    \end{tabular}
\end{table}

\begin{table}[h]
    \centering  \footnotesize 
    \caption{Frequently used acronyms and their meaning}
    \label{tab:Acronyms}
    \begin{tabular}{r|l}\hline
       \textbf{Acronym}  & \textbf{Meaning} \\\hline
       AxI & Approximate Inference\\ 
       ACI & Active Inference\\
       BIC& Bayesian Information Criterion\\
       BN& Bayesian Network\\
       BNL& Bayesian Network Learning\\
       CC & Computing Continuum\\
       CDN & Content Distribution Networks\\
       CPTs& Conditional Probability Tables\\
       CPU & Central Processing Unit\\
       DAG & Directed Acyclic Graph\\
       DL & Deep Learning \\ 
       DNN & Deep Neural Networks\\ 
       EI& Exact Inference\\
       EOSC& Equilibrium-Oriented SLO-Compliance\\
       FE& Free Energy\\
       GPU& Graphics Processing Unit\\
       HCS& Hill-Climb Search\\
       IoT& Internet of Things\\
       KT& Knowledge Transfer\\
       MB& Markov Blanket\\
       MEC & Mobile Edge Computing\\
       ML& Machine Learning\\
       MLE&  Maximum Likelihood Estimation\\
       P2P & Peer-to-Peer\\ 
       PC& Parent-Child\\
       PERL & Parameter Learning\\
       QoE & Quality of Experience \\
       QoS & Quality of Service\\ 
       TL& Transfer Learning\\
       SCM& Structural causal models\\ 
       SLOs & Service Level Objectives\\
       STRL & Structure Learning\\
       SVM& Support Vector Machine\\
       VE& Variable Elimination\\
       WISE& What-If Scenario Evaluator\\
    \hline
    \end{tabular}
\end{table}

\bibliographystyle{elsarticle-num}
\bibliography{praveen}

\begin{thebibliography}{10}
\expandafter\ifx\csname url\endcsname\relax
  \def\url#1{\texttt{#1}}\fi
\expandafter\ifx\csname urlprefix\endcsname\relax\def\urlprefix{URL }\fi
\expandafter\ifx\csname href\endcsname\relax
  \def\href#1#2{#2} \def\path#1{#1}\fi

\bibitem{Beckman2020}
P.~Beckman, et~al., Harnessing the computing continuum for programming our world, in: Fog {Computing}, John Wiley \& Sons, Ltd, 2020, pp. 215--230.

\bibitem{dustdar_distributed_2023}
S.~Dustdar, V.~C. Pujol, P.~K. Donta, On {Distributed} {Computing} {Continuum} {Systems}, IEEE Transactions on Knowledge and Data Engineering 35~(4) (2023) 4092--4105.
\newblock \href {https://doi.org/10.1109/TKDE.2022.3142856} {\path{doi:10.1109/TKDE.2022.3142856}}.

\bibitem{tarneberg_6g_2022}
W.~Tärneberg, et~al., The {6G} {Computing} {Continuum} ({6GCC}): {Meeting} the {6G} computing challenges, in: {International} {Conference} on {6G} {Networking}, 2022.

\bibitem{casamayor-pujol_fundamental_2023}
V.~Casamayor-Pujol, A.~Morichetta, I.~Murturi, P.~K. Donta, S.~Dustdar, Fundamental {Research} {Challenges} for {Distributed} {Computing} {Continuum} {Systems}, Information 14 (2023) 198.
\newblock \href {https://doi.org/10.3390/info14030198} {\path{doi:10.3390/info14030198}}.

\bibitem{sedlak_designing_2023}
B.~Sedlak, V.~C. Pujol, P.~K. Donta, S.~Dustdar, Designing {Reconfigurable} {Intelligent} {Systems} with {Markov} {Blankets}, in: F.~Monti, S.~Rinderle-Ma, A.~Ruiz~Cortés, Z.~Zheng, M.~Mecella (Eds.), Service-{Oriented} {Computing}, Lecture {Notes} in {Computer} {Science}, Springer Nature Switzerland, Cham, 2023, pp. 42--50.
\newblock \href {https://doi.org/10.1007/978-3-031-48421-6_4} {\path{doi:10.1007/978-3-031-48421-6_4}}.

\bibitem{friston_free_2023}
K.~Friston, L.~Da~Costa, N.~Sajid, C.~Heins, K.~Ueltzhöffer, G.~A. Pavliotis, T.~Parr, The free energy principle made simpler but not too simple (May 2023).
\newblock \href {https://doi.org/10.48550/arXiv.2201.06387} {\path{doi:10.48550/arXiv.2201.06387}}.

\bibitem{kokkonen_autonomy_2023}
H.~Kokkonen, L.~Lovén, N.~H. Motlagh, A.~Kumar, J.~Partala, T.~Nguyen, V.~C. Pujol, P.~Kostakos, T.~Leppänen, A.~González-Gil, E.~Sola, I.~Angulo, M.~Liyanage, M.~Bennis, S.~Tarkoma, S.~Dustdar, S.~Pirttikangas, J.~Riekki, Autonomy and {Intelligence} in the {Computing} {Continuum}: {Challenges}, {Enablers}, and {Future} {Directions} for {Orchestration} (Feb. 2023).
\newblock \href {https://doi.org/10.48550/arXiv.2205.01423} {\path{doi:10.48550/arXiv.2205.01423}}.

\bibitem{pearl_causal_2009}
J.~Pearl, Causal inference in statistics: {An} overview, Statistics Surveys 3 (2009) 96--146, publisher: Amer. Statist. Assoc., the Bernoulli Soc., the Inst. Math. Statist., and the Statist. Soc. Canada.
\newblock \href {https://doi.org/10.1214/09-SS057} {\path{doi:10.1214/09-SS057}}.

\bibitem{ganguly_review_2023}
N.~Ganguly, D.~Fazlija, M.~Badar, M.~Fisichella, S.~Sikdar, J.~Schrader, J.~Wallat, K.~Rudra, M.~Koubarakis, G.~K. Patro, W.~Z.~E. Amri, W.~Nejdl, A {Review} of the {Role} of {Causality} in {Developing} {Trustworthy} {AI} {Systems} (Feb. 2023).
\newblock \href {https://doi.org/10.48550/arXiv.2302.06975} {\path{doi:10.48550/arXiv.2302.06975}}.

\bibitem{chen_causeinfer_2019}
P.~Chen, Y.~Qi, D.~Hou, {CauseInfer}: {Automated} {End}-to-{End} {Performance} {Diagnosis} with {Hierarchical} {Causality} {Graph} in {Cloud} {Environment}, IEEE Transactions on Services Computing (2019).
\newblock \href {https://doi.org/10.1109/TSC.2016.2607739} {\path{doi:10.1109/TSC.2016.2607739}}.

\bibitem{lin_microscope_2018}
J.~Lin, P.~Chen, Z.~Zheng, Microscope: {Pinpoint} {Performance} {Issues} with {Causal} {Graphs} in {Micro}-service {Environments}, in: C.~Pahl, M.~Vukovic, J.~Yin, Q.~Yu (Eds.), Service-{Oriented} {Computing}, Lecture {Notes} in {Computer} {Science}, Springer International Publishing, Cham, 2018, pp. 3--20.
\newblock \href {https://doi.org/10.1007/978-3-030-03596-9_1} {\path{doi:10.1007/978-3-030-03596-9_1}}.

\bibitem{tsamardinos_time_2003}
I.~Tsamardinos, C.~F. Aliferis, A.~Statnikov, Time and sample efficient discovery of {Markov} blankets and direct causal relations, Association for Computing Machinery, New York, USA, 2003.
\newblock \href {https://doi.org/10.1145/956750.956838} {\path{doi:10.1145/956750.956838}}.

\bibitem{niculescu-mizil_inductive_2007}
A.~Niculescu-Mizil, R.~Caruana, Inductive {Transfer} for {Bayesian} {Network} {Structure} {Learning}, in: Proceedings of the {Eleventh} {International} {Conference} on {Artificial} {Intelligence} and {Statistics}, PMLR, 2007, pp. 339--346, iSSN: 1938-7228.

\bibitem{vowels_ya_2021}
M.~J. Vowels, N.~C. Camgoz, R.~Bowden, D'ya like {DAGs}? {A} {Survey} on {Structure} {Learning} and {Causal} {Discovery} (Mar. 2021).
\newblock \href {https://doi.org/10.48550/arXiv.2103.02582} {\path{doi:10.48550/arXiv.2103.02582}}.

\bibitem{pujol_towards_2021}
V.~C. Pujol, P.~Raith, S.~Dustdar, Towards a new paradigm for managing computing continuum applications, in: 2021 {IEEE} {Third} {International} {Conference} on {Cognitive} {Machine} {Intelligence} ({CogMI}), 2021, pp. 180--188.
\newblock \href {https://doi.org/10.1109/CogMI52975.2021.00032} {\path{doi:10.1109/CogMI52975.2021.00032}}.

\bibitem{friston_life_2013}
K.~Friston, Life as we know it, Journal of The Royal Society Interface 10~(86) (2013) 20130475.
\newblock \href {https://doi.org/10.1098/rsif.2013.0475} {\path{doi:10.1098/rsif.2013.0475}}.

\bibitem{kirchhoff_markov_2018}
M.~Kirchhoff, T.~Parr, E.~Palacios, K.~Friston, J.~Kiverstein, The {Markov} blankets of life: autonomy, active inference and the free energy principle, Journal of The Royal Society Interface (2018).

\bibitem{friston_reinforcement_2009}
K.~J. Friston, J.~Daunizeau, S.~J. Kiebel, Reinforcement {Learning} or {Active} {Inference}?, PLOS ONE 4~(7) (2009) e6421.
\newblock \href {https://doi.org/10.1371/journal.pone.0006421} {\path{doi:10.1371/journal.pone.0006421}}.

\bibitem{smith_step-by-step_2022}
R.~Smith, K.~J. Friston, C.~J. Whyte, A step-by-step tutorial on active inference and its application to empirical data, Journal of Mathematical Psychology 107 (2022) 102632.
\newblock \href {https://doi.org/10.1016/j.jmp.2021.102632} {\path{doi:10.1016/j.jmp.2021.102632}}.

\bibitem{sajid_active_2021}
N.~Sajid, P.~J. Ball, T.~Parr, K.~J. Friston, Active inference: demystified and compared, Neural Computation 33~(3) (2021) 674--712.
\newblock \href {https://doi.org/10.1162/neco_a_01357} {\path{doi:10.1162/neco_a_01357}}.

\bibitem{parr_active_2022}
T.~Parr, G.~Pezzulo, K.~J. Friston, Active {Inference}: {The} {Free} {Energy} {Principle} in {Mind}, {Brain}, and {Behavior}, The MIT Press, 2022.
\newblock \href {https://doi.org/10.7551/mitpress/12441.001.0001} {\path{doi:10.7551/mitpress/12441.001.0001}}.

\bibitem{camps-valls_discovering_2023}
G.~Camps-Valls, A.~Gerhardus, U.~Ninad, G.~Varando, G.~Martius, E.~Balaguer-Ballester, R.~Vinuesa, E.~Diaz, L.~Zanna, J.~Runge, Discovering causal relations and equations from data, Physics Reports 1044 (2023) 1--68.
\newblock \href {https://doi.org/10.1016/j.physrep.2023.10.005} {\path{doi:10.1016/j.physrep.2023.10.005}}.

\bibitem{heins_pymdp_2022}
C.~Heins, B.~Millidge, D.~Demekas, B.~Klein, K.~Friston, I.~Couzin, A.~Tschantz, pymdp: {A} {Python} library for active inference in discrete state spaces, Journal of Open Source Software (May 2022).
\newblock \href {https://doi.org/10.21105/joss.04098} {\path{doi:10.21105/joss.04098}}.

\bibitem{sedlak_active_2023}
B.~Sedlak, V.~C. Pujol, P.~K. Donta, S.~Dustdar, Active {Inference} on the {Edge}: {A} {Design} {Study} (Nov. 2023).
\newblock \href {https://doi.org/10.48550/arXiv.2311.10607} {\path{doi:10.48550/arXiv.2311.10607}}.

\bibitem{levchuk_active_2019}
G.~Levchuk, K.~Pattipati, D.~Serfaty, A.~Fouse, R.~McCormack, Active {Inference} in {Multiagent} {Systems}: {Context}-{Driven} {Collaboration} and {Decentralized} {Purpose}-{Driven} {Team} {Adaptation}, in: Artificial {Intelligence} for the {Internet} of {Everything}, Academic Press, 2019.

\bibitem{scanagatta2019survey}
M.~Scanagatta, A.~Salmer{\'o}n, F.~Stella, A survey on bayesian network structure learning from data, Progress in Artificial Intelligence 8 (2019) 425--439.

\bibitem{donta_governance_2023}
P.~K. Donta, B.~Sedlak, V.~Casamayor~Pujol, S.~Dustdar, Governance and sustainability of distributed continuum systems: a big data approach, Journal of Big Data 10~(1) (2023) 53.
\newblock \href {https://doi.org/10.1186/s40537-023-00737-0} {\path{doi:10.1186/s40537-023-00737-0}}.

\bibitem{kwisthout2011most}
J.~Kwisthout, Most probable explanations in bayesian networks: Complexity and tractability, International Journal of Approximate Reasoning 52~(9) (2011) 1452--1469.

\bibitem{scutari2019learning}
M.~Scutari, C.~Vitolo, A.~Tucker, Learning bayesian networks from big data with greedy search: computational complexity and efficient implementation, Statistics and Computing 29 (2019) 1095--1108.

\bibitem{darwiche2008bayesian}
A.~Darwiche, Bayesian networks, Foundations of Artificial Intelligence 3 (2008) 467--509.
\newblock \href {https://doi.org/10.1016/S1574-6526(07)03011-8} {\path{doi:10.1016/S1574-6526(07)03011-8}}.

\bibitem{aliferis_short_2010}
C.~Aliferis, et~al., Local {Causal} and {Markov} {Blanket} {Induction} for {Causal} {Discovery} and {Feature} {Selection} for {Classification} {Part} {I}: {Algorithms} and {Empirical} {Evaluation}, Journal of Machine Learning Research 11 (2010) 171--234.

\bibitem{casamayor_pujol_towards_2021}
V.~Casamayor~Pujol, P.~Raith, S.~Dustdar, Towards a new paradigm for managing computing continuum applications, in: {IEEE} 3rd {International} {Conference} on {Cognitive} {Machine} {Intelligence}, {CogMI} 2021, 2021, pp. 180--188.

\bibitem{gao2016efficient}
T.~Gao, Q.~Ji, Efficient markov blanket discovery and its application, IEEE transactions on Cybernetics 47~(5) (2016) 1169--1179.
\newblock \href {https://doi.org/10.1109/TCYB.2016.2539338} {\path{doi:10.1109/TCYB.2016.2539338}}.

\bibitem{zhang_simple_1994}
N.~Zhang, D.~Poole, A simple approach to {Bayesian} network computations, in: Engineering-{Economic} {Systems}, {Stanford} {University}, 1994.

\bibitem{dustdar_principles_2011}
S.~Dustdar, Y.~Guo, B.~Satzger, H.-L. Truong, Principles of {Elastic} {Processes}, Internet Computing, IEEE 15 (2011) 66--71.
\newblock \href {https://doi.org/10.1109/MIC.2011.121} {\path{doi:10.1109/MIC.2011.121}}.

\bibitem{ren2022bayesian}
D.~Ren, J.~Guo, X.~Hao, Bayesian network variable elimination method optimal elimination order construction, in: ITM Web of Conferences, Vol.~45, EDP Sciences, 2022.

\bibitem{ghio_bayes-optimal_2023}
D.~Ghio, A.~L.~M. Aragon, I.~Biazzo, L.~Zdeborová, Bayes-optimal inference for spreading processes on random networks, Physical Review E 108~(4) (2023) 044308.
\newblock \href {https://doi.org/10.1103/PhysRevE.108.044308} {\path{doi:10.1103/PhysRevE.108.044308}}.

\bibitem{lei2012cpu}
G.~Lei, Y.~Dou, W.~Wan, F.~Xia, R.~Li, M.~Ma, D.~Zou, Cpu-gpu hybrid accelerating the zuker algorithm for rna secondary structure prediction applications, in: BMC genomics, Vol.~13, BioMed Central, 2012, pp. 1--11.

\bibitem{wu_online_2017}
Q.~Wu, H.~Wu, X.~Zhou, M.~Tan, Y.~Xu, Y.~Yan, T.~Hao, Online {Transfer} {Learning} with {Multiple} {Homogeneous} or {Heterogeneous} {Sources}, IEEE Transactions on Knowledge and Data Engineering 29~(7) (2017) 1494--1507.
\newblock \href {https://doi.org/10.1109/TKDE.2017.2685597} {\path{doi:10.1109/TKDE.2017.2685597}}.

\bibitem{pujol_intelligent_2023}
V.~C. Pujol, A.~Morichetta, S.~Nastic, Intelligent {Sampling}: {A} {Novel} {Approach} to {Optimize} {Workload} {Scheduling} in {Large}-{Scale} {Heterogeneous} {Computing} {Continuum}, in: 2023 {IEEE} {International} {Conference} on {Service}-{Oriented} {System} {Engineering} ({SOSE}), 2023, pp. 140--149, iSSN: 2642-6587.
\newblock \href {https://doi.org/10.1109/SOSE58276.2023.00024} {\path{doi:10.1109/SOSE58276.2023.00024}}.

\bibitem{vagnoli_updating_2022}
M.~Vagnoli, R.~Remenyte-Prescott, Updating conditional probabilities of {Bayesian} belief networks by merging expert knowledge and system monitoring data, Automation in Construction 140 (2022) 104366.
\newblock \href {https://doi.org/10.1016/j.autcon.2022.104366} {\path{doi:10.1016/j.autcon.2022.104366}}.

\bibitem{vanis_novel_2023}
M.~Vaniš, Z.~Lokaj, M.~Šrotýř, A {Novel} {Algorithm} for {Merging} {Bayesian} {Networks}, Symmetry 15~(7) (2023) 1461.
\newblock \href {https://doi.org/10.3390/sym15071461} {\path{doi:10.3390/sym15071461}}.

\bibitem{murturi_decentralized_2022}
I.~Murturi, S.~Dustdar, A {Decentralized} {Approach} for {Resource} {Discovery} using {Metadata} {Replication} in {Edge} {Networks}, IEEE Transactions on Services Computing 15~(5) (2022) 2526--2537.
\newblock \href {https://doi.org/10.1109/TSC.2021.3082305} {\path{doi:10.1109/TSC.2021.3082305}}.

\bibitem{dustdar_towards_2020}
S.~Dustdar, I.~Murturi, Towards {Distributed} {Edge}-based {Systems}, in: 2020 {IEEE} {Second} {International} {Conference} on {Cognitive} {Machine} {Intelligence} ({CogMI}), IEEE, Atlanta, GA, USA, 2020, pp. 1--9.
\newblock \href {https://doi.org/10.1109/CogMI50398.2020.00021} {\path{doi:10.1109/CogMI50398.2020.00021}}.

\bibitem{sedlak_privacy_2023}
B.~Sedlak, I.~Murturi, P.~K. Donta, S.~Dustdar, A {Privacy} {Enforcing} {Framework} for {Transforming} {Data} {Streams} on the {Edge}, IEEE Transactions on Emerging Topics in Computing (2023).
\newblock \href {https://doi.org/10.1109/TETC.2023.3315131} {\path{doi:10.1109/TETC.2023.3315131}}.

\bibitem{ankan_pgmpy_2023}
A.~Ankan, J.~Textor, pgmpy: {A} {Python} {Toolkit} for {Bayesian} {Networks} (Apr. 2023).

\bibitem{zhang_comprehensive_2023}
Q.~Zhang, X.~Che, Y.~Chen, X.~Ma, M.~Xu, S.~Dustdar, X.~Liu, S.~Wang, A {Comprehensive} {Deep} {Learning} {Library} {Benchmark} and {Optimal} {Library} {Selection}, IEEE Transactions on Mobile Computing (2023) 1--14\href {https://doi.org/10.1109/TMC.2023.3301973} {\path{doi:10.1109/TMC.2023.3301973}}.

\bibitem{linzaer_ultra-light-fast-generic-face-detector-1mb_2022}
Linzaer, Ultra-{Light}-{Fast}-{Generic}-{Face}-{Detector}-{1MB}, https://github.com/Linzaer/Ultra-Light-Fast-Generic-Face-Detector-1MB (Feb. 2022).

\bibitem{rothe_dex_2015}
R.~Rothe, R.~Timofte, L.~V. Gool, {DEX}: {Deep} {EXpectation} of {Apparent} {Age} from a {Single} {Image}, in: 2015 {IEEE} {International} {Conference} on {Computer} {Vision} {Workshop} ({ICCVW}), IEEE, Santiago, Chile, 2015, pp. 252--257.
\newblock \href {https://doi.org/10.1109/ICCVW.2015.41} {\path{doi:10.1109/ICCVW.2015.41}}.

\bibitem{scutari_learning_2010}
M.~Scutari, Learning {Bayesian} {Networks} with the bnlearn {R} {Package}, Journal of Statistical Software 35 (2010) 1--22.
\newblock \href {https://doi.org/10.18637/jss.v035.i03} {\path{doi:10.18637/jss.v035.i03}}.

\bibitem{sudharsan_edge2train_2020}
B.~Sudharsan, J.~G. Breslin, M.~I. Ali, {Edge2Train}: a framework to train machine learning models ({SVMs}) on resource-constrained {IoT} edge devices, in: Proceedings of the 10th {International} {Conference} on the {Internet} of {Things}, {IoT} '20, Association for Computing Machinery, New York, NY, USA, 2020, pp. 1--8.
\newblock \href {https://doi.org/10.1145/3410992.3411014} {\path{doi:10.1145/3410992.3411014}}.

\bibitem{hao_monitoring_2023}
W.~Hao, Z.~Wang, L.~Hong, L.~Li, N.~Karayanni, C.~Mao, J.~Yang, A.~Cidon, Monitoring and {Adapting} {ML} {Models} on {Mobile} {Devices} (May 2023).
\newblock \href {https://doi.org/10.48550/arXiv.2305.07772} {\path{doi:10.48550/arXiv.2305.07772}}.

\bibitem{tariq_answering_2008}
M.~Tariq, A.~Zeitoun, V.~Valancius, N.~Feamster, M.~Ammar, Answering what-if deployment and configuration questions with wise, ACM SIGCOMM Computer Communication Review (2008).
\newblock \href {https://doi.org/10.1145/1402946.1402971} {\path{doi:10.1145/1402946.1402971}}.

\bibitem{goyal_hardware-friendly_2022}
V.~Goyal, R.~Das, V.~Bertacco, Hardware-friendly {User}-specific {Machine} {Learning} for {Edge} {Devices}, ACM Transactions on Embedded Computing Systems 21~(5) (2022) 62:1--62:29.
\newblock \href {https://doi.org/10.1145/3524125} {\path{doi:10.1145/3524125}}.

\bibitem{hsu_learning_2018}
Y.-C. Hsu, Z.~Lv, Z.~Kira, Learning to cluster in order to transfer across domains and tasks, in: Sixth {International} {Conference} on {Learning} {Representations} ({ICLR} 2018), 2018.
\newblock \href {https://doi.org/10.48550/arXiv.1711.10125} {\path{doi:10.48550/arXiv.1711.10125}}.

\bibitem{xing_enabling_2018}
T.~Xing, S.~S. Sandha, B.~Balaji, S.~Chakraborty, M.~Srivastava, Enabling {Edge} {Devices} that {Learn} from {Each} {Other}: {Cross} {Modal} {Training} for {Activity} {Recognition}, in: Proceedings of the 1st {International} {Workshop} on {Edge} {Systems}, {Analytics} and {Networking}, ACM, Munich Germany, 2018, pp. 37--42.
\newblock \href {https://doi.org/10.1145/3213344.3213351} {\path{doi:10.1145/3213344.3213351}}.

\bibitem{sharma_are_2018}
R.~Sharma, S.~Biookaghazadeh, M.~Zhao, Are {Existing} {Knowledge} {Transfer} {Techniques} {Effective} {For} {Deep} {Learning} on {Edge} {Devices}?, in: Proceedings of the 27th {International} {Symposium} on {High}-{Performance} {Parallel} and {Distributed} {Computing}, {HPDC} '18, Association for Computing Machinery, New York, NY, USA, 2018, pp. 15--16.
\newblock \href {https://doi.org/10.1145/3220192.3220459} {\path{doi:10.1145/3220192.3220459}}.

\bibitem{kolcun_case_2020}
R.~Kolcun, D.~A. Popescu, V.~Safronov, P.~Yadav, A.~M. Mandalari, Y.~Xie, R.~Mortier, H.~Haddadi, The {Case} for {Retraining} of {ML} {Models} for {IoT} {Device} {Identification} at the {Edge} (Nov. 2020).
\newblock \href {https://doi.org/10.48550/arXiv.2011.08605} {\path{doi:10.48550/arXiv.2011.08605}}.

\bibitem{nastic_sloc_2020}
S.~Nastic, A.~Morichetta, T.~Pusztai, S.~Dustdar, X.~Ding, D.~Vij, Y.~Xiong, {SLOC}: {Service} {Level} {Objectives} for {Next} {Generation} {Cloud} {Computing}, IEEE Internet Computing 24~(3) (May 2020).
\newblock \href {https://doi.org/10.1109/MIC.2020.2987739} {\path{doi:10.1109/MIC.2020.2987739}}.

\bibitem{furst_elastic_2018}
J.~Fürst, M.~Fadel~Argerich, B.~Cheng, A.~Papageorgiou, Elastic {Services} for {Edge} {Computing}, in: 2018 14th {International} {Conference} on {Network} and {Service} {Management} ({CNSM}), 2018, pp. 358--362.

\bibitem{TRAN2024304}
M.-N. Tran, Y.~Kim, Optimized resource usage with hybrid auto-scaling system for knative serverless edge computing, Future Generation Computer Systems 152 (2024) 304--316.
\newblock \href {https://doi.org/10.1016/j.future.2023.11.010} {\path{doi:10.1016/j.future.2023.11.010}}.

\bibitem{hazra2022cooperative}
A.~Hazra, P.~K. Donta, T.~Amgoth, S.~Dustdar, Cooperative transmission scheduling and computation offloading with collaboration of fog and cloud for industrial iot applications, IEEE Internet of Things Journal 10~(5) (2023) 3944--3953.

\bibitem{karjee_energy_2021}
J.~Karjee, S.~Praveen~Naik, N.~Srinidhi, Energy {Profiling} based {Load}-{Balancing} {Approach} in {IoT}-{Edge} for {Split} {Computing}, 2021 IEEE 18th India Council International Conference (INDICON) (2021) 1--6\href {https://doi.org/10.1109/INDICON52576.2021.9691607} {\path{doi:10.1109/INDICON52576.2021.9691607}}.

\bibitem{lim_load_2020}
J.~Lim, D.~Lee, A {Load} {Balancing} {Algorithm} for {Mobile} {Devices} in {Edge} {Cloud} {Computing} {Environments}, Electronics 9~(4) (2020) 686.
\newblock \href {https://doi.org/10.3390/electronics9040686} {\path{doi:10.3390/electronics9040686}}.

\bibitem{duan_novel_2022}
Z.~Duan, C.~Tian, N.~Zhang, M.~Zhou, B.~Yu, X.~Wang, J.~Guo, Y.~Wu, A novel load balancing scheme for mobile edge computing, Journal of Systems and Software 186 (2022) 111195.
\newblock \href {https://doi.org/10.1016/j.jss.2021.111195} {\path{doi:10.1016/j.jss.2021.111195}}.

\bibitem{rao_load_2004}
A.~Rao, K.~Lakshminarayanan, S.~Surana, R.~Karp, I.~Stoica, Load {Balancing} in {Structured} {P2P} {Systems}, Vol. 2735, 2004, journal Abbreviation: Lecture Notes in Computer Science Publication Title: Lecture Notes in Computer Science.
\newblock \href {https://doi.org/10.1007/978-3-540-45172-3_6} {\path{doi:10.1007/978-3-540-45172-3_6}}.

\bibitem{menino_novel_2021}
V.~H. Menino, A {Novel} {Approach} to {Load} {Balancing} in {P2P} {Overlay} {Networks} for {Edge} {Systems}, 2021.

\end{thebibliography}

\end{document}